\documentclass[%
 reprint,
 amsmath,amssymb,
 aps,
]{revtex4-2}

\usepackage{graphicx}% Include figure files
\usepackage{dcolumn}% Align table columns on decimal point
\usepackage{bm}% bold math
\usepackage{hyperref}% add hypertext capabilities
\usepackage[mathlines]{lineno}% Enable numbering of text and display math
\usepackage{lipsum}  
\usepackage{epsfig, amssymb, amsmath, amsthm, braket, physics, xcolor}
\usepackage{cleveref}
\usepackage[normalem]{ulem}
\usepackage{subcaption} % Required for creating subfigures
\usepackage{mathtools}
\newtheorem{theorem}{Theorem}

\newtheorem{lemma}[theorem]{Lemma}

\theoremstyle{definition}
\newtheorem{observation}{Observation}
\begin{document}

\preprint{APS/123-QED}

\title{Adiabatic Dynamics of Entanglement}
\author{Einar Gabbassov$^{1,3,4}$}
\email{egabbass@uwaterloo.ca}
\author{Achim Kempf$^{1,2,3,4}$}

\affiliation{$^1$Department of Applied Mathematics, University of Waterloo, Waterloo, ON N2L 3G1, Canada}

\affiliation{$^2$Department of Physics, University of Waterloo, Waterloo, ON N2L 3G1, Canada}

\affiliation{$^3$Perimeter Institute for Theoretical Physics, Waterloo, ON N2L 2Y5, Canada}

\affiliation{$^4$ Institute for Quantum Computing, University of Waterloo, Waterloo, ON N2L 3G1, Canada}
%\date{}% It is always \today, today,

\begin{abstract}
We show that, during adiabatic evolution generated by a time-dependent Hamiltonian, any changes in entanglement can be attributed to a succession of avoided crossings in the Hamiltonian’s instantaneous energy spectrum, at which its eigenvalues swap their associated eigenvectors. These swaps mediate the generation and redistribution of entanglement in multipartite systems. The efficiency of this redistribution depends on the narrowness of the avoided level crossings and thus constrains the speed of adiabatic evolution. Moreover, we relate the ruggedness of the energy landscape, which directly affects the hardness of a computational problem, to the extent to which entanglement must be redistributed among subsystems. This enables an analysis of computational complexity and quantum advantage in terms of the required entanglement redistribution. Applied to adiabatic quantum computation, our findings directly relate the computation’s speed to its utilization of entanglement as a resource. The same principles extend to gate-based discretized adiabatic quantum algorithms, including those for Hamiltonian simulation and combinatorial optimization.
\end{abstract}

\maketitle

%\tableofcontents

\section{\label{sec:Intro}Introduction}
Entanglement, a phenomenon without a classical analog, is an essential resource for quantum technologies. In this work, we analyze its dynamics during adiabatic evolution and show how this resource is generated and redistributed within a multipartite quantum system.

We find that any change in entanglement during adiabatic evolution can be attributed to avoided energy level crossings where eigenvalues swap their associated eigenvectors. We show that these exchanges constitute the fundamental mechanism responsible for modifying the system’s entanglement structure. Furthermore, the extent of these changes depends on the narrowness of the avoided crossing: the narrower it is, the more significant the entanglement change. This, in turn, constrains the permissible speed of the adiabatic evolution, which is limited by the size of the spectral gap at such narrowly avoided crossings. Therefore, our results provide new insights into the relationship between an adiabatic quantum computation's maximum speed and its use of entanglement.

We also relate the amount of entanglement required to the ruggedness of the energy landscape, which affects the hardness of a computational problem. Rugged landscapes are characterized by many well-separated low-energy local minima and typically correspond to classically hard problems. In such cases, the ground state must explore and coherently superpose computational basis states that are widely separated in Hamming distance. This leads to intermediate quantum states with high multipartite entanglement. For example, in optimization problems where local minima are far apart, the adiabatic evolution must transiently pass through superpositions of these distant configurations, thereby generating and redistributing entanglement across many subsystems. These entanglement manipulations are governed by the mechanics outlined above: eigenvector swaps at narrowly avoided level crossings.

These findings are applicable to adiabatic quantum computing \cite{born1928beweis, kato1950adiabatic, amin2009consistency, farhi2000quantum, albash2018adiabatic, amin2008thermally}, a model known to be equivalent to standard gate-based quantum computation up to a polynomial overhead. Also, this work extends naturally to gate-based digitized adiabatic algorithms such as the quantum approximate optimization algorithm (QAOA) \cite{farhi2014quantum, bucher2025efficient} and the discretized adiabatic quantum algorithm (DAQC) used in Hamiltonian simulations \cite{gabbassov2025lagrangian, granet2025practicality, hatomura2025benchmarking, granet2024hamiltonian}. Consequently, any change in entanglement and associated computational slowdown in continuous evolution will also manifest in the gate-based counterparts.

\subsection{Setup}
In order to analyze the basic mechanics underlying the generation and redistribution of entanglement during adiabatic evolution, we consider a setup consisting of three systems, $\tilde{A}$, $A$, and $B$. This tripartite configuration provides the minimal and general framework for analyzing entanglement transfer using the formalism of quantum channels, which capture how entanglement is redistributed between subsystems. Two systems suffice to generate or destroy entanglement; three systems are required to meaningfully track its redistribution.

\begin{figure}[h]
    \centering
    \includegraphics[scale=0.66]{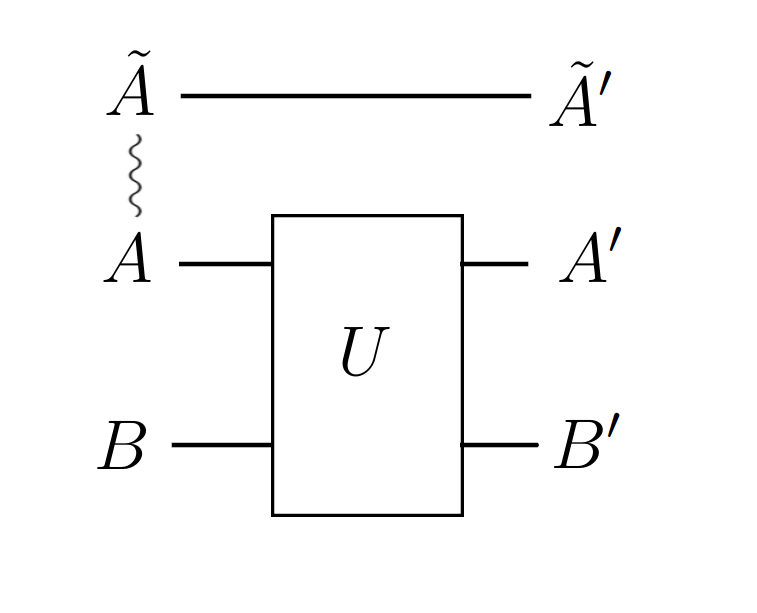}
    \caption{Systems $\tilde{A}$ and $A$ are prepared in an entangled state. The time evolution operator, $U$, is adiabatically generated by a Hamiltonian which acts nontrivially only on system $AB$.}
    \label{fig:interaction_setup}
\end{figure}

We take the Hilbert space of these systems to be finite-dimensional, which is justified as long as the systems possess finite size and a finite energy budget. Systems $\tilde{A}$ and $A$ are assumed to be initially entangled. During the subsequent unitary time evolution, systems $A$ and $B$ interact adiabatically while system $\tilde{A}$ remains a non-interacting ancilla, as illustrated in \Cref{fig:interaction_setup}. This setup allows us to analyze how the interaction generates entanglement between $A$ and $B$, and also how $\tilde{A}$, which was initially entangled with $A$, can become entangled with system $B$ and/or with system $AB$. To this end, we will consider a quantum channel from $A$ to $A'$ and the complementary channel from $A$ to $B'$, which together formalize how entanglement is redistributed during the evolution. We will use R\'enyi entropy-based coherent information to analytically quantify the extent of entanglement redistribution among partitions. 

Intuitively, a tripartite setup captures the redistribution of entanglement between subsystems, or the loss of entanglement due to interaction with an external environment. For example, $\tilde{A}$, $A$, and $B$ could each be qubit systems, where $B$ acts as a target that acquires entanglement from $A$, which is initially entangled with $\tilde{A}$. Alternatively, $\tilde{A}$ and $A$ could form an entangled pair, with $B$ representing an environment that extracts entanglement from $\tilde{A}A$ through its interaction with $A$.

\subsection{Methods}\label{sec:methods}
We consider the dynamics generated by a generic Hamiltonian of the form $H_0+g(t) H_{int}$, where the function $g(t)$ is called an \textit{adiabatic schedule}  that increases from zero to $g(T)>0$ slowly enough for the evolution to be adiabatic. Our overall task is to understand the evolution of the entanglement in the system $\tilde{A}AB$. For this, we will decompose the evolution into a sequence of more manageable segments by employing the spectral and adiabatic theorems. First, we use the spectral theorem to decompose $H_{int}$ into rank-one projectors, $H_{int}=\sum_n \lambda_n \ketbra{v_n}$. We then reschedule the adiabatic evolution by sequentially adiabatically adding, or `activating', these projectors. We start with $H_0$ and then one by one we adiabatically add $\mu_n \vert v_n\rangle\langle v_n\vert$ by 
dialing the prefactor $\mu_n$ of each projector $\ketbra{v_n}$ from $\mu_n=0$ to $\mu_n=g(T)\lambda_n$. In order to arrive at the final Hamiltonian $H_0 + g(T) H_{int}$, 
each of these adiabatic activations of a rank-one projector takes the form 
\begin{equation} H(\mu) = D + \mu \ketbra{v} \label{Deq}
\end{equation} 
where $D$ is the Hamiltonian incorporating previous activations, $\ketbra{v}:=\ketbra{v_n}$ is the projector that is next to be activated, and $\mu:=\mu_n$ is a schedule function adiabatically swept from the initial $\mu=0$ to its final value $\mu=g(T)\lambda_n$.

This technique may be interpreted as a piecewise scheduling protocol, in which rank-one projectors are sequentially activated and remain active thereafter. The cumulative effect of these activations reconstructs the full interaction Hamiltonian, thereby faithfully implementing the complete adiabatic evolution. Although this schedule is physically sound and captures the full dynamics of the process, it is not intended as a practical computational scheme. Rather, it serves as a mathematically convenient framework for deriving theoretical insights into the behaviour of entanglement.

Having introduced the sequential activations of projectors, we can study the evolution of entanglement within a single activation. Specifically, for each activation of a projector, we track the evolution of the eigenvalues $s_n(\mu)$ and eigenvectors $\vert s_n(\mu)\rangle$ in $H(\mu)\vert s_n(\mu)\rangle = s_n(\mu)\vert s_n(\mu)\rangle$ as $\mu$ is dialed in from $0$ to its final value. 
This is a useful decomposition because it allows us to recruit the recent results in \cite{vsoda2022newton}, which nonperturbatively describe the evolution of the eigenvalues \(s_n(\mu)\) and eigenstates \(|s_n(\mu)\rangle\) of Hermitian operators $H(\mu)$ for all $D$ and $\vert v\rangle$, in finite-dimensional Hilbert spaces.

By applying these methods to the interactions in the multipartite system $\tilde{A}, A, B$, we show that, during each activation of a new projector, the generation and transfer of entanglement decompose into a succession of avoided level crossings at which, crucially, two eigenvalues exchange eigenvectors. These eigenvector exchanges constitute the fundamental mechanism by which entanglement is generated and redistributed.

\subsection{Structure of the paper}

\begin{itemize}
    \item \Cref{sec:results_intuition} provides an intuitive overview of key findings, highlighting conceptual points and using visualization. 

    \item \Cref{sec:background} introduces several lemmas that serve as basic mathematical preliminaries.
    
    \item \Cref{sec:ent_trans} establishes conditions for maximizing the transfer or preservation of entanglement during the adiabatic interactions of $A$ and $B$.
    
    \item \Cref{sec:main_theorems,sec:permutations} demonstrate that complete entanglement transfer, or preservation, induces a clearly structured  evolution of the eigenspectrum and eigenvectors of the total time-dependent Hamiltonian of the system $\tilde{A}AB$. Further, a link is established between the entanglement dynamics and the permutation group.
    
    \item \Cref{sec:eigengap} establishes a direct connection between the maximum speed of adiabatic quantum computation and the generation and transfer of entanglement. Specifically, it relates the narrowness of avoided level crossings to the efficiency of entanglement dynamics during the evolution.       
    \item \Cref{sec:AQC_Optimization} shows that the computational hardness of, for example, problems of combinatorial optimization, in the sense of the ruggedness of its cost function landscape and the corresponding Hamming distance of low energy candidate solutions, can be directly translated into the need for large amounts of temporary multi-partite entanglement during the corresponding adiabatic computation.  
    \item \Cref{sec:entanglement_measures} introduces and applies measures of coherent information that are useful for non-perturbative quantifying the transfer of entanglement among subsystems.
\end{itemize}

\section{Intuitive overview of the results \label{sec:results_intuition}}
As indicated above, our strategy is to decompose a given generic adiabatic dynamics into a succession of adiabatic `activations' of rank one projectors, and then for each activation to track the evolution of the eigenvectors and eigenvalues. Here, each activation is described by Hamiltonian of the form $H(\mu) = D + \mu \ketbra{v}$, where $\mu$ is adiabatically dialed from $\mu=0$ to its final value $\mu=g(T)\lambda_n$, and  $D$ is the Hamiltonian at the end of the prior activation. With the notation $D\vert d_n\rangle = d_n\vert d_n\rangle$, we have $d_n=s_n(0)$ and $\vert d_n\rangle = \vert s_n(0)\rangle$. 

During each of these activations, i.e., as a $\mu$ is adiabatically swept from $\mu=0$ to its final value $\mu =g(T)\lambda_n$, a succession of avoided level crossings occurs, see \cite{vsoda2022newton}. As we show in the present paper, if the system is subdivided into subsystems, it is at these avoided level crossings that the generation and transfer of entanglement among the subsystems occurs. This is because at each avoided level crossing, a pair of eigenvalues swaps their eigenvectors. 

These key dynamics are most clearly seen in those cases where \(|v\rangle\) is not parallel but almost parallel to an initial eigenvector $\vert s_n(0)\rangle =\ket{d_n}$ of the initial Hamiltonian $D$.
In these cases, which we will call \emph{edge cases}, any changes of the entanglement among \(\tilde{A}\), \(A\), and \(B\) while adiabatically sweeping $\mu$ from $\mu=0$ to $\mu=g(T)\lambda_n$ always occur suddenly at specific values of $\mu$ that we will call \emph{critical $\mu$ values}. 
It is at these critical $\mu$ values, two energy eigenvalues, say \(s_n(\mu)\) and \(s_{n+1}(\mu)\) narrowly avoid each other and swap their eigenvectors, \(|s_n(\mu)\rangle\) and \(|s_{n+1}(\mu)\rangle\), thereby changing the structure of entanglement. In the non-edge cases, characterized by generic $\ket{v}$, the behaviour remains qualitatively the same, except that the level crossings are avoided with greater gaps, and the eigenvector swaps are less sudden.

The fact that in edge cases, i.e., for narrow gaps, each involved eigenvector evolves into an almost orthogonal and therefore very different eigenvector over a very small range of values of $\mu$ implies that these $\mu$-values need to be traversed particularly slowly for the evolution to remain adiabatic. When the system is a multi-partite system then, as we will show, this implies a relation between the narrowness of avoided level crossings and the efficiency of entanglement transfers. In this way, an adiabatic quantum computation's utilization of the resource of entanglement can be related to the maximum speed of that computation.    

Before we present the mathematical details, it will be useful to establish intuition and visualizations for the dynamics of the generation and redistribution of entanglement that we discussed above. To this end, we will (A) visualize the dynamics of eigenvalues and eigenvectors under adiabatic activation for a total system, (B) apply this to visualize the generation of entanglement among subsystems $A$ and $B$, and finally (C) visualize the transfer of entanglement among subsystems by visualizing how system $\tilde{A}$, which was initially entangled with $A$ can become entangled with $B$ or $AB$ instead.     

\subsection{Visualization of the dynamics of eigenvectors and eigenvalues during an adiabatic activation}
%\subsection{Single systems}
As outlined above, we decompose the general problem into segments of adiabatic evolution, each of which is described by a Hamiltonian of the form of (\ref{Deq}). 
%\begin{equation}
%H(\mu) = D + \mu \ketbra{v}
%\end{equation}
%Here, $D$ is an arbitrary non-degenerate starting Hamiltonian, to which is added a multiple of a rank-one projector $\ketbra{v}$ by adiabatically sweeping the value of $\mu$. 
%We denote the eigenvalues and eigenvectors of $D$ and $H(\mu)$ by $d_k$, $\vert d_k\rangle$ and $s_k(\mu),\vert s_k(\mu)\rangle$ respectively, i.e., we have $d_k=s_k(0)$ and $\vert d_k\rangle=\vert s_k(0)\rangle$.  

\begin{figure}[t]
    \centering
     \includegraphics[scale=0.35]{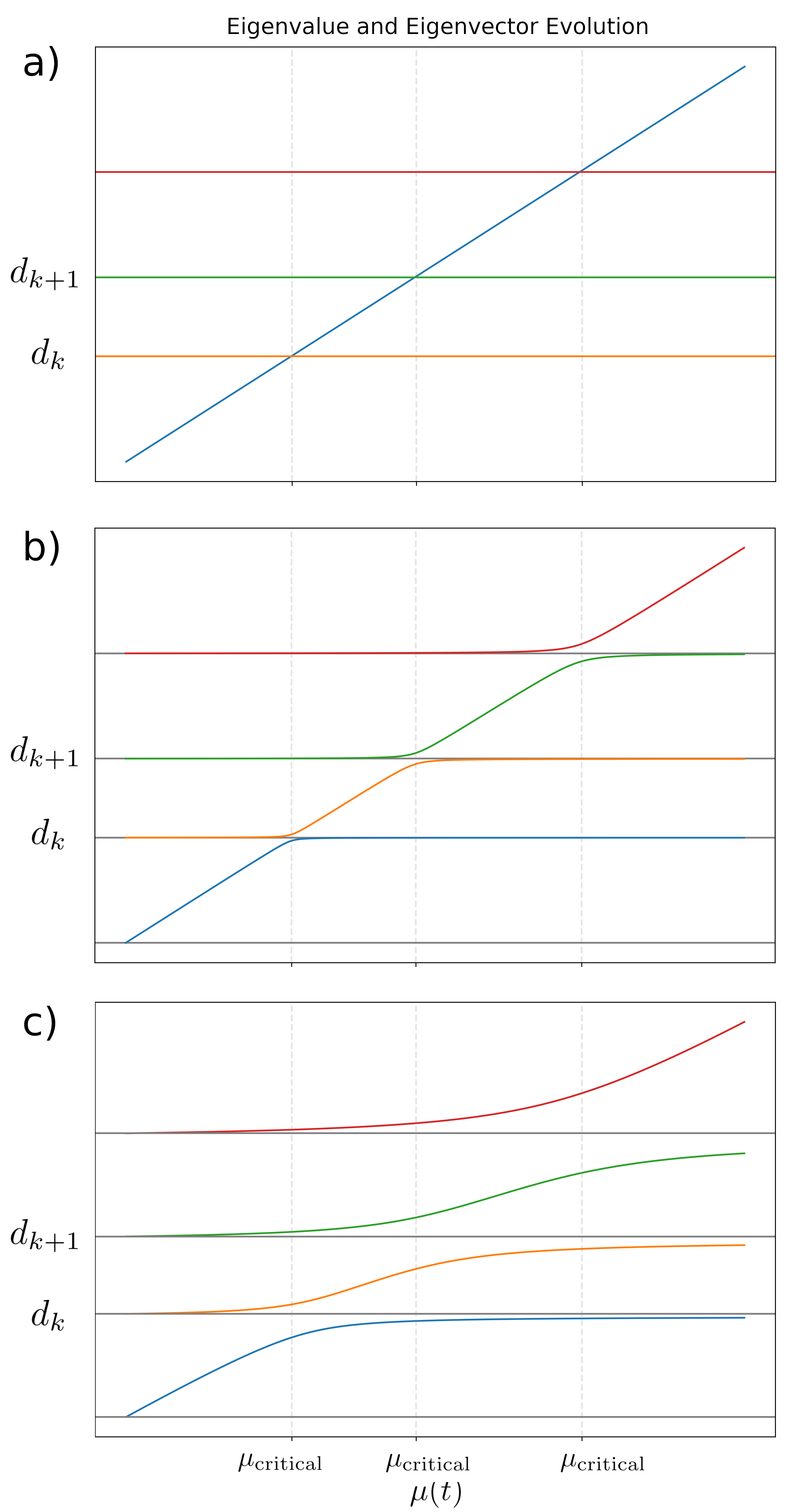}
    \caption{The dynamics of the eigenvalues of $H(\mu)$ for a generic choice of $D$ and with $\vert v\rangle$ chosen to be a) a trivial case, b) an edge case and c) a non-edge case. The vertical grey bars indicate where avoided level crossings are.}   
%    of three cases:}     
%a) A trivial case, where $\vert v\rangle=\vert d_i\rangle$ for some $i$. With increasing $\mu$, only the eigenvalue $d_{i}$ moves, crossing all other energy levels. b) An edge case, i.e., $\vert v\rangle \approx \vert d_i\rangle$. As $\mu$ runs through the reals, all level crossings are avoided and each eigenvalue exhibits a three-stage evolution: stasis, quick smooth transition to linear growth, and quick smooth transition to stasis. c) A non-edge case. The level crossings are less narrowly avoided.}
    \label{fig:one_partite_fig_2}
\end{figure}

We begin by visualizing the dynamics of the eigenvalues. 
%as $\mu$ runs through all reals. 
In the simplest case, $\vert v\rangle$ equals one of the eigenvectors of $D$, i.e., $\vert v\rangle = \vert d_i\rangle=\vert s_i(0)\rangle$. The adiabatic evolution is trivial in this case, as all eigenvectors stay the same and also all eigenvalues are constant, $s_n(\mu)=s_n(0)=d_n$, except for $n=i$ where we have $s_i(\mu)=s_i(0)+\mu$. Since $s_i(\mu)$ is the only eigenvalue that changes, it crosses the levels of all other eigenvalues, see the blue line in \Cref{fig:one_partite_fig_2}.a.

Now let us consider what we call an edge case, i.e., a case where $\vert v\rangle$ is not equal to but close to one of the eigenvectors of $D$, i.e.: $\vert v\rangle \approx \vert d_i\rangle=\vert s_i(0)\rangle$ for a fixed $i$. The adiabatic behaviour is now very different: as illustrated in \Cref{fig:one_partite_fig_2}.b, all level crossings are narrowly avoided as $\mu$ runs through the reals. We call the values of $\mu$ at which these level crossing avoidances occur the \emph{critical $\mu$ values}. The eigenvalue that is increasing while $\mu$ runs through $\mu=0$ is the eigenvalue $s_i(\mu)$. 

The non-edge cases, i.e., when $\vert v\rangle$ is generic and not close to an eigenvector of $D$, the qualitative behaviour is similar, see \cite{vsoda2022newton}: as $\mu$ runs from $\mu=-\infty$ to $\mu=+\infty$, all eigenvalues $s_k(\mu)$ strictly monotonically increase and all level crossings are avoided, though generally less narrowly. This behaviour is illustrated in \Cref{fig:one_partite_fig_2}.c.
% Each eigenvalue converges to the starting value of the next higher eigenvalue $s_{k+1}(\mu)$.  

We now address the dynamics of the eigenvectors $\vert s_k(\mu)\rangle$. 
In the trivial cases, as in \Cref{fig:one_partite_fig_2}.a, the eigenvectors do not change. In the edge cases, $\vert v\rangle \approx \vert d_i\rangle=\vert s_i(0)\rangle$ for a fixed $i$, as illustrated in \Cref{fig:one_partite_fig_2}.b, let us consider the $k$'th eigenvalue, $d_k=s_k(0)$ of $D$. Initially, as $\mu$ is running from $\mu=0$ to larger values, $s_k(\mu)$ and its eigenvector remain approximately constant: $s_k(\mu)\approx s_k(0)= d_k$ and $\vert s_k(\mu)\rangle\approx \vert s_k(0)\rangle = \vert d_k\rangle$. Then, at a critical $\mu$ value, the eigenvalue $s_k(\mu)$ starts to grow, see \Cref{fig:one_partite_fig_2}.b. At the same time, at this critical $\mu$ value, the eigenvector $\vert s_k(\mu)\rangle$ changes quickly from $\vert s_k(\mu)\rangle\approx \vert s_k(0)\rangle=\vert d_k\rangle$ to $\vert s_k(\mu)\rangle\approx \vert v \rangle$. 

\begin{figure}[t]
    \centering
    \includegraphics[scale=0.35]{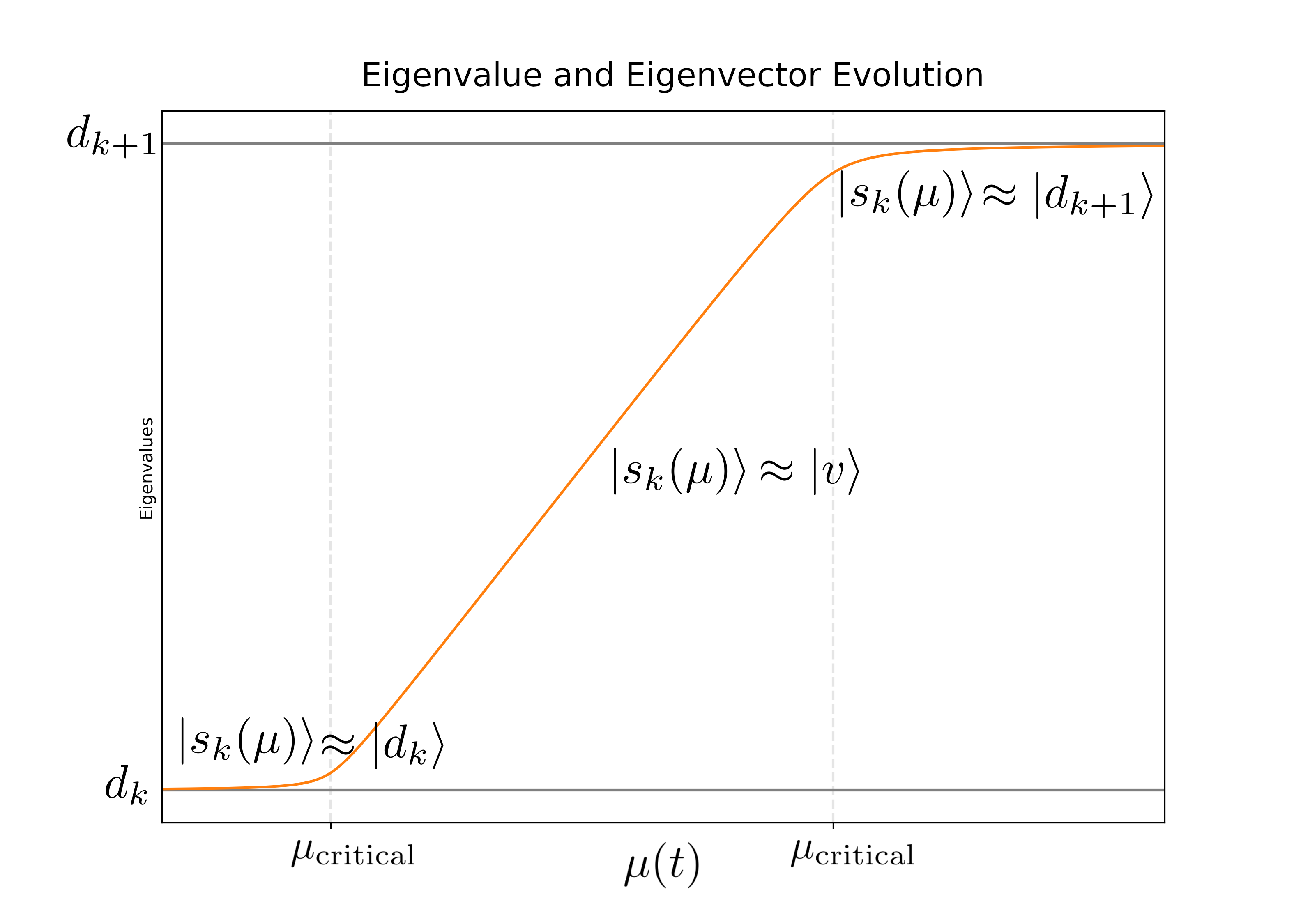}
    \caption{The evolution of the eigenvalue $s_k(\mu)$ and eigenvector $\vert s_k(\mu)\rangle$ as a function of the parameter $\mu$. Annotations along the eigenvalue curve specify the eigenvector corresponding to $s_k(\mu)$. 
    As $\mu$ runs from left to right, it passes two critical $\mu$ values, the eigenvector $\vert s_k(\mu)\rangle$ first transitions from $\vert s_k(\mu)\rangle \approx \vert s_k(0)\rangle=\ket{d_k}$ to $\vert s_k(\mu)\rangle \approx\ket{v}$ and then to $\vert s_k(\mu)\rangle \approx\ket{d_{k+1}}$.}
    \label{solevec}
\end{figure}

As $\mu$ grows further it eventually reaches the next critical $\mu$ value, where the eigenvalue $s_k(\mu)$ changes its eigenvector again, now from $\vert s_k(\mu)\rangle\approx \vert v \rangle$ to $\vert s_k(\mu)\rangle\approx \vert s_{k+1}(0) \rangle=\vert d_{k+1}\rangle$. This is illustrated in \Cref{solevec}. Notice that for $k=i$, we have $\vert s_i(\mu)\rangle \approx \vert v\rangle$ as $\mu$ passes through $\mu=0$. 

\begin{figure}[t]
    \centering
    \includegraphics[scale=0.68]{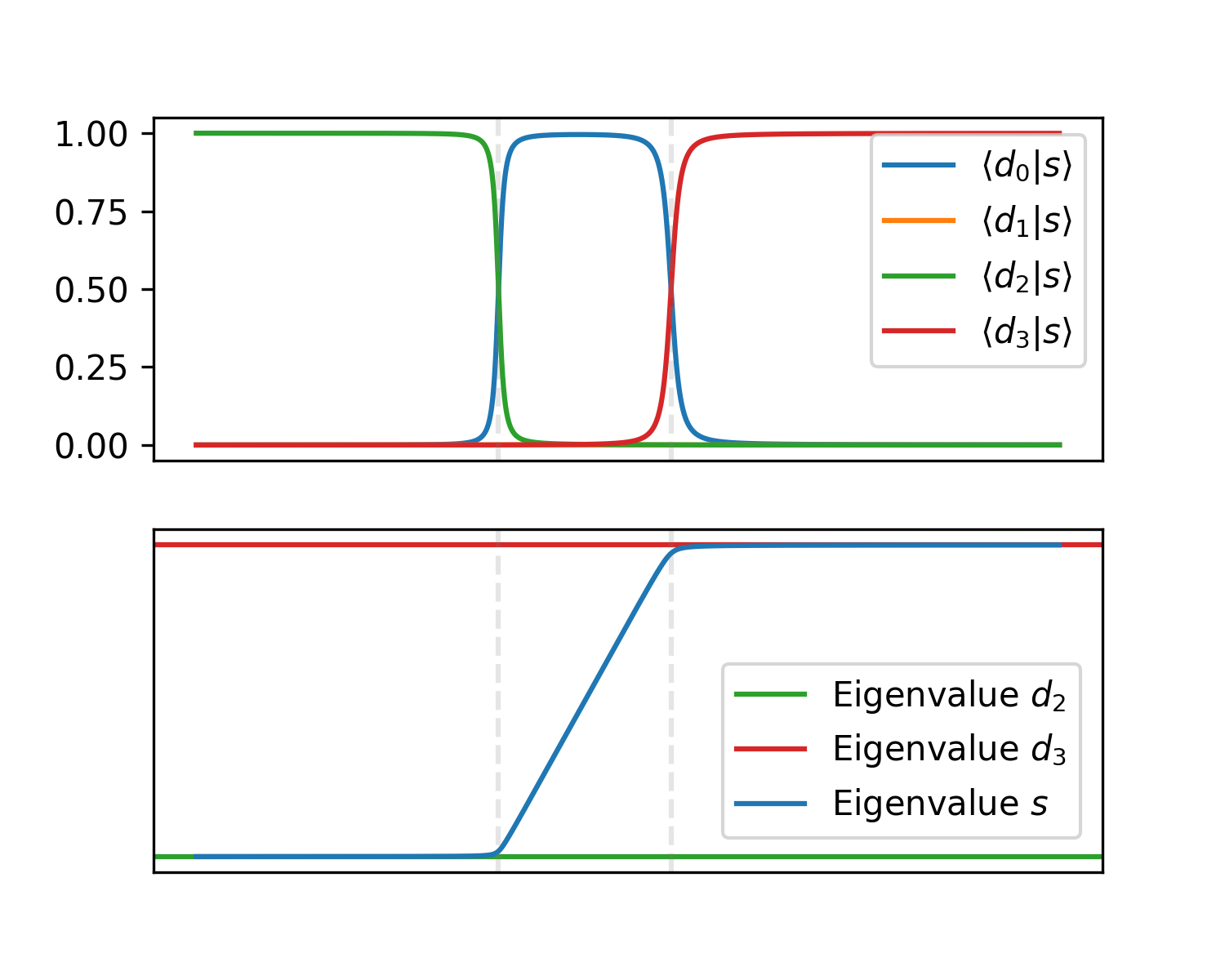}
    \caption{Example of an edge case with $\ket{v} \approx \ket{d_0}$, illustrating the eigenvalue $s_k(\mu)$ changing its eigenvector $\vert s_k(\mu)\rangle$ when passing through critical $\mu$ values (indicated by grey vertical dashed lines). }
    \label{solevec2}
\end{figure}

\begin{figure}[t]
    \centering
    \includegraphics[scale=0.35]{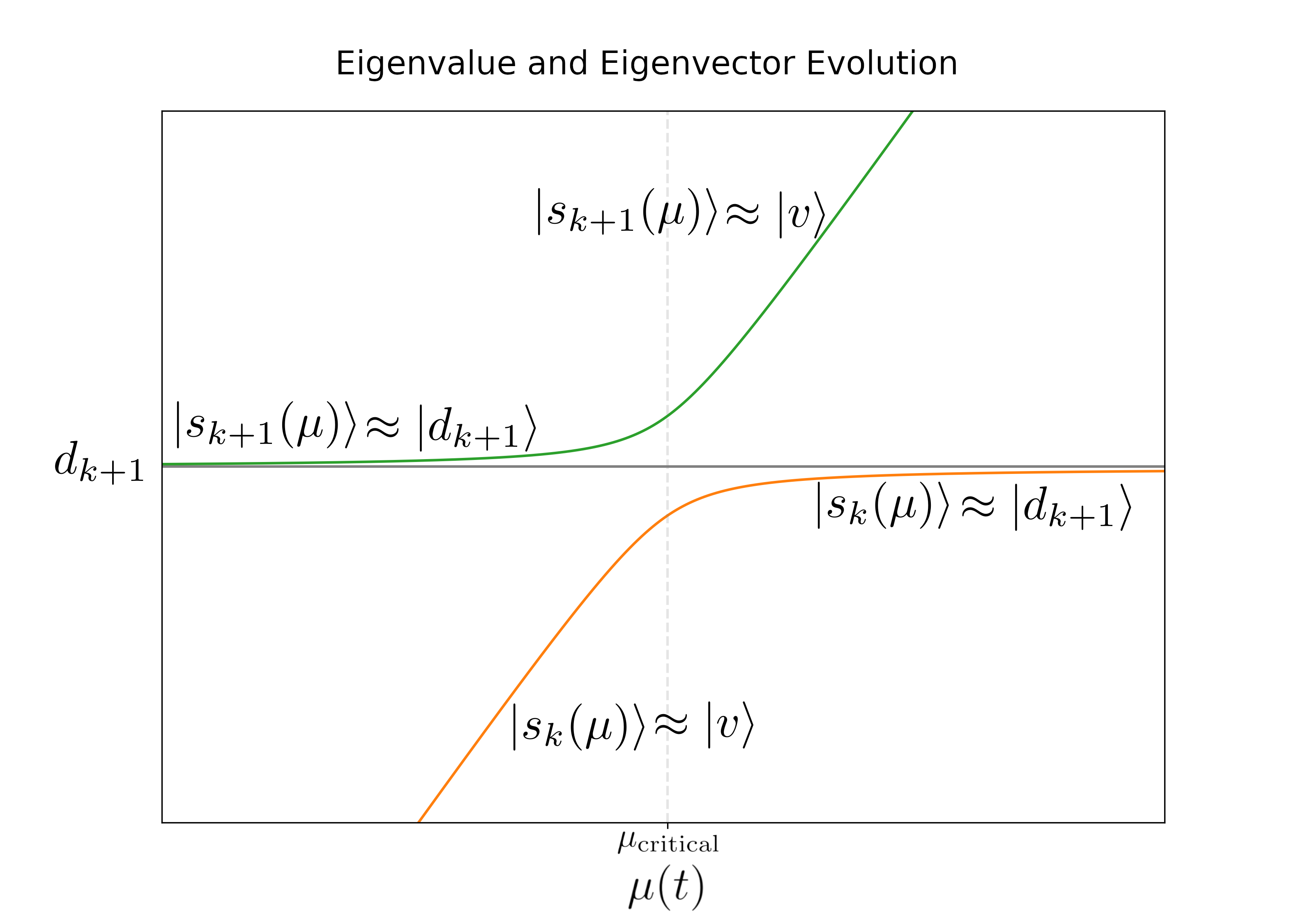}
    \caption{At a critical $\mu$ value, the eigenvalues $s_k(\mu)$ and $s_{k+1}(\mu)$ of $H(\mu)=D+\mu \vert v\rangle\langle v\vert$ trade their eigenvectors.}
    \label{two-vecs}
\end{figure}

The suddenness with which an eigenvalue changes its eigenvector as $\mu$ moves through a critical point is illustrated in \Cref{solevec2}.
The larger picture is, as shown in \Cref{fig:one_partite_fig_2}, that at each critical $\mu$ value two eigenvalues narrowly avoid crossing. \Cref{two-vecs} visualizes what this means for the eigenvectors $\vert s_k(\mu)\rangle$ and $\vert s_{k+1}(\mu)\rangle$ of two eigenvalues $s_k(\mu)$ and $s_{k+1}(\mu)$ at a narrow crossing avoidance: the eigenvalues quickly trade their eigenvectors. 

In later sections, we will show that the extent of crossing avoidance and suddenness of eigenvector change is determined by a single parameter $\epsilon$.
In non-edge cases, as in \Cref{fig:one_partite_fig_2}.c, the situation is qualitatively the same. All level crossings are avoided, and the eigenvalues partially trade their eigenvectors, although the level crossings can be avoided less narrowly and can also occur at values other than the eigenvalues $d_k$. Further, during a phase in which an eigenvalue $s_n(\mu)$ quickly grows, its eigenvector $\vert s_n(\mu)\rangle$ need no longer approximate $\vert v\rangle$. See \cite{vsoda2022newton} for the details. In comparison, the edge cases are characterized by very narrow level crossing avoidances that occur suddenly, and the associated eigenvector swaps are correspondingly sudden. Hence, the edge cases offer sharper tools for analyzing or controlling entanglement dynamics. We will, therefore, pay particular attention to the edge cases.

\subsection{Visualization of entanglement generation in the bi-partite system $AB$}
By applying these findings to a system that is divided into two subsystems, $A$ and $B$, we can track the dynamics of the entanglement between $A$ and $B$ during adiabatic evolution. The analysis can be applied to any activation of a projector, i.e., with any starting Hamiltonian $D$. However, for the purpose of gaining intuition, it is convenient to consider an initial Hamiltonian $D$ that is of the typical form of a free Hamiltonian, $D=H_A\otimes I + I\otimes H_B$. We denote the spectra of $H_A$ and $H_B$ by $\{a_i\},\{b_j\}$ and we assume them to be generic, i.e., such that the spectrum of $D$ is non-degenerate. We denote the eigenbases of $A$ and $B$ by $\{\vert a_i\rangle \},\{\vert b_j\rangle\}$.  

\begin{figure}[t]
    \centering
    \includegraphics[scale=0.35]{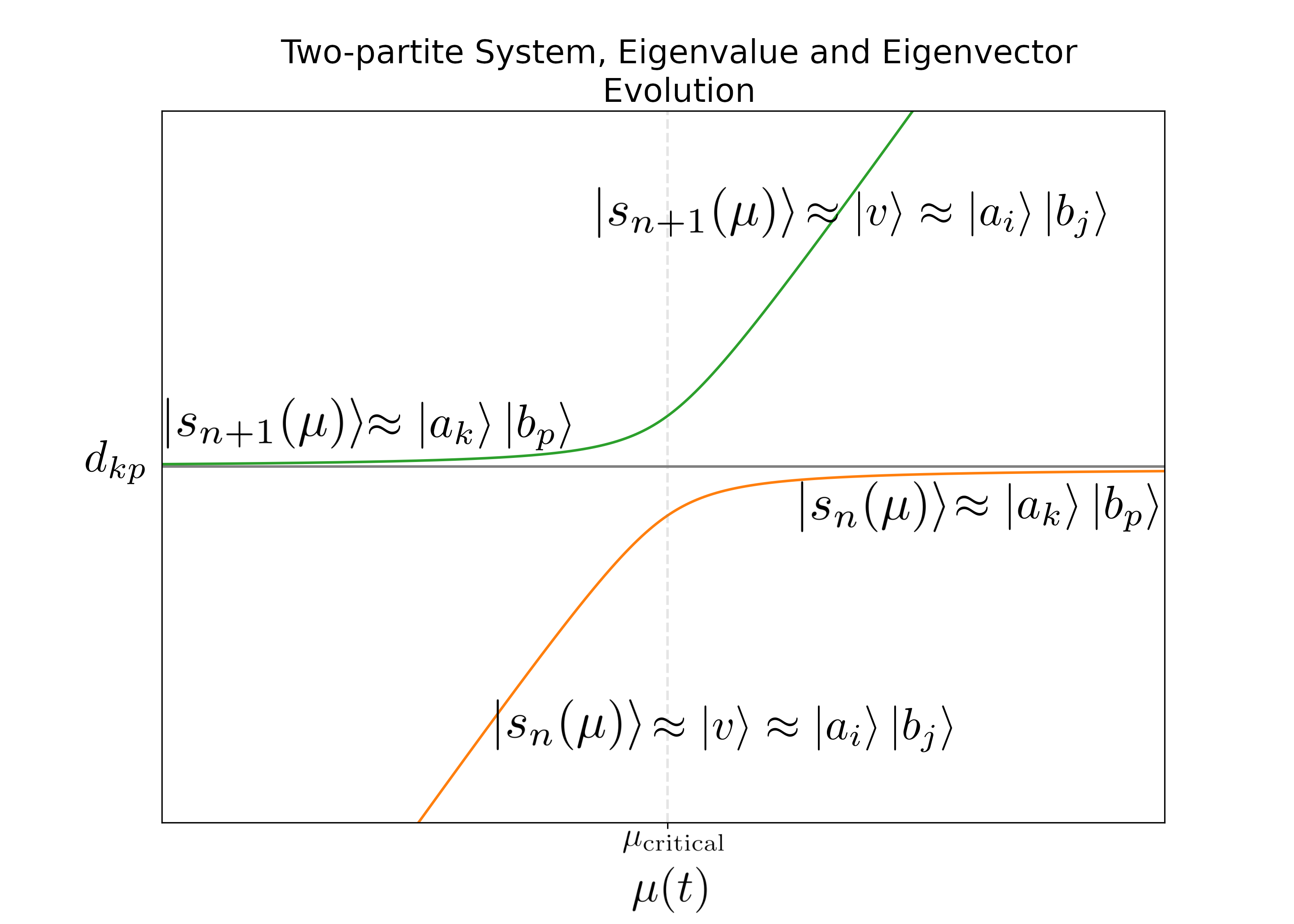}
    \caption{Eigenvector swap at a critical $\mu$ value, in a two-partite system $AB$.}
   \label{two_partite1}
\end{figure}

Further, when activating a projector $\ketbra{v}$,  we will, for clarity, focus on edge cases, $\vert v\rangle \approx \vert a_i\rangle\vert b_j\rangle$, for fixed $i$ and $j$. The dynamics at each critical $\mu$ value can then be deduced from \Cref{two-vecs}: as two neighbouring energy eigenvalues narrowly avoid crossing at a critical $\mu$ value, they trade their eigenvectors: prior to reaching the critical $\mu$ value, the eigenvector $\vert s_k(\mu)\rangle$ of the eigenvalue $s_k(\mu)$ is $\vert s_k(\mu)\rangle \approx\ket{v}$, while the eigenvector $\vert s_{k+1}(\mu)\rangle$ associated with $s_{k+1}(\mu)$ is $\vert s_{k+1}(\mu)\rangle\approx \ket{d_{k+1}}$. Passing the critical $\mu$ value, the eigenvectors switch: the eigenvector $\vert s_k(\mu)\rangle$ belonging to the eigenvalue $s_k(\mu)$ transitions to $\vert s_k(\mu)\rangle\approx \ket{d_{k+1}}$, and the eigenvector $\vert s_{k+1}(\mu)\rangle$ belonging to the eigenvalue $s_{k+1}(\mu)$ transitions to $\vert s_{k+1}(\mu)\rangle \approx \ket{v}$. The application to the two-partite system $AB$ is illustrated in \Cref{two_partite1},
where we can read off two observations on the dynamics of the entanglement in system $AB$: 
\begin{observation}\label{observation_1}
Let us consider the edge case where $\ket{v}$ is chosen to be almost parallel to the unentangled eigenstate of the initial Hamiltonian; i.e. $\ket{v} \approx \ket{a_i}\ket{b_j}$. Suppose the system is initialized in an unentangled eigenstate $\ket{\psi(0)} = \ket{a_k}\ket{b_p}$ of the initial Hamiltonian $D$. As $\mu$ is adiabatically increased, this state evolves into $\ket{\psi(\mu)} \approx \ket{v}$, which is also a product state. Eventually, the system encounters an avoided level crossing at the critical value $\mu_{\text{critical}}$. Precisely at this value, the system's state becomes a superposition of $\ket{v} \approx \ket{a_i}\ket{b_j}$ and another product eigenstate, say $\ket{a_{k+1}}\ket{b_p}$, associated with the nearby avoided eigenvalue. Therefore, at $\mu_{\text{critical}}$ the state of the system is an entangled state. This otherwise temporary entanglement can be made permanent by freezing the value of $\mu$ at the critical value.
\end{observation}
\begin{observation}\label{observation_2}
    In edge cases, any initially unentangled state that is not an energy eigenstate can become entangled during an avoided level crossing, and this entanglement persists after the avoided level crossing. \rm Assume, e.g., that system $AB$ was prepared in an unentangled state that is a superposition of energy eigenstates, such as the state $(\vert a_{k}\rangle+\vert a_m\rangle)\vert b_{p}\rangle$. During the time evolution, at a critical $\mu$ value, this state can become permanently entangled as, e.g., the state $\vert a_{k}\rangle\vert b_{p}\rangle$ can be swapped, for example, for the state $\vert a_k\rangle\vert b_{p+1}\rangle$, so that after the avoided level crossing our system is in the entangled state $\vert a_k\rangle\vert b_{p+1}\rangle + \vert a_m\rangle\vert b_{p}\rangle$. Unlike in the case of Observation 1 above, here the newly generated entanglement between $A$ and $B$ persists for values of $\mu$ larger than the critical $\mu$. In this sense, entanglement between $A$ and $B$ can be said to be redistributed during an eigenvector swap at an avoided level crossing.  
\end{observation}

\subsection{Visualization of entanglement transfer in the tri-partite system $\tilde{A}AB$}
\begin{figure}[t]
    \centering
    \includegraphics[scale=0.35]{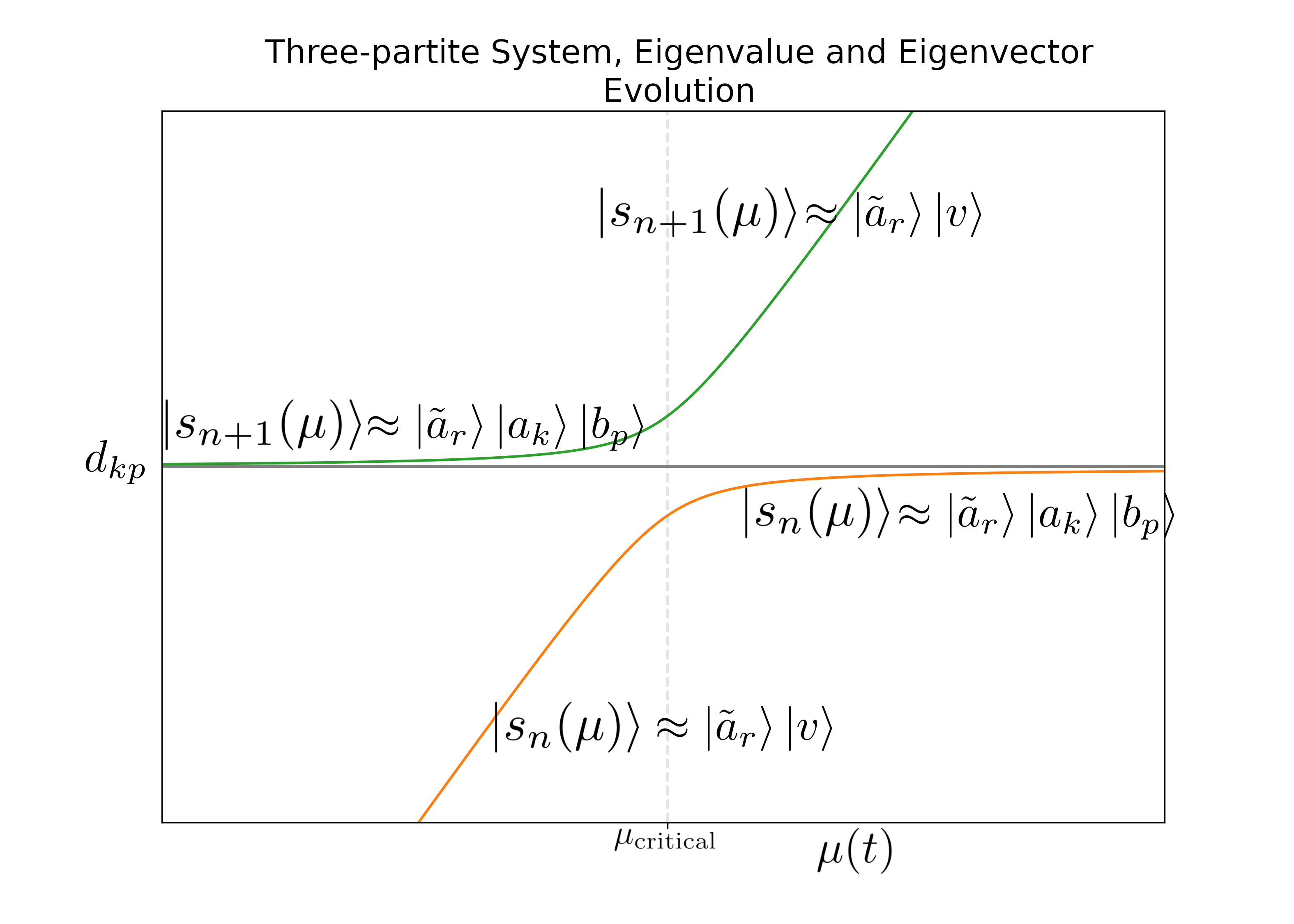}
    \caption{Example of an edge case in the three-partite system $\tilde{A}AB$. As two eigenvalues avoid a level crossing, they swap their eigenvectors, which can induce changes to the entanglement in system $\tilde{A}AB$.}
    \label{fig:three_partite}
\end{figure}

Similarly, in the tripartite system $\tilde{A}AB$, let us consider initial Hamiltonians of the form $D = H_{\tilde{A}} \otimes I \otimes I + I \otimes H_A \otimes I + I \otimes I \otimes H_B$ and activate a projector though $H(\mu)=D+ \mu (I \otimes \ketbra{v})$ which acts on the  Hilbert space $\mathcal{H}_A \otimes \mathcal{H}_B$. Similar to the two-partite case, the structure of the entanglement in the tri-partite system $\tilde{A}AB$ can only change when eigenvalues avoid crossing and thereby swap their eigenvectors. Again, this is particularly clear in edge cases, as is illustrated in Fig.\ref{fig:three_partite},
where we can read off the basic dynamics of the entanglement in system $\tilde{A}AB$. 

First, since the ancilla $\tilde{A}$ does not participate in the interaction, we obtain analogs of Observations 1 and 2 for system $AB$, if the ancilla $\tilde{A}$ is prepared unentangled. 
For example, since the result of any narrowly avoided level crossing is that the two eigenvalues swap their eigenvectors, any 
unentangled eigenstate of $D$, such as $\vert s_{n+1}(\mu)\rangle \approx \vert \tilde{a}_r\rangle\vert a_k\rangle\vert b_p\rangle$ 
ultimately evolves into another unentangled eigenstate of $D$, such as $\vert s_{n+1}(\mu)\rangle\approx \vert \tilde{a}_r\rangle\vert v\rangle$ and therefore no entanglement has been generated after an avoided level crossing. Except, during the brief period when the swap occurs, i.e., close to a critical $\mu$ value, an initially unentangled state  such as here $\vert s_{n+1}(\mu)\rangle \approx \vert \tilde{a}_r\rangle\vert a_k\rangle\vert b_p\rangle$ will be a linear combination of the form: 
$\vert s_{n+1}(\mu)\rangle \approx \alpha \vert \tilde{a}_r\rangle\vert a_k\rangle\vert b_p\rangle+ \beta \vert \tilde{a}_r\rangle\vert v\rangle$. We notice that if $\vert v\rangle \approx \vert a_i\rangle\vert b_j\rangle$ with $i\neq k$ and $j\neq p$, then $AB$ will be entangled around the critical $\mu$ value. As anticipated, since the ancilla is here a mere spectator, this matches \Cref{observation_1} above for the bipartite case. Clearly, also \Cref{observation_2} for the bipartite system $AB$ above has an immediate generalization to the tripartite system, if the ancilla system $\tilde{A}$ is chosen to be initially unentangled. 

Finally, we analyze the fundamental processes in which an interaction induces a transfer of entanglement.

To this end, we assume that $\tilde{A}$ and $A$ are initially entangled and we use our findings above to study how an interaction between $A$ and $B$ can turn the initial entanglement between $\tilde{A}$ and $A$ into entanglement between $\tilde{A}$ and either $B$ or $AB$. 
Consider, for example, the initial state of the form 
\begin{equation}
\vert \psi_0\rangle := \frac{1}{\sqrt{2}}\left( \vert \tilde{a}_r\rangle\vert a_k\rangle\vert b_p\rangle + \vert \tilde{a}_s\rangle\vert a_l\rangle\vert b_p\rangle\right)
\end{equation}
with $r\neq s$ and $k\neq l$. Here, systems $\tilde{A}$ and $A$ are entangled, and system $B$ is unentangled.
Then, during the evolution, one of the two vectors on the RHS, say $\vert \tilde{a}_r\rangle\vert a_k\rangle\vert b_p\rangle$ will be the first to be swapped for $\vert v\rangle$ at some critical $\mu$ value. After passing this critical $\mu$ value, the initial state $\vert \psi_0\rangle$ will therefore have adiabatically evolved into:    
\begin{equation}
\vert \psi_1\rangle := \frac{1}{\sqrt{2}}\left( \vert \tilde{a}_r\rangle\vert v\rangle + \vert \tilde{a}_s\rangle\vert a_l\rangle\vert b_p\rangle\right)
\end{equation}
We can now easily read off, for example, that by choosing $\vert v\rangle \approx \vert a_l\rangle\vert b_q\rangle$ with $q\neq p$, the state $\vert \psi_1\rangle$ possesses entanglement only between $\tilde{A}$ and $B$ while $A$ becoming unentangled:
\begin{equation}
\vert \psi_1\rangle \approx \frac{1}{\sqrt{2}}\left( \vert \tilde{a}_r\rangle \vert a_l\rangle\vert b_q\rangle
+ \vert \tilde{a}_s\rangle\vert a_l\rangle\vert b_p\rangle\right)
\end{equation}
For this choice of $\vert v\rangle$, the interaction between $A$ and $B$, therefore, transferred the entanglement that $A$ initially possessed with $\tilde{A}$ to $B$, while passing the critical $\mu$ value. 

Similarly, it is clear that the choice of $\vert v\rangle \approx \vert a_t\rangle\vert b_q\rangle$ with $t\neq l$ and $q\neq p$, transfers the entanglement that $\tilde{A}$ initially possessed with $A$ to system $AB$, by yielding instead the tri-partite entangled state:
\begin{equation}\label{eq:entanglement_distribution}
\vert \psi_1\rangle \approx \frac{1}{\sqrt{2}}\left( \vert \tilde{a}_r\rangle \vert a_t\rangle\vert b_q\rangle
+ \vert \tilde{a}_s\rangle\vert a_l\rangle\vert b_p\rangle\right)
\end{equation}
For this choice of $\vert v\rangle$, the interaction between $A$ and $B$ therefore transferred the entanglement that $A$ initially possessed with $\tilde{A}$ to the system $AB$. 

The adiabatic dynamics of entanglement in $n$-partite systems for large $n$ can be analyzed similarly, and we will use this fact in \cref{AQCsection} where we will apply these results to adiabatic quantum computation. 

\subsection{Application to adiabatic quantum computing for combinatorial optimization}
These results can be applied to adiabatic quantum computing (AQC), to address, for example, combinatorial optimization problems. In what follows, we relate the entanglement generation and its subsequent loss to the hardness of an optimization problem. In optimization, we seek $\min(f(x))$ for $x \in \{0,1\}^n$ by searching for the ground state of a problem Hamiltonian $H_p = \sum_x f(x) \ketbra{x}{x}$. To this end, we analyze an AQC protocol that uses a schedule $H(t) = H_0 + \mu(t) H_p$, where the coupling $\mu(t)$ increases monotonically. As $\mu(t)$ grows, terms $\mu(t)f(x)\ketbra{x}{x}$ corresponding to high-cost states $\ket{x}$ dominate earlier. This effectively activates the projectors sequentially from those of highest cost on down. The adiabatic evolution, therefore, forces the evolving ground state $\ket{0(t)}$ to become orthogonal first to the highest-cost basis states $\ket{x}$ and the evolution then systematically eliminates successively less and less poor candidate solutions.

At later stages of the adiabatic evolution, the ground state $\ket{0(t)}$ is, therefore, a superposition of the remaining low-cost candidate solutions $\{\ket{x_{low}}\}$. 

Interestingly, we can relate the amount of entanglement in such superpositions to the classical hardness of the problem being solved. Consider a case where the energy landscape is classically hard in the sense that low-energy configurations $\{\ket{x_{\text{low}}}\}$, defined below some energy threshold, are typically separated by large Hamming distances. A quantum state that superposes basis states with large Hamming distances typically exhibits entanglement, since it does not factorize across qubit subsystems. For example, consider the case where the instantaneous ground state $\ket{0(t)}$ is a linear superposition of computational basis states with pairwise Hamming distances of at most two:
\begin{equation*}
    \ket{0(t)} = ac \ket{000} + ad \ket{001} + bc \ket{010} + bd \ket{011}
\end{equation*}
This state factorizes completely as a product state:
\begin{equation*}
    \ket{0(t)} = \ket{0} \otimes (a\ket{0} + b\ket{1}) \otimes (c\ket{0} + d\ket{1})
\end{equation*}
In contrast, if we use the same non-zero coefficients but superpose basis states that are more widely separated in Hamming space, we obtain genuine tri-partite entanglement:
\begin{equation*}
    \ket{0(t)} = ac \ket{000} + ad \ket{111} + bc \ket{011} + bd \ket{001}
\end{equation*}
Due to the larger pairwise Hamming distances between the basis states, the superposition above cannot be factorized, as all three subsystems are multipartite entangled. The most extreme case occurs when the superposition involves basis states with the maximum pairwise Hamming distance, resulting in maximal multipartite entanglement. A canonical example is the GHZ state, given by $1/\sqrt{2}(\ket{000} + \ket{111})$, which exhibits tripartite entanglement and cannot be decomposed into a product across any bipartition. Such a state represents two local minima, which are separated by a distance of 3. Hence, moving from one local minima to another requires at least three local moves (bit flips) over higher-energy barriers. We see that large Hamming distances between local minima imply classical hardness, as we need to traverse the energy landscape over long distances and do more work to avoid being trapped in local minima. From the quantum mechanics perspective, we get states that are multipartite entangled.

Generally, the poor factorizability directly implies that the intermediate ground state $\ket{0(t)}$ possesses high multipartite entanglement. Conversely, smoother landscapes (classically easier problems), where regions of low-energy states are nearby (short Hamming distances), result in intermediate states with lower entanglement. Therefore, a key insight is that classical problem hardness, characterized by a rugged cost landscape with large Hamming distance local minima, necessitates the development of significant temporary multipartite entanglement in the instantaneous ground state $\ket{0(t)}$ at intermediate stages of the AQC process. 

Let us here recall that one usually chooses the initial state to be an equal superposition of all computational basis states, which can be written as the tensor product of plus states. Since this initial state is a product state, it is unentangled. Also, the final state, assuming it is unique, is unentangled. This implies that, for optimization problems with local minima that possess large Hamming distances, the adiabatic evolution must build up and then destroy a large peak of entanglement. 

We have shown that all changes in entanglement occur at avoided level crossings, and that the narrowness of these crossings determines the efficiency of these changes. This insight provides new tools for exploring how the hardness of a problem, such as an optimization task, relates to the magnitude and character of the entanglement peak during adiabatic quantum computation (AQC), and how this peak, in turn, necessitates a slowdown due to narrow avoided crossings.

\section{\label{sec:background} Mathematical preparations}
As motivated above, we consider a time-dependent Hamiltonian that acts on an $N$-dimensional Hilbert space $\mathcal{H}$ and takes the form:
\begin{equation}\label{eq:general_hamiltonian}
    H(t) := H_0 + g(t)H_{int}
\end{equation}
Here, $H_0$ and $H_{int}$ are Hermitian operators acting on $\mathcal{H}$. We refer to $H(0) = H_0$ as the initial Hamiltonian, and for some suitably chosen final time $T > 0$, we call $H(T)$ the final Hamiltonian. 
The coupling function $g(t)$ is assumed to monotonically increase from $g(0)=0$ to a final value $g(T)$, slowly enough for the evolution to be adiabatic. As anticipated above, we break down this problem into more manageable parts. To this end, we invoke the spectral theorem and rewrite $H_{int}$ as a weighted sum of rank one projectors, $H_{int} = \sum_k \lambda_k \ketbra{v_k}$. Instead of adiabatically adding all of $g(T) H_{int}$ simultaneously to $H_0$, as in (\ref{eq:general_hamiltonian}), we  adiabatically add the weighted projectors $g(T) \lambda_k \ketbra{v_k}$ one by one to $H_0$. 

Each activation of a projector is described by a time-dependent Hamiltonian of the form:
\begin{equation}\label{eq:general_hamiltonian_rank1_proj}
    H(t) := D + \mu(t)\ketbra{v}{v}
\end{equation}
Here, for a fixed index $k$, we define $\ketbra{v}:=\ketbra{v_k}$. For the purpose of activating an individual projector, it is convenient to denote the starting and the final time of the activation of this new projector by $t=0$ and $t=t_f$, respectively. We choose $\mu(t)$ to be a function that monotonically runs from $\mu(0)=0$ to $\mu(t_f)=g(T)\lambda_k$ slowly enough for the evolution to be adiabatic. We now focus on the dynamics during the addition of an individual projector.

Before we get to the main results, we present a useful lemma which shows that for any real value $s$, there is an eigenstate $\ket{s}$ of the Hamiltonian $H(t)$ in (\ref{eq:general_hamiltonian_rank1_proj}) for which the eigenvalue nonperturbatively an explicit expression can be given and for which $s$ is the eigenvalue. 
\begin{lemma}\label{lemma:eigenstate_s}
   Let $H(t)$ in (\ref{eq:general_hamiltonian_rank1_proj}) be the Hamiltonian of a quantum system with a non-degenerate spectrum. Let $\ket{v}$ be an arbitrary generic vector and $d_k$ be the eigenvalues of $D$. For any eigenvalue $s := s(\mu(t)) \neq d_k$ of $H(t)$, the corresponding eigenstate $\ket{s}$ is given by
    \begin{align}\label{eq:formula_ket_s_in_lemma}
        \ket{s} &= \gamma \big( sI - D \big)^{-1}\ket{v},
    \end{align}
   where $\gamma$ is a normalization constant defined as
   \begin{align}
       \gamma = \frac{1}{\| (sI - D)^{-1}\ket{v} \|}.
   \end{align}
\end{lemma}
The lemma states that the eigenstate $\ket{s}$ can be explicitly expressed as a function of $\ket{v}$, the initial Hamiltonian $D$, and the eigenvalue $s$. The proof of this lemma and of the subsequent propositions in this paper are in the appendices. 
The lemma can be combined with the adiabatic theorem \cite{born1928beweis, amin2009consistency}:

\begin{lemma}\label{lemma:eigenstate_s_adiabatic}
   Let $H(t)$ in (\ref{eq:general_hamiltonian_rank1_proj}) be the Hamiltonian of a quantum system whose spectrum is non-degenerate, and let $\ket{v}$ be arbitrary. Let $u := t/t_f \in[0,1]$ and $\tilde{\mu}(u) := \mu(u t_f)$, then the rescaled version of $H(t)$ is:
   \begin{equation}
       \tilde{H}(u):= D + \tilde{\mu}(u) \ketbra{v}
   \end{equation}
   Assume that the system is initially in the $n$'th--level energy eigenstate $\ket{\tilde s_n(0)}=\ket{d_n} $ of $\tilde H(0)=D$. Suppose that it undergoes adiabatic evolution until a final time $t_f$, which is chosen in accordance with the condition of the adiabatic theorem \cite{amin2009consistency}:
   \begin{align}\label{eq:adiabatic_condition_for_t}
        t_f &\gg \max_{u \in [0,1]}\frac{ \left |\braket{\tilde s_m(u)}{v} \braket{v}{\tilde s_n(u)}  d\tilde \mu/du \right|}{\tilde s_{nm}(u)^2}, \ \ \ n \neq m 
        % &= \max_{u \in [0,1]} \frac{|d\tilde \mu / du| \gamma_n \gamma_m}{\tilde s_{nm}^2 \tilde \mu(\tilde s_m(u)) \tilde \mu(\tilde s_n(u))}, 
    \end{align}
    Here, $\tilde s_{nm}(u) = \tilde s_n(u) - \tilde s_m(u)$ and $\tilde s_k(u)$ is an eigenvalue of $\tilde H(u)$. Then, the evolved state of the system at time $t_f$ will in effect remain in the instantaneous eigenstate $\ket{s_n(\mu(t_f))}$, that is
    \begin{equation}
        \left \|U(t_f)\ket{d_n} - \ket{s_n(\mu(t_f))} \right \| \approx 0,
    \end{equation}
    where
    \begin{align}
        U(t_f) &:=  \mathcal{T} \exp \left \{ -\textrm{i} \int_0^{t_f}H(t)dt \right \},
    \end{align}
    and $\mathcal{T}$ denotes the time-ordering operator.
%The lemma states that if the condition in (\ref{eq:adiabatic_condition_for_t}) is satisfied, then the initial eigenstate $\ket{d_n}$ of $D$ evolves into the eigenstate $\ket{s_n}$ of the final Hamiltonian $H(t_f)$. 
In the limit of $t_f \rightarrow \infty$, we have the exact relation:
\begin{equation}\label{eq:eigenstate_s_and_evolution_relation}
U(t_f)\ket{d_n} = \ket{s_n(\mu(t_f))} =  \gamma (s_n(\mu(t_f))I - D)^{-1}\ket{v}
\end{equation}
\end{lemma}
The statements of \Cref{lemma:eigenstate_s_adiabatic} can be easily derived by applying the adiabatic theorem in \cite{amin2009consistency} to the result of \Cref{lemma:eigenstate_s}.

In this paper, we always assume that $t_f$ is chosen sufficiently large to ensure that the time evolution operator $U(t_f)$ evolves the initial eigenstate $\ket{d_n}$ into the eigenstate $\ket{s_n(t_f)}$ to any desired target accuracy, so that, in effect, we can write  $U(t_f)\ket{d_n} = \ket{s_n(t_f)}$.

Remark: The reason for using the unitless variable $u$ in condition (\ref{eq:adiabatic_condition_for_t}) traces to the observation that traditional conditions for adiabaticity had missed the fact that exceptional non-adiabaticity can build up through resonances from seemingly adiabatic oscillatory time dependence in the Hamiltonian, see \cite{amin2009consistency}. The condition that the Hamiltonian can be written as a function of the unitless time variable $u$ excludes such oscillatory behaviour since it would set a new scale. This, in turn, renders condition (\ref{eq:adiabatic_condition_for_t}) robust. For a detailed review, see \cite{albash2018adiabatic}.

\subsection{\label{sec:relation_mu_s}The relationship between the coupling strength $\mu(t_f)$ and the final eigenvalue $s$.}
It would be desirable to be able to express the eigenvalues, $s_n(\mu)$ of a Hamiltonian $H(\mu)=D+\mu \vert v\rangle\langle v \vert$ as an explicit expression in radicals, e.g., by solving explicitly for the roots, $s_n(\mu)$, of the characteristic polynomial of $H(\mu)$. As Galois theory showed, this is impossible for Hilbert space dimensions larger than 4, see \cite{artin1998galois}. 
In this section, we present a lemma, see \cite{vsoda2022newton}, that describes a method to bypass this limitation by inverting the problem, i.e., by calculating $\mu(s)$, and by then using the Lagrange inversion theorem to obtain the $s_n(\mu)$ not in terms of radicals but as a power series whose coefficients are all explicitly known. 

We start by noticing that it is possible to give an explicit expression for $\mu$ as a function of the eigenvalue $s$ for all real $s$. Concretely, working in the eigenbasis $\{\ket{d_k}\}_{k=0}^{N-1}$ of the initial Hamiltonian $D$, we have:
\begin{equation}\label{eq:function_mu}
    \mu(s) = \left(\sum_{k=0}^{N-1}\frac{|v_{k}|^2}{s - d_{k}} \right)^{-1}
\end{equation}
In the equation above, $v_k := \braket{v}{d_k}$ and the $d_k$ are the eigenvalues of the initial Hamiltonian $D$.

Therefore, one can fix $s$ and then use (\ref{eq:function_mu}) to obtain the desired coupling strength $\mu$ so that $H(\mu)$ has $s$ as an eigenvalue.
\begin{lemma}\label{lemma:explicit_eigenvalue_s}
    Let $s$ be an arbitrary real number. Consider the Hamiltonian defined by $H(s) = D + \mu(s)\ketbra{v}{v}$, where $\mu(s)$ is as in (\ref{eq:function_mu}) and $\ket{v}$ is arbitrary. Then, $s$ is an eigenvalue of $H(s)$.
    % Furthermore, if $\mu(t_f)$ is set equal to $\mu(s)$ for some final time $t_f$, the resulting final Hamiltonian $H(t_f)$ also has $s$ as an eigenvalue.
\end{lemma}
Allowing $s$ to change over time, we will, for simplicity, write $H(t)=H(s(t))$. Incorporating the principles from \Cref{lemma:eigenstate_s_adiabatic} and \Cref{lemma:explicit_eigenvalue_s}, a comprehensive non-perturbative framework can be formulated for the determination of final eigenstates and eigenvalues of a time-dependent Hamiltonian $H(t)$. Initially, one selects a real number $s$ and employs \Cref{lemma:explicit_eigenvalue_s} to define the final Hamiltonian $H(t_f)$. Subsequently, \Cref{lemma:eigenstate_s} provides the exact form of the final eigenstate associated with $H(t_f)$.

Finally, we can use the Lagrange inversion theorem to obtain, see \cite{vsoda2022newton}:
\begin{eqnarray}
s_r(\mu) &=& d_r \label{exact} \\ & &-\sum_{n=1}^\infty \frac{\mu^n}{n}\!\!\!\!\!\!
\sum_{\substack{k_1,...,k_n=0\\ \sum_{i=1}^n k_i=n-1}}^{n-1}\sum_{\substack{p_1,...,p_n=1\\ p_1\neq r,...,p_n\neq r}}^N\prod_{i=1}^n\frac{\vert v_{p_i}\vert^2-\frac{\delta_{k_i,0}}{N-1}}{(d_r-d_{p_i})^{k_i}} \nonumber
 \end{eqnarray}
%By the Lagrange inversion theorem the radius of convergence is finite. 
%Then, we need the coefficients $v^{(2)}_n$ of $\vert v^{(2)}\rangle$ in the eigenbasis $\{\vert s^{(1)}_r(\mu^{(1)})\rangle\}_{n=1}^N$. Using \cref{evecs}, we obtain them exactly: 
%\be
%v^{(2)}_n = \langle s_n(\mu^{(1)})\vert v^{(2)}\rangle =\sum_{r=1}^N\langle s_n(\mu^{(1)})\vert s_r\rangle\langle s_r\vert v^{(2)}\rangle \label{step2}
%\ee
Recall that we start with $H_0$ and then during the adiabatic evolution toward the final Hamiltonian $H_0 + g(T) H_{int}$, we adiabatically add the projectors $\mu_n \vert v_n\rangle\langle v_n\vert$ one by one by 
dialing their prefactor $\mu_n$  from $\mu_n=0$ to $\mu_n=g(T)\lambda_n$. The exact result (\ref{exact}) now allows us to perform all these successive  additions of projectors explicitly, thereby nonperturbatively yielding the running eigenvectors and eigenvalues. 
%In instances where pre-selecting the eigenvalue $s$ is impractical, an alternative strategy is to establish a fixed value for $\mu(t_f)$ and then employ numerical methods in conjunction with (\ref{eq:function_mu}).

\section{\label{sec:ent_trans}Transmission and Preservation of Entanglement in Tripartite Systems}
In this section, our goal is to identify the elementary principles that govern entanglement transmission or preservation under addition of the initial Hamiltonian $D$ and the coupling term $\mu(t)\ketbra{v}{v}$. Therefore, we investigate the dynamics of entanglement within a tri-partite system situated in the Hilbert space $ \mathcal{H}_{\tilde{A}} \otimes \mathcal{H}_{A} \otimes \mathcal{H}_{B}$. We assume here that system $A$ is initially prepared entangled with an ancillary system $\tilde{A}$, such that $\tilde{A}$ purifies $A$. The system $B$ is assumed to be initially pure and unentangled with both $A$ and $\tilde{A}$. Subjected to a time-dependent Hamiltonian as in (\ref{eq:tripartite_hamiltonian}), the composite pure system $\tilde{A}AB$ then undergoes adiabatic evolution while $\tilde{A}$ remains inert. The interaction between $A$ and $B$ induced by the term $\mu(t)\ketbra{v}{v}$ can then not only entangle the systems $A$ and $B$ but can also transfer entanglement that $A$ initially possessed with $\tilde{A}$ to $B$ or $AB$. The setup is illustrated in \Cref{fig:interaction_setup}. For example, $\tilde{A}$ and $A$ might represent qubits in a quantum processor, while $B$ could describe another qubit of the processor or an environment. 

We assume that the following Hamiltonian interaction governs the adiabatic evolution:
\begin{equation}\label{eq:tripartite_hamiltonian}
    H(t) = I \otimes H_A \otimes I + I \otimes I \otimes H_B + \mu(t) I \otimes \ketbra{v}{v}
\end{equation}
Here, $H_A$ and $H_B$ denote the Hamiltonians of systems $A$ and $B$ respectively. We define the initial Hamiltonian $D$ as $D :=  H_A \otimes I + I \otimes H_B$. Subsequently, the total Hamiltonian in (\ref{eq:tripartite_hamiltonian}) simplifies to:
\begin{equation}
    H(t) = I \otimes D + \mu(t) I \otimes \ketbra{v}{v}
\end{equation}
For this system, let $\{\ket{a_i}\}$ and $\{\ket{b_i}\}$ represent the sets of eigenstates corresponding to systems $A$ and $B$ respectively. Additionally, we introduce $\{\ket{\tilde{a}_i}\}$ as an arbitrary basis for the ancillary system $\tilde{A}$.
% \begin{figure}[h]
%     \centering
%     \includegraphics{figures/tri-system-interaction-scheme.png}
%     \caption{Initially, the systems $\tilde{A}$ and $A$ are pre-entangled, whereas $B$ is independent. The adiabatic evolution operator $U(t_f)$ is generated by a Hamiltonian $H(t)$, which acts non-trivially on $AB$.}
%     \label{fig:interaction_setup}
% \end{figure}
We let the system $\tilde{A}A$ start in a pure state. For some fixed $i$ and $j$, we define the state of $\tilde{A}A$ to be
\begin{equation}\label{eq:max_ent_state_tildeAA}
    \ket{\psi_{\tilde{A}A}} = \alpha_{ii}\ket{\tilde{a}_i}\ket{a_i} + \alpha_{jj}\ket{\tilde{a}_j}\ket{a_j},
\end{equation}
where $\alpha_{ii}$ and $\alpha_{jj}$ are complex coefficients. We may assume, for example, an entangled state $\ket{\psi_{\tilde{A}A}}$ with $|\alpha_{ii}|^2 = |\alpha_{jj}|^2 = 1/\sqrt{2}$. Then the joint initial state of $\tilde{A}AB$ is:
\begin{equation}\label{eq:max_ent_state_tildeAAB}
    \ket{\psi_{\tilde{A}AB}} = \left(\alpha_{ii}\ket{\tilde{a}_i}\ket{a_i} + \alpha_{jj}\ket{\tilde{a}_j}\ket{a_j}\right)\ket{b_0}
\end{equation}

Now, we are ready to establish the conditions for maximizing the transfer or preservation of entanglement during the course of adiabatic evolution. That is, we show that for an appropriate choice of $\ket{v}$ and $\mu(t_f)$, it is possible to transmit entanglement in the sense that $\tilde{A}$'s initial entanglement with $A$ turns into entanglement of $\tilde{A}$ with $B$, while $A$ becomes pure and therefore disentangled from both $\tilde{A}$ and $B$. Conversely, we show how to reverse the entanglement transmission so that $\tilde{A}$ is again entangled with $A$, and $B$ returns to being in a pure state.

Furthermore, we find that transferring entanglement is computationally expensive due to the need for a slower rate of adiabatic evolution. As we will discuss in \Cref{sec:eigengap}, selecting the state $\ket{v}$ to maximize entanglement transfer leads to reduced energy gaps of the Hamiltonian in (\ref{eq:tripartite_hamiltonian}), which is important because the minimum duration, $t_f$, of the evolution is determined by the minimum energy gap, see, e.g., (\ref{eq:adiabatic_condition_for_t}). We find that highly efficient entanglement transfer necessitates a significantly prolonged evolution time.

Another interesting finding is that entanglement transfer and preservation are not continuous processes. We show that entanglement relations between systems can abruptly reverse to their original state even if $\mu(t_f)$ and $\ket{v}$ are changed in a continuous manner.

Let us first start with the trivial choice of $\ket{v}$. Let $\ket{v}$ be an eigenstate of the Hamiltonian $D = H_A \otimes I + I \otimes H_B$. That is, $\ket{v} = \ket{a_p}\ket{b_k}$ for some fixed $p$ and $k$. In this case, the initial Hamiltonian $I \otimes D$ commutes with the interaction Hamiltonian $\mu(t)I \otimes \ketbra{v}{v}$ and the final state
$\ket{\psi_{\tilde{A}'A'B'}}$ only acquires relative phases:
\begin{equation}
    \ket{\psi_{\tilde{A}'A'B'}} = \alpha_{ii} \omega_{ii} \ket{\tilde{a}_i}\ket{a_i}\ket{b_0} + \alpha_{jj}\omega_{jj}\ket{\tilde{a}_j}\ket{a_j}\ket{b_0}
\end{equation}
where $|\omega_{ii}|, |\omega_{jj}|=1$ for some fixed $i$ and $j$.
Importantly, $\tilde{A}A$ remains entangled and pure, and $B$ is unentangled and pure. Therefore, the entanglement structure remained unchanged. In this trivial case, we see that $\ket{v}$, being an eigenstate of $D$, cannot affect the entanglement characteristics of any of the systems.

It is reasonable to assume that $\ket{v}$ that maximizes the transmission or preservation of entanglement must be a linear combination of specific eigenstates of $D$. Indeed, this is the case. However, surprisingly, such $\ket{v}$ must be arbitrarily close to exactly one of the eigenstates of $D$ and yet not be equal to it. Let us use the shorthand notation $\mu_{t_f} := \mu(t_f)$ and define $v_{kp}$ as the coefficients of $\ket{v}$ in the eigenbasis $\{\ket{a_k}\ket{b_p}\}_{kp}$. That is,
\begin{equation}
    v_{kp} := \braket{v}{a_k b_p},
\end{equation}
With this notation, we now present the following theorem, which provides a construction of the edge case $\ket{v}$ discussed earlier.
\begin{theorem}\label{theorem:maximizing_transfer_preservation}
    Let $n = \dim A$ and $m = \dim B$. For some fixed indices $i < j$, suppose the eigenvalues of $D$ are ordered as
    \begin{equation}
        d_{i0} < d_{i1} < \cdots < d_{j0} < d_{j1}.
    \end{equation}
    Let system $\tilde{A}A$ be entangled and $B$ be in a pure, unentangled state:
    \begin{equation}\label{eq:init_max_ent_state_tildeAAB}
        \ket{\psi_{\tilde{A}AB}} = \alpha_{ii}\ket{\tilde{a}_i}\ket{a_i}\ket{b_0} + \alpha_{jj}\ket{\tilde{a}_j}\ket{a_j}\ket{b_0}
    \end{equation}
    Fix an index $\ell \geq 0$ such that $d_{i0} \leq d_{i\ell} < d_{j0}$. For $\epsilon > 0$, if  $\ket{v}$ is such that 
    \begin{align}\label{eq:choice_of_v}
        v_{kp} = 
        \begin{cases}
            1 - (nm - 1) \epsilon^2 & \text{for } k = i, p = \ell, \\
            \epsilon & \text{otherwise},
        \end{cases}
    \end{align}
   and
   \begin{equation}\label{eq:mu_open_interval}
       \mu_{t_f} \in \left(\frac{d_{j0}-d_{i\ell}}{|v_{i\ell}|^2}, \frac{d_{j1}-d_{i\ell}}{|v_{i\ell}|^2}\right).
   \end{equation}
  Then, as $\epsilon \to 0$, the entanglement of the evolved system is such that $\tilde{A}$ becomes entangled with $B$, and $A$ becomes pure and unentangled.
    Furthermore, if 
    \begin{equation}\label{eq:values_of_mu_larger}
        \mu_{t_f} > \frac{d_{j1}-d_{i\ell}}{|v_{i\ell}|^2},
    \end{equation}
    then as $\epsilon \to 0$, the entanglement of the evolved system is such that $\tilde{A}$ becomes again entangled with $A$, and $B$ becomes again pure and unentangled.
\end{theorem}
The theorem, therefore, provides us with a construction of interaction Hamiltonians $\mu(t_f) \ketbra{v}$ to accomplish transfer or preservation of entanglement with the edge case $\ket{v}$. Furthermore, the theorem explicitly identifies the values of $\mu(t_f)$ at which $A$ transfers its entanglement with $\tilde{A}$ to $B$, as detailed in (\ref{eq:mu_open_interval}). It also specifies the precise values of $\mu(t_f)$ required to revert the entanglement relationships back to their original configuration.

We note that since the edge case $\ket{v}$ in (\ref{eq:choice_of_v}) has the dominant component $v_{i\ell} \approx 1$ for $\epsilon \ll 1$, it is close to an eigenstate $\ket{a_i}\ket{b_\ell}$ of the initial Hamiltonian $D$. However, $\ket{v}$ significantly differs from $\ket{a_i}\ket{b_\ell}$ in its effect on the final system's entanglement structure. We see that entanglement transfer or preservation is not continuous relative to the change of $\ket{v}$. For example, one might construct $\ket{v}$ with sufficiently small $\epsilon$ and some finite $\mu_{t_f}$ such that $\tilde{A}$ becomes entangled with $B$ and $A$ becomes pure and unentangled. However, if $\epsilon$ reaches its limit $0$, i.e., $\epsilon=0$, then $\ket{v} = \ket{a_i}\ket{b_\ell}$ and the structure of entanglement abruptly reverts to its original state. This is because, as we have seen, eigenstates of $D$ do not change the entanglement structure of the initial system, i.e., $A$ remains entangled with $\tilde{A}$ and $B$ remains pure and unentangled.

The construction of $\ket{v}$ in \Cref{theorem:eigenstate_swap} enables maximal changes in entanglement and explicitly identifies the values of $\mu$ at which these changes occur. While slight variations of this construction are possible, they generally come at the cost of less transparent and significantly more cumbersome mathematical expressions. Nevertheless, the core principles for the edge case scenario remain unchanged: $\ket{v}$ must have a dominant component aligned with an eigenvector $\ket{d}$ of the initial Hamiltonian $D$, satisfying $|\langle v | d \rangle| = 1 - O(\epsilon^2)$. For any other eigenvector $\ket{d'}$ of $D$, we must have $|\langle v | d' \rangle| = O(\epsilon)$, where $\epsilon \ll 1$.

\section{\label{sec:main_theorems}Entanglement Transfer/Preservation Induces Structured Eigenssystem Dynamics.}
The dynamics of eigenvalues and associated eigenstates of $H(t_f)$, under the influence of $\mu(t_f)\ketbra{v}{v}$ as constructed in \Cref{theorem:maximizing_transfer_preservation}, play a crucial role in understanding entanglement transfer and preservation in quantum systems. In this section, we explore these dynamics in detail, showing that each eigenstate of the initial Hamiltonian $D$ undergoes a three-stage adiabatic evolution.

Initially, in the first stage of the evolution, an eigenstate of $D$ remains nearly static, exhibiting minimal change. The second stage is marked by the eigenstate closely aligning with $\ket{v}$, reflecting a significant shift in its characteristics. Finally, in the third stage, the eigenstate transitions towards a neighbouring higher energy eigenstate. Similar behaviour can be observed for the eigenvalues. Each eigenvalue undergoes exactly the same three-stage evolution. In the first stage, an eigenvalue remains almost static. During the second stage, the eigenvalue increases at a speed close to $1$. Importantly, only one eigenvalue at a time has a velocity near $1$ while the rest remain almost static. During the third stage, the eigenvalue approaches a higher energy eigenvalue and slows down without ever crossing it. The dynamics of this behaviour are illustrated in \Cref{solevec}. This three-stage progression elucidates the mechanisms behind the subtle shifts in eigenstate positions and energies, providing a comprehensive understanding of how entanglement properties are influenced and manipulated through the choice of $\mu(t_f)\ketbra{v}$ in \Cref{theorem:maximizing_transfer_preservation}.

The influence of positive semidefinite $\mu(t_f)\ketbra{v}{v}$ with an arbitrary $\ket{v}$ is thoroughly documented in the literature \cite{vsoda2022newton, bunch1978rank}. Generally, the neighbouring eigenvalues $s_{k} < s_{k+1}$ of $H(t_f)$ interlace with the neighbouring eigenvalues  $d_{k} < d_{k+1}$ of $D$. This yields the interlacing inequality:
\begin{equation}\label{eq:interlacing_inequality}
    d_{k} \leq s_{k} \leq d_{k+1} \leq s_{k+1}
\end{equation}
As $\mu_{t_f}$ increases from negative infinity to positive infinity, the eigenvalues of $H(t_f)$ increase monotonically. Absence of orthogonality between $\ket{v}$ and any eigenstate $\ket{d_k}$ of $D$ ensures the prohibition of energy level-crossings for all finite $\mu_{t_f}$. Conversely, if $\ket{v}$ is orthogonal to a particular eigenstate $\ket{d_j}$, then the associated eigenvalue $s_{j}$ remains fixed, permitting potential level-crossings. This scenario has been previously encountered in \Cref{sec:ent_trans}; when $\ket{v}$ coincided with an eigenvector of $D$. In this scenario, when $\mu_{t_f}$ approaches positive infinity, a single eigenvalue increases, sequentially crossing all higher energy levels. Furthermore, we have seen that such $\ket{v}$ does not change the entanglement structure for any values of $\mu_{t_f}$.

On the other hand, the optimal selection of $\ket{v}$ in Theorem~\ref{theorem:maximizing_transfer_preservation} maximizes entanglement transmission or preservation for finite values $\mu_{t_f}$, it also imposes a strict pattern in the evolution of eigenvalues $s_k$ and eigenstates $\ket{s_k}$ of $H(t)$. As per the formulation of $\ket{v}$ in (\ref{eq:choice_of_v}) with $\epsilon>0$ sufficiently small, no level-crossings occur for all finite $\mu_{t_f}$ due to the non-zero components $v_{k} = \braket{v}{d_k} \geq \epsilon$ with any eigenstate $\ket{d_k}$ of $D$. Furthermore, considering the eigenvalue velocities total to unity \cite{vsoda2022newton, bunch1978rank}, at any moment, precisely one eigenvalue exhibits a velocity of $|1 - (nm-1)\epsilon^2|^2 \approx 1$, with the rest advancing minimally with the maximum velocity proportional to $\epsilon^2 \approx 0$. Specifically, as $\mu_{t_f}$ increases, a solitary eigenvalue $s_{k}$ ascends to the next larger eigenvalue $s_{k+1}$ at a velocity of $|1-(nm-1)\epsilon^2|^2$. Due to the eigenvalue $s_{k}$ ascending to the almost static larger eigenvalue $s_{k+1}$, the energy gap becomes narrower, yet no level-crossing is happening. The level-crossing avoidance can be attributed to an eventual deceleration in eigenvalue velocity which changes from $|1-(nm-1)\epsilon^2|^2$ to near zero, and the larger eigenvalue $s_{k+1}$ acquiring the velocity of $|1-(nm-1)\epsilon^2|^2$ and hence avoiding the crossing with the eigenvalue $s_{k}$.

Consequently, we can categorize the behaviour of each eigenvalue $s_{k}$ of $H(t)$ and its corresponding eigenvector into three distinct evolutionary phases, characterized by three specific intervals of $\mu(t)$. During the initial interval $\mathcal{I}_1 := (-\infty, \mu_1)$, the eigenvalue $s_{k}$ is $\epsilon^2$--close to $d_{k}$ with the maximum velocity proportional to $\epsilon^2$. In the subsequent interval $\mathcal{I}_2 := (\mu_1, \mu_2)$, the eigenvalue $s_{k}$ experiences an increased velocity of $|1-(nm-1)\epsilon^2|^2$, indicating a more rapid ascent. Finally, in the interval $\mathcal{I}_3 := (\mu_2, +\infty)$, the eigenvalue $s_{k}$ is $\epsilon^2$--close to $d_{k+1}$ and resumes back to a slower pace. These intervals collectively span nearly the entire spectrum of the real line. This comprehensive coverage allows us to predict the eigenstate $\ket{s_{k}}$ associated with any given value of $\mu(t)$ within these intervals. This observation is formally presented in the following theorem.
\begin{theorem}\label{theorem:eigenstates_dynamics}
    Let $N$ be the dimension of the Hilbert space that $H(t)$ acts on and $\{d_j\}_{j=0}^{N-1}$ be the increasing sequence of the eigenvalues of $D$.
    
    Fix index $i$ and let $\epsilon > 0$. Suppose $\ket{v}$ is a vector characterized by its components $v_{j} = \braket{v}{d_{j}}$ as follows:
    \begin{equation}\label{eq:choice_of_v_dynamics}
        v_{j} = 
        \begin{cases}
            1 - (N - 1) \epsilon^2 & \text{if } j = i\\
            \epsilon & \text{otherwise}
        \end{cases}
    \end{equation}
    For $k> i$, if $\mu_{t_f}$ lies within the interval
    \begin{equation}
        \left( -\infty, \frac{d_{k}-d_{i}}{|v_{i}|^2} \right),
    \end{equation}
    then the eigenstate $\ket{d_k}$ of $D$ adiabatically evolves as
    \begin{equation}\label{eq:conversion_dk_to_dk}
        U_{t_f}\ket{d_k} \rightarrow \ket{d_k} \text{ as } \epsilon \rightarrow 0.
    \end{equation}
    If $\mu_{t_f}$ lies within the interval
    \begin{equation}\label{eq:open_interval_mu}
       \left( \frac{d_{k}-d_{i}}{|v_{i}|^2},\frac{d_{k+1} - d_{i}}{|v_{i}|^2} \right),
    \end{equation}
    then
    \begin{equation}\label{eq:conversion_dk_to_di}
        U_{t_f}\ket{d_k} \rightarrow \ket{d_i} \text{ as } \epsilon \rightarrow 0.
    \end{equation}
    If $\mu_{t_f}$ lies within the interval
    \begin{equation}\label{eq:third_interval}
       \left( \frac{d_{k+1} - d_{i}}{|v_{i}|^2}, +\infty \right),
    \end{equation}
    then
    \begin{equation}\label{eq:conversion_dk_to_dk+1}
        U_{t_f}\ket{d_k} \rightarrow \ket{d_{k+1}} \text{ as } \epsilon \rightarrow 0.
    \end{equation}
\end{theorem}
The endpoints of the interval in (\ref{eq:open_interval_mu}) deserve special attention. These are the ``critical points" around which eigenstate transitions are happening. For $\epsilon$ sufficiently small, consider the critical point $\mu_{k+1} := (d_{k+1} - d_{i})/|v_{i}|^2$ in (\ref{eq:open_interval_mu}).
Slightly before this critical point, say at $\mu_{k+1} - \Delta \mu$, the state $\ket{d_k}$ evolves into $\ket{d_i}$. However, right after the critical point at $\mu_{k+1} + \Delta \mu$, the state $\ket{d_k}$ evolves into the next energy level eigenstate $\ket{d_{k+1}}$, while the state $\ket{d_{k+1}}$ evolves into $\ket{d_i} \approx \ket{v}$. Essentially, in the limit of $\epsilon$ tending to zero, this evolution behaviour can be characterized as a swap: $\ket{d_k}$ and $\ket{d_{k+1}}$ swap places while $\mu(t)$ passes the critical point. To emphasize this remarkable fact, we present an alternative theorem that focuses on the eigenstate transitions around the critical points.
\begin{theorem}[Alternative]\label{theorem:eigenstates_dynamics_alter}
    Let $\{d_j\}_{j=0}^{N-1}$ be an increasing sequence of eigenvalues of $D$.
     
    Fix index $i$ and let $\epsilon \in (0,1)$. Suppose $\ket{v}$ is a vector characterized by its components $v_{j}= \braket{v}{d_j}$ as follows:
    \begin{equation}\label{eq:choice_of_v_alternative}
        v_{j} = 
        \begin{cases}
            1 - (N - 1) \epsilon^2 & \text{if } j = i \\
            \epsilon & \text{otherwise}
        \end{cases}
    \end{equation}
    For each $k \geq i$, define $\mu_{k+1}$ as:
    \begin{align}\label{eq:critical_mu_k+1}
        \mu_{k+1} := \frac{d_{k+1}-d_{i}}{|v_{i}|^2}
    \end{align}
    Let $U(t)$ be the adiabatic evolution operator of $H(t)$. Then, in the limit as $\epsilon \to 0$, the following transitions occur:
    \begin{align}
        &U\left( t(\mu_{k+1} - \Delta \mu) \right)\ket{d_k} \rightarrow \ket{d_i}\\
        &U\left( t(\mu_{k+1} + \Delta \mu) \right)\ket{d_k} \rightarrow \ket{d_{k+1}} \\
        &U\left( t(\mu_{k+1} - \Delta \mu) \right)\ket{d_{k+1}} \rightarrow \ket{d_{k+1}} \\
        &U\left( t(\mu_{k+1} + \Delta \mu) \right)\ket{d_{k+1}} \rightarrow \ket{d_{i}}
    \end{align}
    In the above, $\Delta \mu$ is chosen such that
    \begin{equation}\label{eq:delta_mu}
        0 < \Delta \mu < \min\{\mu_{k+2} - \mu_{k+1}, \mu_{k+1}-\mu_{k}\},
    \end{equation}
    where $\mu_{k} < \mu_{k+1} < \mu_{k+2}$.
\end{theorem}

\Cref{theorem:eigenstates_dynamics} and its alternative formulation \Cref{theorem:eigenstates_dynamics_alter}  illustrate explicit dynamics of eigenstate transitions around the critical points $\mu_{\text{critical}} \equiv \mu_k$. Both theorems can be further restated in an equivalent but succinct statement which describes a swap of states around the critical points. To get an intuition for the upcoming statements, it is useful to examine \Cref{two-vecs} where we see the mechanics of a swap. Specifically, for $\mu < \mu_{\text{critical}}$, the eigenstate $\ket{s_{k}(\mu)} \approx \ket{v}$, whereas $\ket{s_{k+1}(\mu)} \approx \ket{d_{k+1}}$. However, for $\mu > \mu_{\text{critical}}$, we have $\ket{s_{k}(\mu)} \approx \ket{d_{k+1}}$ and $\ket{s_{k+1}(\mu)} \approx \ket{v}$.
\begin{theorem}[Eigenstates Swap]\label{theorem:eigenstate_swap}
    Let $d_{k+1}$ be an eigenvalue of $D$. Suppose that $\ket{v}$ is constructed as in (\ref{eq:choice_of_v_alternative}).
    Let $\mu_{k+1}$ and $\Delta \mu >0$ be the critical point and increment of $\mu_{k+1}$ as described in (\ref{eq:critical_mu_k+1}) and (\ref{eq:delta_mu}), respectively. Then, in the limit $\epsilon \rightarrow 0$, we have the following relation:
    \begin{align*}
        \ket{s_{k}(\mu_{k+1} - \Delta \mu)} = \ket{v} \ \text{ and } \ \ket{s_{k+1}(\mu_{k+1} - \Delta \mu)} = \ket{d_{k+1}} \\
        \ket{s_{k}(\mu_{k+1} + \Delta \mu)} = \ket{d_{k+1}} \ \text{ and } \ \ket{s_{k+1}(\mu_{k+1} + \Delta \mu)} = \ket{v}
    \end{align*}
    Here, for the adiabatic evolution operator $U(t(\mu))$ of $H(t)$, we have:
    \begin{equation}
        \ket{s_k(\mu)} := U(t(\mu))\ket{d_k}
    \end{equation} 
\end{theorem}

% We develop the intuition behind the theorem above by recalling that $\mu(t)$ is a monotonous and hence invertible function of time. This allows us to identify ``critical time points" $t(\mu_{k'p'})$, which are time moments when eigenstate transitions occur. For instance, let us assume we have chosen $\ket{v}$ with its dominant component being $\ket{a_i}\ket{b_j}$ and suppose we select a small time interval $\Delta t$ such that $t(\mu_{k'p'}) \pm \Delta t = t(\mu_{k'p'} \pm \Delta \mu)$. In the asymptotic sense of $\epsilon$ tending to zero, this implies that just before the critical time at $t(\mu_{k'p'}) - \Delta t$, the eigenstate $\ket{a_k}\ket{b_p}$ approaches $\ket{a_i}\ket{b_j}$. Then immediately after the critical time at $t(\mu_{k'p'}) + \Delta t$, the same eigenstate $\ket{a_k}\ket{b_p}$ approaches the higher-energy state $\ket{a_k'}\ket{b_p'}$, while $\ket{a_k'}\ket{b_p'}$ in turn approaches $\ket{a_i}\ket{b_j}$. Essentially, this evolution behaviour can be characterized as swapping because the $\ket{a_k}\ket{b_p}$ and $\ket{a_k'}\ket{b_p'}$ swap states around the critical time points $t(\mu_{k'p'})$.

\section{Maximum Entanglement Transfer is Computationally Expensive }\label{sec:eigengap}
In the context of entanglement transfer or preservation, the selection of the edge case $\ket{v}$ as defined in (\ref{eq:choice_of_v}) plays a crucial role. While this choice maximizes the transfer or preservation of entanglement, it also leads to the minimization of energy gaps within the Hamiltonian. Consequently, the evolution time $t_f$, which inversely correlates with the smallest energy gap according to (\ref{eq:adiabatic_condition_for_t}), is significantly extended in scenarios of nearly complete entanglement transfer.

This section provides a lower bound estimation for the energy gap between any two consecutive energy levels. To achieve maximal entanglement transfer or preservation, the state $\ket{v}$ is configured such that one of its components equals $1 - (\dim A \cdot \dim B - 1) \epsilon^2$, while the remaining components are set to $\epsilon$. The limit of maximal entanglement corresponds to $\epsilon \rightarrow 0$.

We will demonstrate that for sufficiently small $\epsilon > 0$, the minimum energy gap $g_{min}$ between any two adjacent energy levels is at the very least proportional to $\epsilon^2$.
\begin{theorem}\label{theorem:min_gap_theorem}
    Let $\ket{v}$ be constructed as in (\ref{eq:choice_of_v}).
    For index $k$, let $s_k(\mu(t)) < s_{k+1}(\mu(t))$ be any two consecutive eigenvalues of $H(t)$.
    Then, there exists $C_0>0$ such that to second order in $\epsilon$, we have: 
    \begin{equation}
        s_{k+1}(\mu(t)) - s_k(\mu(t)) \geq C_0 \epsilon^2  \text{ for all } t>0
    \end{equation}
\end{theorem}

\Cref{theorem:min_gap_theorem} highlights that the minimum energy gap $g_{min}$ is directly influenced by the value of $\epsilon$. This discovery is pivotal in establishing a relationship between the time complexity of adiabatic computation and the dynamics of entanglement transfer or preservation. Let \( \ket{v_\epsilon} \) denote the edge-case vector constructed in (\ref{eq:choice_of_v}), and let \( \ket{v} \) denote an arbitrary vector. By combining the result of \Cref{theorem:min_gap_theorem} with the adiabaticity condition given in (\ref{eq:adiabatic_condition_for_t}), we obtain the following bound:
\begin{align}
    t_f &\gg \max_{u \in [0,1]} \frac{ \left| \frac{d\tilde{\mu}}{du} \braket{\tilde{s}_m(u)}{v_\epsilon} \braket{v_\epsilon}{\tilde{s}_n(u)} \right|}{(C_0 \epsilon^2)^2} \label{eq:evolution_time_bound} \\
    &\geq \max_{u \in [0,1]} \frac{ \left| \frac{d\tilde{\mu}}{du} \braket{\tilde{s}_m(u)}{v} \braket{v}{\tilde{s}_n(u)} \right|}{\tilde{s}_{nm}(u)^2}, \quad n \neq m \label{eq:evolution_time_standard_condition}
\end{align}

Thus, the edge-case choice $\ket{v_\epsilon}$ with $\epsilon \ll 1 $ not only optimizes entanglement transfer or preservation but also increases the minimum required adiabatic evolution time. In particular, the right-hand side of (\ref{eq:evolution_time_bound}) provides an upper bound on the minimal evolution time over all Hamiltonians where $\ket{v}$ is chosen arbitrarily. That is, (\ref{eq:evolution_time_bound}) upper bounds the standard adiabatic condition in (\ref{eq:evolution_time_standard_condition}) for arbitrary choices of $\ket{v} $. This establishes a lower bound on $t_f$ required for successful entanglement transfer or preservation. Given that $\langle \tilde{s}_m(u) | v_\epsilon \rangle = O(\epsilon)$ and $ \langle \tilde{s}_n(u) | v_\epsilon \rangle = 1 - O(\epsilon^2)$ (or vice versa), the expression in (\ref{eq:evolution_time_bound}) implies a scaling of $t_f = O(1/\epsilon^3)$.

To conclude, we have shown that entanglement change, such as the transfer of entanglement, necessitates a slowdown. This follows from the fact that entanglement changes are due to eigenvalues swapping their eigenvectors as shown in \cref{theorem:eigenstates_dynamics,theorem:eigenstates_dynamics_alter,theorem:eigenstate_swap}. 
The swaps occur at avoided energy level crossings. To achieve complete swap and hence maximal entanglement transfer, the Hamiltonian's energy gaps must be minimized at these crossings. The minimum time required for an adiabatic evolution, however, is proportional to the square of the smallest energy gap. Therefore, a smaller gap, which is necessary for efficient entanglement transfer, requires a significantly prolonged evolution time to maintain adiabaticity, thus slowing down the computation. 

It is important to note that the redistribution of entanglement and the associated slowdown in adiabatic evolution depend solely on the narrowness of the avoided level crossings. This implies that it is immaterial whether we activate and manipulate a single rank-one projector or multiple projectors simultaneously. For a complete redistribution of entanglement to occur, two eigenvalues must approach one another within a distance of order $\epsilon^2$, as implied by \Cref{corollary:epsilon_square_convergence}. Therefore, the critical bottleneck remains the same: the need for eigenvalue proximity at the scale of $\epsilon^2$, which governs both the extent of entanglement transfer and the evolution time via the adiabatic condition.

We complete this section by recalling the interlacing inequalities delineated in (\ref{eq:interlacing_inequality}), which we, for convenience, restate below:
\begin{equation}
    d_{k} \leq s_{k} \leq d_{k+1} \leq s_{k+1}
\end{equation}
Given the construction of $\ket{v}$  that maximizes entanglement transfer or preservation, the eigenvalue $s_{k}$ converges to either $d_k$ or $d_{k+1}$ as $\epsilon$ tends to zero.
\begin{lemma}\label{corollary:epsilon_square_convergence}
    Let $\ket{v}$ be constructed as in (\ref{eq:choice_of_v}) and suppose that for $k \geq i,$ we have
    \begin{equation}
        \mu(t) > \frac{d_{k+1} - d_{i}}{|v_{i}|^2}.
    \end{equation}
    Then, there exists $C_1>0$ such that
    \begin{equation}
        d_{k+1} - s_{k}(\mu(t)) < C_1 \epsilon^2.
    \end{equation}
    Consequently,
    \begin{equation}
        \frac{\epsilon}{d_{k+1} - s_{k}(\mu(t))} > \frac{1}{C_1\epsilon}.
    \end{equation}
    Furthermore, if
    \begin{equation}
        0 \leq \mu(t) < \frac{d_{k}-d_{i}}{|v_{i}|^2},
    \end{equation}
    then there exists $C_2 \geq 0$ such that
    \begin{equation}
        s_{k} - d_{k} \leq C_2 \epsilon^2.
    \end{equation}
\end{lemma}
The lemma above will be pivotal in understanding the following section, where we demonstrate the entanglement transfer, preservation and orderly evolution of the eigenspectrum of $H(t)$ with narrowing eigengaps.

\subsection{\label{sec:qubit_example}Qubits Example}
To elucidate the principles and develop intuition behind Theorem~\ref{theorem:maximizing_transfer_preservation} and Theorem~\ref{theorem:eigenstates_dynamics}, we analyze a model wherein the systems $\tilde{A}, A,$ and $B$ are represented as qubits. Let $D = H_A \otimes I + I \otimes H_B$ be the initial Hamiltonian of systems $A$ and $B$. Suppose $D$ is a $4 \times 4$ matrix with distinct, positive eigenvalues:
\begin{equation}
    d_{00} = 5, d_{01} = 10, d_{10} = 12, d_{11} = 17
\end{equation}
Working in the eigenbasis $\{\ket{a_k}\ket{b_p}\}_{kp}$ of $D$, we let the initial state of the composite system be
\begin{equation}\label{eq:init_max_ent_state_qubit}
    \ket{\psi_{\tilde{A}AB}} = \alpha_{00}\ket{\tilde{a}_0}\ket{a_0}\ket{b_0} + \alpha_{11}\ket{\tilde{a}_1}\ket{a_1}\ket{b_0},
\end{equation}
where $|\alpha_{00}|=|\alpha_{11}|$,  and $\{\ket{\tilde{a}_k}\}$ is an arbitrary basis of $\tilde{A}$. Note that $\tilde{A}$ and $A$ are maximally entangled and $B$ is unentangled and pure. Selecting $\ket{v}$ in accordance with (\ref{eq:choice_of_v}) and setting $\epsilon=0.02$ yields:
\begin{equation}\label{eq:choice_ket_v_qubits}
    \ket{v} = \left(1-(nm-1)\epsilon^2\right)\ket{a_0}\ket{b_0} + \epsilon \sum_{kp\neq 00} \ket{a_k}\ket{b_p}
\end{equation}
We remark that $\ket{v}$ has a dominant component
\begin{align}\label{eq:velocity_by_v00}
    v_{00} &= 1-(nm-1)\epsilon^2 \nonumber \\
    &= 1 - (2\times 2 - 1) \times 0.02^2 \approx 0.9988.
\end{align}
For comparative analysis, we introduce another state $\ket{v'}$ constituted by random, non-zero coefficients, normalized to unity.

\begin{figure}[h]
    \centering
    \includegraphics[scale=0.15]{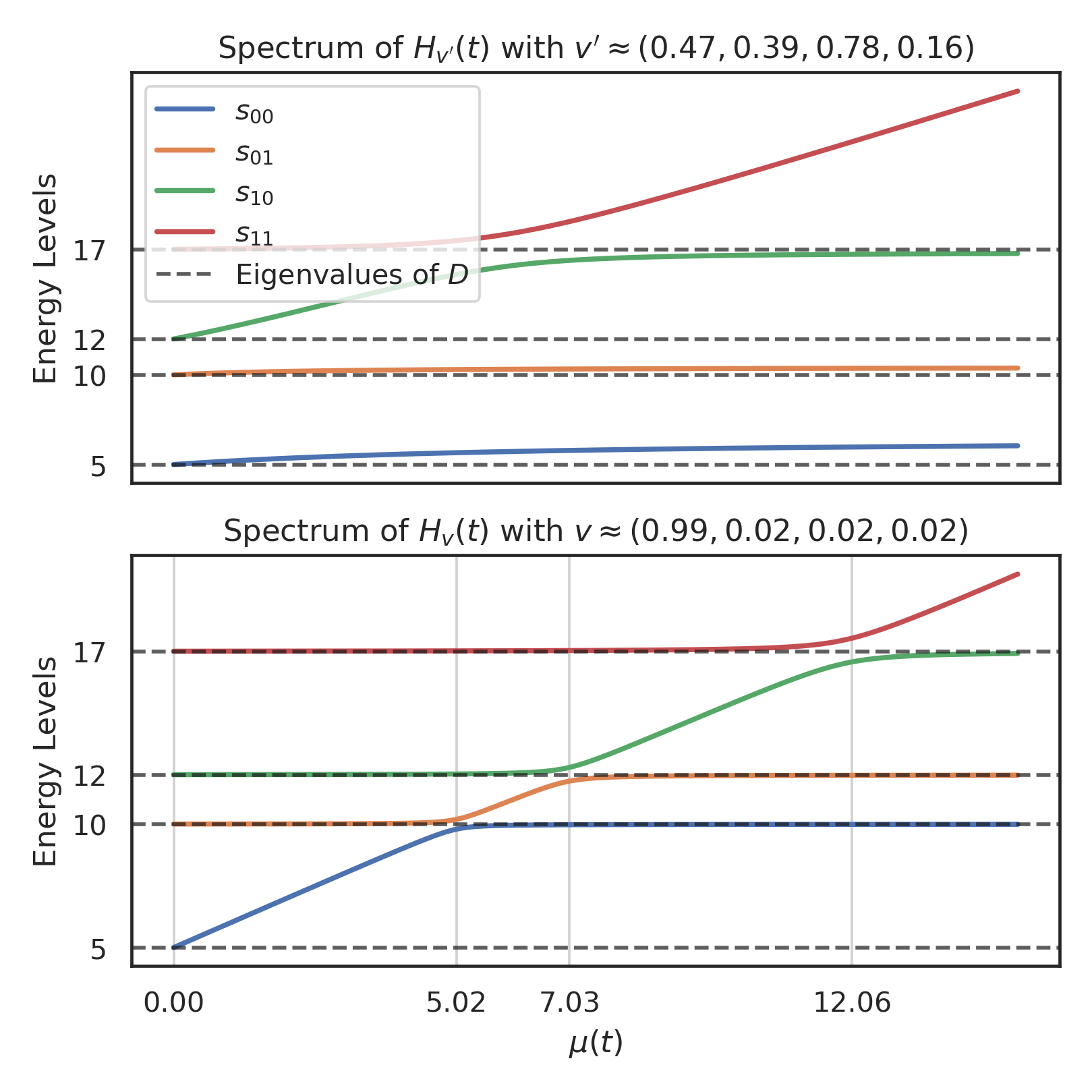}
    \caption{The generic dynamics of eigenvalues of $H_{v'}(t)$} (top) vs. the edge case dynamics of $H_v(t)$ (bottom).
    \label{fig:energy_levels_qubits}
\end{figure}

This establishes two scenarios: one where $\ket{v'}$ is arbitrarily selected, and a second where $\ket{v}$ is an edge case adhering to Theorem~\ref{theorem:maximizing_transfer_preservation}. The respective Hamiltonians for each scenario are:
\begin{align}
    H_{v'}(t) &= I \otimes D + \mu(t) I \otimes \ketbra{v'}{v'} \nonumber \\
     H_v(t) &= I \otimes D + \mu(t) I \otimes \ketbra{v}{v}
\end{align}
In \Cref{fig:energy_levels_qubits}, we present the evolution of the energy spectra as a function of the coupling parameter $\mu(t)$ for both outlined scenarios. Upon qualitative analysis, it is observed that the eigenvalues of $H_{v'}(t)$ ascend at various velocities, with various energy gaps. Conversely, the eigenvalues of $H_v(t)$ manifest a highly ordered progression. Specifically, for any given value of $\mu(t)$, a solitary eigenvalue ascends at a velocity of $|v_{00}|^2 \approx 1$, while other eigenvalues progress at a rate near zero. As was mentioned before, this distinct dynamical pattern adheres to the conservation of eigenvalue velocities (all velocities add to unity). Importantly, we see that energy gaps become narrower, yet no level-crossings are happening.

The systematic progression of eigenvalues associated with $H_v(t)$ facilitates an intuitive characterization of the time evolution of the state $\ket{\psi_{\tilde{A}AB}}$ in (\ref{eq:init_max_ent_state_qubit}). Let $I \otimes U(t_f)$ denote the adiabatic evolution operator prescribed by the Hamiltonian $H_v(t)$. Then the final evolved state is:
\begin{align}\label{eq:evolved_qubit_state}
    \ket{\psi_{\tilde{A}'A'B'}} = 
    \alpha_{00}&\ket{\tilde{a}_0}U(t_f)\ket{a_0}\ket{b_0} \nonumber \\
    &+\alpha_{11}\ket{\tilde{a}_1}U(t_f)\ket{a_1}\ket{b_0}
\end{align}
Let us consider the first term $U(t_f)\ket{a_0}\ket{b_0}$ of the equation above. Applying Lemma~\ref{lemma:eigenstate_s} and recalling \Cref{theorem:eigenstates_dynamics} we deduce that
\begin{align}\label{eq:first_term_evolved}
    &U(t_f)\ket{a_0}\ket{b_0} =  \eta_{00}\gamma_{00} (s_{00}I - D)^{-1}\ket{v} \nonumber \\
    &=  \eta_{00}\gamma_{00} \left(\frac{v_{00}}{s_{00}-d_{00}}\ket{a_0}\ket{b_0} + \frac{\epsilon}{s_{00}-d_{01}}\ket{a_0}\ket{b_1}\right) + O(\epsilon), \nonumber \\
\end{align}
where $\eta_{00}$ is a dynamical phase.
From the bottom subplot of \Cref{fig:energy_levels_qubits}, we see that on the interval $\mathcal{I}_1:=(5.02, +\infty)$, the lowest eigenvalue $s_{00}$ converges to the eigenvalue $d_{01}=10$ of $D$. Indeed, by \Cref{corollary:epsilon_square_convergence}, the coefficient $\epsilon/(s_{00}-d_{01}) \approx 1/\epsilon $ is a large number that dominates other coefficients. Hence,
\begin{equation}\label{eq:first_term_evolved_simplified}
    U(t_f)\ket{a_0}\ket{b_0} \approx \eta_{00}\ket{a_0}\ket{b_1}.
\end{equation}

Let us now examine the evolution of the second term $U(t_f)\ket{a_1}\ket{b_0}$ of the equation in (\ref{eq:evolved_qubit_state}). Consider the interval $\mathcal{I}_2 := (7.03, 12.06)$ which can be computed using eigenvalues of $D$ and (\ref{eq:mu_open_interval}) in Theorem~\ref{theorem:maximizing_transfer_preservation}. Then for $\mu_{t_f}$ in $\mathcal{I}_2$, by Lemma~\ref{lemma:eigenstate_s}, we have:
\begin{equation}
    U(t_f)\ket{a_1}\ket{b_0} = \eta_{10}\gamma_{10}(s_{10}I -D)^{-1}\ket{v}
\end{equation}
Since on the interval $\mathcal{I}_2$, the eigenvalue $s_{10}$ is not close to any of the eigenvalues and the coefficient $v_{00} \approx 0.9988$ [\ref{eq:velocity_by_v00}] of $\ket{a_0}\ket{b_0}$ dominates, we have:
\begin{align*}
    &U(t_f)\ket{a_1}\ket{b_0} = \eta_{10}\gamma_{10}(s_{10}I -D)^{-1}\ket{v} \nonumber \\
    &= \eta_{10}\gamma_{10}\left( \frac{v_{00}}{s_{10}-d_{00}}\ket{a_0}\ket{b_0} \nonumber + \sum_{kp\neq 00}\frac{\epsilon}{s_{10}-d_{kp}}\ket{a_k}\ket{b_p} \right) \nonumber \\
    &= \eta_{10}\gamma_{10}\frac{v_{00}}{s_{10}-d_{00}}\ket{a_0}\ket{b_0} + O(\epsilon)
\end{align*}
Therefore, we have:
\begin{equation}\label{eq:secomnd_term_evolved}
    U(t_f)\ket{a_1}\ket{b_0} \approx \eta_{10}\ket{a_0}\ket{b_0} \approx \eta_{10}\ket{v}
\end{equation}
Substituting the results in (\ref{eq:first_term_evolved_simplified}) and (\ref{eq:secomnd_term_evolved}) into the final state in (\ref{eq:evolved_qubit_state}), we get:
\begin{equation}
    \ket{\psi_{\tilde{A}'A'B'}} \approx \alpha_{00}\eta_{00}\ket{\tilde{a}_0}\ket{a_0}\ket{b_1} + \alpha_{11}\eta_{10}\ket{\tilde{a}_1}\ket{a_0}\ket{b_0}
\end{equation}
We immediately see that $\ket{a_0}$ can be factored out. Therefore, the system $A'$ is disentangled and pure, whereas system $\tilde{A}'$ and $B'$ are maximally entangled. Hence, for $\mu_{t_f} \in \mathcal{I}_2$ and  the choice of $\ket{v}$ in (\ref{eq:choice_ket_v_qubits}), we achieved maximum transfer of entanglement.

A similar analysis on the interval $\mathcal{I}_3 :=(12.06, +\infty)$ which is computed according to (\ref{eq:values_of_mu_larger}) in Theorem~\ref{theorem:maximizing_transfer_preservation} shows that the entanglement between $\tilde{A}'$ and $A'$ is reverted back, while $B'$ becomes disentangled and pure. To see this, we inspect the bottom subplot of \Cref{fig:energy_levels_qubits} and note that on the interval $\mathcal{I}_3$, we have $s_{00}$ and $s_{10}$ being close to $d_{01}=10$ and $d_{11}=17$ respectively. Hence:
\begin{align}\label{eq:terms_evolved_third_interval}
    U(t_f)\ket{a_0}\ket{b_0} &\approx \eta_{00}\ket{a_0}\ket{b_1},\nonumber \\
    U(t_f)\ket{a_1}\ket{b_0} &\approx \eta_{10}\ket{a_1}\ket{b_1}
\end{align}
Therefore, the final state is:
\begin{equation}
    \ket{\psi_{\tilde{A}'A'B'}} \approx \alpha_{00}\eta_{00}\ket{\tilde{a}_0}\ket{a_0}\ket{b_1} + \alpha_{11}\eta_{10}\ket{\tilde{a}_1}\ket{a_1}\ket{b_1}
\end{equation}
We readily see that $\ket{b_1}$ can be factored out, and $\tilde{A}'$ and $A'$ are maximally entangled. That is, the original entanglement is maximally preserved.

This example illustrates the complex yet ordered nature of adiabatic state evolution and its implications for the entanglement characteristics of a quantum system. From knowledge of the eigenvalues of the matrix $D$, one can compute specific intervals using (\ref{eq:open_interval_mu}) for the 
coupling parameter $\mu_{t_f}$ where entanglement transfer or preservation occurs.

\subsection{\label{sec:pathological_example}Construction of Pathological Hamiltonians Using Edge Cases}

\begin{figure}[t]
    \centering
    \includegraphics[scale=0.65]{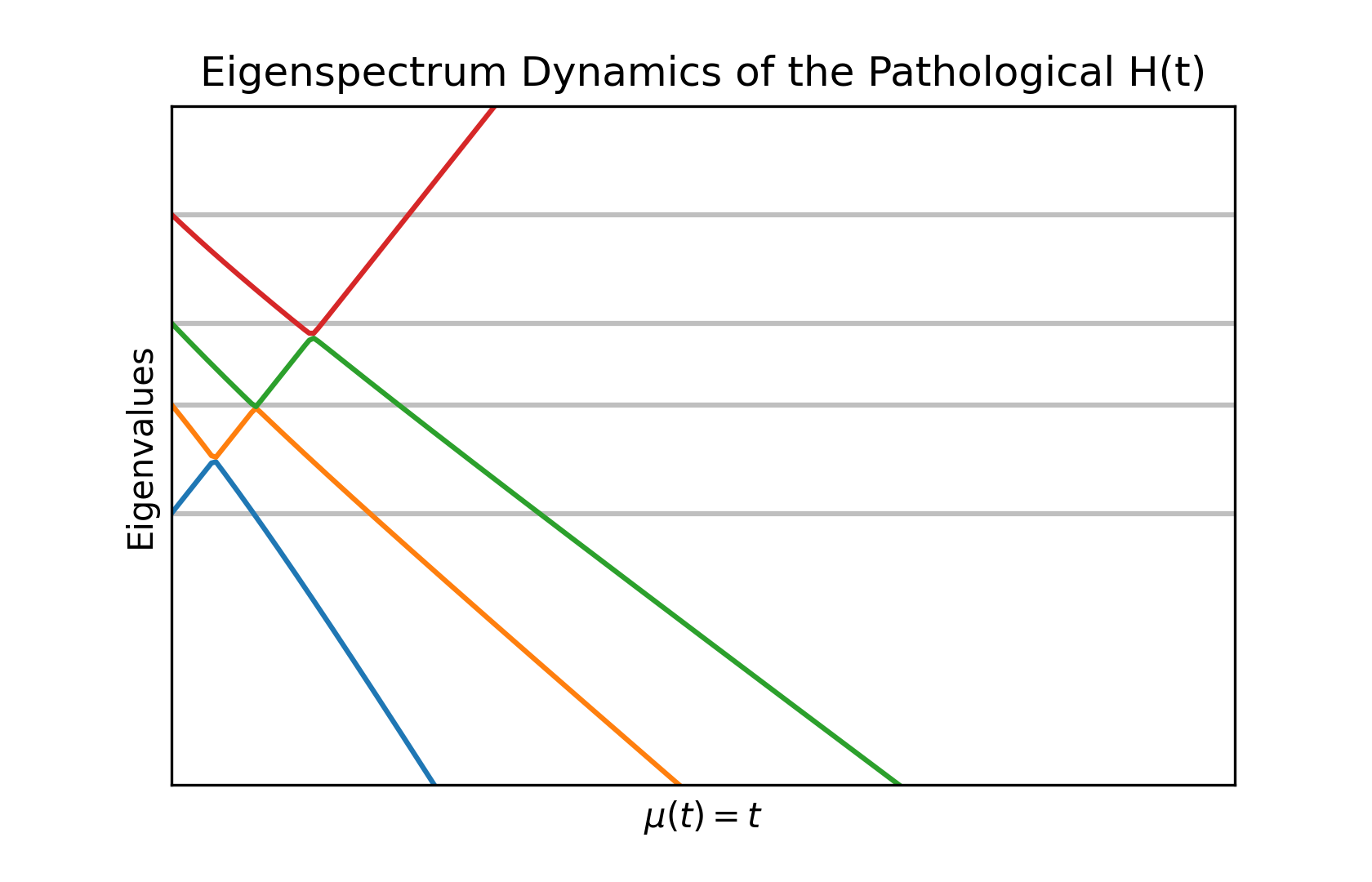}
    \caption{The dynamics of eigenvalues of the total Hamiltonian $H(t)$ comprised of edge cases. The spectrum is comprised of extremely narrow avoided level crossings, which are tightly clustered over a short time interval. The coloured lines represent trajectories of eigenvalues of $H(t)$, while the horizontal grey lines represent eigenvalues of the initial Hamiltonian $D$. \label{fig:pathological}}
\end{figure}

In this section, we demonstrate how edge cases can be utilized to construct a highly pathological interaction Hamiltonian $H_{int}$. Specifically, we will construct $H_{int}$ that features narrowly avoided level crossings that are clustered within a short time interval. This structure is designed to promote a cascade of undesirable diabatic transitions across the energy spectrum. Consequently, a system initialized in the ground state is highly likely to evolve to the highest energy state, representing a worst-case scenario and a complete loss of adiabatic control.
Our construction begins  with a $4\times 4$ initial Hamiltonian $D$ with eigenvectors $\{\ket{d_k}\}_{k=1}^4$. Let the adiabatic schedule be a linear function $g(t) = t$ where $t \in [0,1]$. Let us construct four edge cases using \Cref{theorem:maximizing_transfer_preservation}. For, $k=1, \ldots, 4$ and $\epsilon \ll 1$, the edge cases are:
\begin{align}
    \ket{v_k} = \left ( 1-3\epsilon^2 \right )\ket{d_k} + \epsilon\sum_{j=1, j\neq k}^4\ket{d_j}
\end{align}
Let us now define $H_{int}$ to be comprised of all $\ket{v_k}$ as follows: 
\begin{equation}
    H_{int} = \sum_{k=1}\lambda_k |v_k\rangle \langle v_k|
\end{equation}
Here, the parameters $\lambda_k$ denote the eigenvalues of $H_{int}$. One can choose $\lambda_k$ to shape the trajectories of eigenvalues of the total Hamiltonian $H(t) := D + t H_{int}$ without affecting the narrowness of the avoided level crossings. The trajectories of eigenvalues of $H(t)$ are illustrated in \Cref{fig:pathological}. There, we have chosen $\lambda = (70, -65, -55, -50)$ and $\epsilon = 10^{-1}$. The figure clearly reveals a cluster of extremely narrow avoided level crossings confined to a short time interval.

\section{\label{sec:AQC_Optimization} Application to Adiabatic Quantum Computation for Combinatorial Optimization}
\label{AQCsection}
The framework developed in this paper, analyzing entanglement dynamics through sequential projector activations and eigenvector swaps at avoided crossings, offers new perspectives on Adiabatic Quantum Computation (AQC), particularly on its application to combinatorial optimization \cite{farhi2000quantum, albash2018adiabatic}.

\subsection{Adiabatic Evolution and Problem Hardness}
\label{sec7}
In combinatorial optimization, the aim is to find an $n$-tuple, $x$, of binary variables, $x \in \{0,1\}^n$, which minimizes a cost function $f(x)$. We choose $f$ to be positive with the absolute minimum set to $0$. 
%\begin{align*}
 %   &\min_x f(x), ~~~
 %   \text{subject to } x \in \{0,1\}^n
%\end{align*}

In AQC, the problem is translated into finding the ground state of a problem Hamiltonian $H_p$. The eigenstates of $H_p$ are the computational basis states $\ket{x}$ and their eigenvalues $f(x)$ are the value of the cost function for $x$. 

The simplest AQC protocol adiabatically evolves the system from an easily-prepared ground state of a suitably chosen initial Hamiltonian $H_0$ (e.g., $H_0 = -\sum_i \sigma_x^{(i)}$) to the desired ground state of $H_p$ using a time-dependent Hamiltonian, with the schedule $H_{simple}(s) = (1-s)H_0 + s H_p$, where $s=t/T$ runs from $0$ to $1$.

For our purposes, it will be more convenient to adopt the equivalent schedule obtained by rescaling the Hamiltonian by $1/(1-s)$. We then obtain  $H(t) = H_0 + \mu(t) H_p$, where $\mu(t)=1/(T/t-1)$ and where $t$ runs from just above $0$ to just under $T$. Any other schedule $\mu(t)$ could be adopted in which the coupling $\mu(t)$ increases monotonically and adiabatically from $\mu(0)=0$ towards infinity. 
For later reference, we here already mention that, as was shown in \cite{vsoda2022newton}, in this limit the eigenvalues, $E_i(\mu)$, of $H(\mu)$ obey 
$\lim_{\mu\rightarrow\infty}E_i(\mu)=E_i^*$ where the $E_i^*$ were identified with the interlacing eigenvalues of the Cauchy interlacing theorem. 

If $\mu(t)$ grows adiabatically, the adiabatic theorem guarantees that the ground state $\ket{0(t)}$ of $H(t)$ adiabatically evolves towards the ground state of $H_p$. We can analyze this schedule using the methods of this paper by considering the spectral decomposition $H_p = \sum_x f(x) \ketbra{x}{x}$. The Hamiltonian becomes $H(t) = H_0 + \sum_x \mu(t)f(x) \ketbra{x}$. As $\mu(t)$ increases, the terms $\mu(t)f(x) \ketbra{x}$ associated with larger costs $f(x)$ become dominant earlier. This process implicitly implements the sequential activation of projectors discussed in \Cref{sec:methods}, by `activating' first the projector corresponding to the highest cost candidate solution and then successively activating the projectors of candidate solutions of lower and lower cost.

As analyzed in \cite{vsoda2022newton} and discussed above, when a projector $\vert v\rangle\langle v\vert$ becomes activated, then if its prefactor, say $\mu$, is let go to infinity, the vector $\vert v\rangle$ becomes an eigenvector of the total Hamiltonian. 

Let us now apply this to adiabatic quantum computing. We begin by considering the activation of any one of the projectors, say $\ketbra{x}$. When the prefactor $\mu(t)f(x)$ of the projector $\ketbra{x}$ eventually becomes very large, the computational basis state $\ket{x}$ gets closer and closer to being an eigenstate of the total Hamiltonian $H(t)$. This also means that all the other instantaneous eigenstates of $H(t)$ become orthogonal to the computational basis state $\vert x\rangle$. 
%(The activation's sequence of more or less narrowly avoided level crossings swaps all components of $\vert x\rangle$ that the lower states previously possessed up in energy to reach $\vert x\rangle$). 
Crucially, this means that with each of the successive activations of a projector $\ketbra{x}$, the instantaneous ground state $\vert 0(t)\rangle$ of $H(t)$ becomes orthogonal to that computational basis state $\vert x\rangle$.   

Consequently, as $\mu(t)$ increases, $\ket{0(t)}$ becomes progressively orthogonal to the computational basis states $\ket{x}$, starting with those of the highest costs $f(x)$. In effect, the AQC process systematically rules out the energetically highest, i.e., the worst candidate solution first, then the next worst, and so on, guiding the system towards the true ground state $\vert x_{GS}\rangle$ of $H_p$. 
Its eigenvalue, $f(x_{GS})$, is by construction zero, and therefore the ground state's projector has no effect.

\begin{figure}[h]
    \centering
    \includegraphics[scale=0.55]{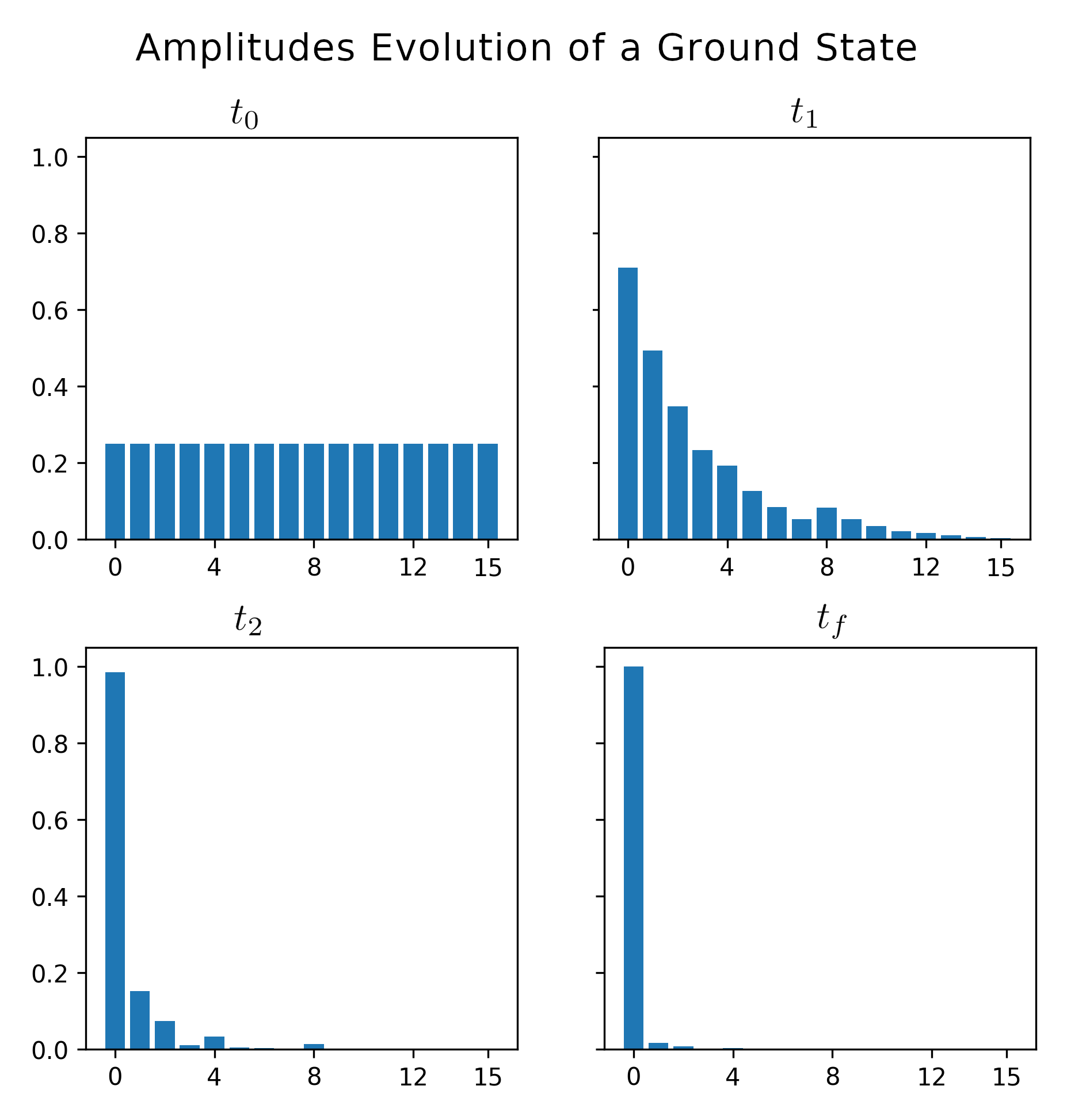}
    \caption{
    Ground state amplitudes $|\langle i | 0(t)\rangle|$ in the 16‐dimensional computational basis $i=0,\ldots, 15$ at four time slices $(t_0, t_1, t_2, t_f)$. As $t$ increases, amplitudes on higher‐energy states $i >0$ are suppressed, and the final state becomes concentrated on the ground state, $\ket{0000}$. Note that at $t_0$ the initial state is a product state -- a uniform superposition of all bitstrings which can be written as $\ket{+}^{\otimes 4}$. The final state, at $t_f$, is also a product state, which is a single bitstring representing the optimal solution.}
    \label{fig:ground_state_dynamics}
\end{figure}

In \Cref{fig:ground_state_dynamics}, we illustrate the dynamics of the evolving ground state $\ket{0(t)}$ 
for an example Hamiltonian $H_p$ where we solve $\min f(x)$ over  $x \in \{0,1\}^4$. 
The plots show the evolution of the magnitude of the amplitudes $c_x(t)$ of the ground state $\ket{0(t)} = \sum_x c_x(t) \ket{x}$ of the Hamiltonian $H(t) = H_0 + \mu(t)\sum_x f(x)\ketbra{x}$. Each integer $y$ on the horizontal axis encodes a bitstring, e.g., $y=0 \equiv \ket{0000}$, $y=15 \equiv \ket{1111}$. By design, we have $f(y) < f(y+1)$. Hence, the optimum solution is $\ket{0000}$. At $t_0$, the ground state is a uniform superposition of all candidate solution states, and it is a product state. As $\mu(t)$ increases, projectors $f(x)\ketbra{x}$ with successively smaller and smaller cost $f(x)$ are activated. Each activation of a projector $\ketbra{x}$ rotates the ground state $\ket{0(t)}$ to become orthogonal to $ \ket{x}$, i.e., suppressing $c_x(t)$. Indeed, we see that already at $t_1$, the highest cost candidates have near-zero amplitudes. As we will discuss below, the amount of entanglement in the ground state $\vert 0(t)\rangle$ at such an intermediate time directly depends on the size of the Hamming distances between the candidate states that are left in the ground state. By time $t_f$, the ground state is essentially the product state $\ket{0000}$ which minimizes $f(x)$.

\subsection{Entanglement as a Resource for Hard Problems}

We can now shed new light on the relation between the hardness of, for example, a classical problem of combinatorial optimization and the amount of entanglement that arises in the corresponding adiabatic quantum computation. Consider an intermediate stage of the quantum computation, where $\mu(t)$ is large enough that $H(t)$ is in large part dominated by projectors from $H_p$, but not so large that the instantaneous ground state $\ket{0(t)}$ is already essentially the ground state of $H_p$. At this stage, $\ket{0(t)}$ is primarily a superposition of low-cost computational basis states $\{\ket{x_{low}}\}$ which are candidates for the optimal solution.

The structure of the optimization problem landscape dictates the nature of this superposition:
\begin{itemize}
    \item \textbf{Rugged Landscape (Classically Hard) implies more entanglement:} If a set, $L$, of low-cost candidate solutions are separated by large Hamming distances, they represent distinct potential solutions corresponding to different local minima in the classical cost function. The ground state \begin{equation}
        \ket{0(t)} = \sum_{x \in L} c_x(t) \ket{x}
            \end{equation}
        is then a superposition of basis states of large Hamming distance. Such a superposition is generally less factorizable across qubit subsystems, implying a higher degree of multipartite entanglement in $\ket{0(t)}$.
    \item \textbf{Smooth Landscape (Classically Easier), implies less entanglement:} If the low-cost candidates are clustered together with small Hamming distances, they represent similar potential solutions. The ground state $\ket{0(t)}$ is a superposition of ``nearby" basis states. These states share many qubit states (e.g., $\ket{000...0}$ and $\ket{100...0}$ share all but the first qubit), allowing for greater factorizability and implying lower multipartite entanglement in $\ket{0(t)}$.
\end{itemize}

This leads to a crucial insight: classical hardness of a combinatorial optimization problem, in the sense of a rugged cost landscape with low-cost states that are distant in Hamming distance, necessitates that the AQC ground state, $\ket{0(t)}$, traverses states with high multipartite entanglement. The typical AQC process starts with an unentangled ground state of $H_0$, develops an ``entanglement peak" during the intermediate stages, and finally converges to a product state of $H_p$, if the global minimum is unique, or possibly into an entangled state if the ground state of $H_p$ is degenerate. We conclude that the magnitude of this entanglement peak directly correlates with the classical difficulty of the problem, since large Hamming distances directly translate into large multi-partite entanglement.

\section{\label{sec:entanglement_measures} Non-perturbative Measures of Entanglement and Coherent Information}
In this section, we focus on quantifying entanglement transfer or preservation in cases where the interaction term $\ketbra{v}{v}$ is chosen arbitrarily. As previously discussed in \Cref{sec:qubit_example} and illustrated in the top subplot of \Cref{fig:energy_levels_qubits}, an arbitrary selection of $\ket{v}$ results in a complex and unordered eigenspectrum dynamics in $H(t)$. As a result, this leads to only partial entanglement transfer or preservation. To accurately assess these effects, we introduce a series of non-perturbative equations. These equations serve as a generalization of the previously presented results. 

Our first inquiry centers on quantifying the residual entanglement within $\tilde{A}A$ following its interaction with $B$ during the course of adiabatic evolution, as governed by the Hamiltonian outlined in (\ref{eq:tripartite_hamiltonian}). Our second inquiry focuses on quantifying the entanglement transfer from $A$ to $B$, also following the course of the adiabatic evolution. For the first inquiry, we will work with the direct channel from $A$ to the adiabatically evolved $A'$. For the second inquiry, we will work with a complementary channel from $A$ to the adiabatically evolved $B'$. We denote the direct and complementary channels as ${A \rightarrow A'}$ and ${A \rightarrow B'}$, respectively. 

A quantum channel's quantum capacity, i.e., its ability to transmit entanglement, is usually defined through an optimization over the so-called coherent information \cite{gyongyosi2018survey}, for parallel uses of the channel with entangled input allowed. The coherent information for the channel ${A \rightarrow A'}$, the so-called direct channel, reads:
\begin{equation}\label{eq:coherent_inf}
    I^d = S(A') - S\left((\tilde{A}A)'\right)
\end{equation}
The coherent information for the complementary channel $A \rightarrow B'$, reads:
\begin{equation}
    I^c = S(B') - S\left((\tilde{A}B)'\right)
\end{equation}
Here, $S(\cdot)$ denotes the von Neumann entropy, and its argument denotes a subsystem for which the entropy is computed. Essentially, the coherent information compares joint and marginal von Neumann entropies and finding it positive proves the existence of entanglement. 
The von Neumann entropy is entanglement monotone if the total system is in a pure state, but its calculation requires the diagonalization of the density operator. This is where the generalization to $\alpha$-R\'enyi entropy proves useful. Defined for a positive real number $ \alpha \neq 1 $ as
\begin{equation}
    S_\alpha(\rho) := \frac{1}{1-\alpha} \log \text{Tr}(\rho^\alpha).
\end{equation}
The $ \alpha $-R\'enyi entropy encompasses the von Neumann entropy as a special case when $ \alpha $ approaches 1. Importantly, the $2$-R\'enyi entropy $ S_2(\rho) = -\log \text{Tr}(\rho^2) $ is directly related to the easy-to-calculate purity $\Tr(\rho^2)$ of a state, which does not require diagonalization of the density operator.

Furthermore, recent work has shown that comparing purities of combined and marginal systems tends to be significantly more effective than using von Neumann entropies for witnessing entanglement \cite{schneeloch2023negativity}. Motivated by this insight, and the ease of obtaining explicit analytical expressions for $2$-R\'enyi entropy, we introduce the concept of purity-based coherent information. 

This alternative formulation for the direct channel $A \rightarrow A'$ mirrors the conventional definition as given in (\ref{eq:coherent_inf}) and is expressed as:
\begin{equation}\label{eq:purity_coherent_info}
    PI^d := P\left((\tilde{A}A)'\right) - P\left(A'\right)
\end{equation}
Similarly, the purity-based coherent information for the complementary channel $A \rightarrow B'$ is defined as:
\begin{equation}\label{eq:purity_coherent_info_complement}
    PI^c = P\left((\tilde{A}B)'\right) - P\left(B'\right)
\end{equation}
Just as coherent information, purity-based coherent information, when positive, indicates the presence of quantum correlations between the subsystems. In contrast to the von Neumann entropy, which requires diagonalization of the density matrix, purities can be computed much more easily, analytically or determined empirically by creating interference between two identical instances of the quantum system \cite{ekert2002direct, brun2004measuring}.

Let us briefly examine a couple of trivial scenarios to illuminate the implications of (\ref{eq:purity_coherent_info}). Assume that after the adiabatic interaction, $(\tilde{A}A)'$ is in a pure, maximally entangled state. Under this condition, the subsystem $A'$ is maximally mixed. In other words, $P((\tilde{A}A)') = 1$ and $P(A') < 1$. This leads to the result that  $PI^d > 0$, affirming the statement that $(\tilde{A}A)'$ features entanglement. In a complementary manner, if $PI^d > 0$ and $(\tilde{A}A)'$ is found to be in a pure state, we can once again deduce that $(\tilde{A}A )'$ is indeed entangled. Generally, whenever $1 \geq P(({A}A)') >  P(A')$ it is always possible to infer that $(\tilde{A}A)'$ features entanglement \cite{schneeloch2023negativity}.

It is straightforward to transform the equations above into the 2-R\'{e}nyi-coherent information. First, we take the negative logarithms of $P((\tilde{A}A)')$ and $P(A')$ in (\ref{eq:purity_coherent_info}). This yields the conditional R\'enyi entropy $S_2(\tilde{A}'|A')$. Then, we multiply both sides of the equation by the negative one to get to the definition of coherent information. Therefore, the $2$-R\'{e}nyi-coherent information for the direct channel $A \rightarrow A'$ is defined as follows:
\begin{align}\label{eq:renyi_coherent_info}
    RI^d &:= \log P\left((\tilde{A}A)'\right) - \log P\left(A'\right) \\ \nonumber
    &= S_2(A') - S_2\left((\tilde{A}A)'\right)
\end{align}
Similarly, we define the $2$-R\'{e}nyi-coherent information for the complementary channel $A \rightarrow B'$ as:
\begin{align}\label{eq:renyi_coherent_info_complement}
    RI^c &:= \log P\left((\tilde{A}B)'\right) - \log P\left(B'\right) \\ \nonumber
    &= S_2(B') - S_2\left((\tilde{A}B)'\right)
\end{align}
Given the monotonic behaviour of the logarithm function, it can be readily inferred that a positive value for $RI^d$ serves as a witness for quantum correlations such as entanglement in $(\tilde{A}A)'$. This leads us to establish the following equivalence relation
\begin{equation*}   
    P\left((\tilde{A}A)'\right) >  P\left(A'\right) \iff S_2(A') > S_2\left((\tilde{A}A)'\right),
\end{equation*}
or equivalently,
\begin{equation*}
    RI^d > 0 \iff PI^d > 0.
\end{equation*}
Similarly, we can show that
\begin{equation}
    RI^c > 0 \iff PI^c > 0.
\end{equation}
In light of this equivalence, our subsequent analysis will focus on purities and $PI^d$ with $PI^c$, which offer the advantage of yielding more straightforward and transparent mathematical expressions. The next section presents the analytical non-perturbative expression for computing (\ref{eq:purity_coherent_info}) and (\ref{eq:renyi_coherent_info}).

\subsection{Coherent information of direct and complementary channels}
In this section, we derive analytical expressions for the purity-based coherent information of the direct channel $A \rightarrow A'$ and complementary channel $A \rightarrow B'$ after the adiabatic interaction of $A$ and $B$, as governed by the Hamiltonian outlined in (\ref{eq:tripartite_hamiltonian}) where the interaction term $\ketbra{v}$ is arbitrary. Throughout this section, let us assume that the system $\tilde{A}A$ is initialized in a pure entangled state while system $B$ starts as a pure and unentangled state. Under these premises, we derive several theorems and corollaries that characterize the evolution of entanglement from its initial value to arbitrary subsequent values.

We let the system $\tilde{A}A$ start in a pure state and choose an arbitrary basis $\{\ket{\tilde{a}_k}\}$ for the Hilbert space $\mathcal{H}_{\tilde{A}}$ of the system $\tilde{A}$. For some fixed $i$ and $j$, we define the state of $\tilde{A}A$ to be
\begin{equation}\label{eq:max_ent_state_tildeAA_coh_inf}
    \ket{\psi_{\tilde{A}A}} = \alpha_{ii}\ket{\tilde{a}_i}\ket{a_i} + \alpha_{jj}\ket{\tilde{a}_j}\ket{a_j},
\end{equation}
where $\alpha_{ii}$ and $\alpha_{jj}$ are complex coefficients. We may assume that $|\alpha_{ii}|^2 = |\alpha_{jj}|^2$, or in other words, the initial state $\ket{\psi_{\tilde{A}A}}$ starts out entangled. Then the joint initial state of $\tilde{A}AB$ is:
\begin{equation}\label{eq:max_ent_state_tildeAAB_for_purity}
    \ket{\psi_{\tilde{A}AB}} = \alpha_{ii}\ket{\tilde{a}_i}\ket{a_i}\ket{b_0} + \alpha_{jj}\ket{\tilde{a}_j}\ket{a_j}\ket{b_0}
\end{equation}
Since $\ket{a_i}$ and $\ket{b_0}$ are eigenstates of the initial Hamiltonian $D$, by \Cref{lemma:eigenstate_s}, the adiabatically evolved state is
\begin{align}\label{eq:evolved_max_ent_state_tildeAAB}
    \ket{\psi_{\tilde{A}'A'B'}} &=\alpha_{ii}\ket{\tilde{a}_i}U(t_f)(\ket{a_i}\ket{b_0}) \nonumber \\
    &\null \quad\quad\quad +\alpha_{jj}\ket{\tilde{a}_j}U(t_f)(\ket{a_j}\ket{b_0}) \\
    &=\alpha_{ii}\eta_{i0}\ket{\tilde{a}_i}\ket{s_{i0}} + \alpha_{jj}\eta_{j0}\ket{\tilde{a}_j}\ket{s_{j0}},
\end{align}
where $\eta_{i0}$ and $\eta_{j0}$ are dynamical phases. We show that the purity-based coherent information of the direct channel $A \rightarrow A'$ can be computed as follows.

\begin{theorem}\label{theorem:purity_and_renyi_coherent_info}
    Consider a tripartite quantum system $\tilde{A}AB$ initially prepared in a state described by (\ref{eq:max_ent_state_tildeAAB_for_purity}). Assume that the system undergoes adiabatic evolution according to the Hamiltonian $H(t)$, as defined by (\ref{eq:tripartite_hamiltonian}) with $\ketbra{v}$ being arbitrary. Then the purity-based coherent information for the direct channel $A \rightarrow A'$ is given by
    \begin{align}
        PI^d &= \nu \Tr\Big[ \Tr_B[\ketbra{s_{i0}}{s_{j0}}]  \Tr_B[\ketbra{s_{j0}}{s_{i0}}] \Big] \\ \nonumber
        &- \nu \Tr\Big[ \Tr_B[\ketbra{s_{i0}}{s_{i0}}]  \Tr_B[\ketbra{s_{j0}}{s_{j0}}] \Big],
    \end{align}
    where $\nu = 2 |\alpha_{ii}|^2 |\alpha_{jj}|^2$.
\end{theorem}
Similarly, the purity-based coherent information of the complementary channel $A \rightarrow B'$ is given by the following theorem.
\begin{theorem}\label{theorem:purity_and_renyi_coherent_info_complement}
    Under the assumptions stated in \Cref{theorem:purity_and_renyi_coherent_info}, the purity-based coherent information for the complementary channel $A \rightarrow B'$ is given by
    \begin{align}
        PI^c &= \nu \Tr\Big[ \Tr_A[\ketbra{s_{i0}}{s_{j0}}]  \Tr_A[\ketbra{s_{j0}}{s_{i0}}] \Big] \nonumber \\ 
        &- \nu \Tr\Big[ \Tr_A[\ketbra{s_{i0}}{s_{i0}}]  \Tr_A[\ketbra{s_{j0}}{s_{j0}}] \Big],
    \end{align}
    where $\nu = 2 |\alpha_{ii}|^2 |\alpha_{jj}|^2$.
\end{theorem}
% The supremum of $PI^c$ occurs when $(\tilde{A}B)'$ is pure and $B'$ is maximally entangled with $\tilde{A}'$ and hence maximally mixed. Thus, we have:
% \begin{align}\label{eq:sup_PI^c}
%     \sup PI^c &= \sup \left\{ P\left((\tilde{A}B)'\right) - P\left(B'\right) \right\} \nonumber \\
%     &= 1 - \frac{1}{\min\left\{\dim \tilde{A}, \dim B\right\}}.
% \end{align}

We conclude this section by presenting two important lemmas that are used to derive explicit formulas for the aforementioned coherent information measures in \Cref{theorem:purity_and_renyi_coherent_info} and \Cref{theorem:purity_and_renyi_coherent_info_complement}. Given the relation in \Cref{lemma:eigenstate_s}, we have:
\begin{equation}\label{eq:definition_of_ket_s}
    \ket{s} = \gamma (sI -D)^{-1}\ket{v}
\end{equation}
Then, we can explicitly compute the purity of the reduced density matrix of $\ketbra{s}{s}$.
\begin{lemma}\label{lemma:purity_formula}
    Let $\rho_{AB} := \ketbra{s}{s}$ and $\rho_A := \Tr_B \left[\rho_{AB}\right]$, then the purity of $\rho_A$ is given by
    \begin{align}\label{eq:purity}
        P(\rho_A) = \gamma^4 \sum_{ijkp} \frac{v_{ij}}{(s-d_{ij})} \frac{v^*_{kj}}{(s-d_{kj})} \frac{v_{kp}}{(s-d_{kp})}\frac{v^*_{ip}}{(s-d_{ip})},
    \end{align}
    where $v_{ij} = \bra{v}\ket{a_i, b_j}$ is a coefficient matrix of $\ket{v}$ in the eigenbasis $\{\ket{a_i} \ket{b_j}\}_{ij}$ of $D$ and
    \begin{equation}\label{eq:gamma_normalization}
        \gamma = \left(\sum_{ij} \left|\frac{v_{ij}}{s - d_{ij}}\right|^2\right)^{-1/2}.
    \end{equation}
\end{lemma}
We have presented a non-perturbative, explicit formula for calculating the purity of a reduced density matrix corresponding to the eigenstates of the final Hamiltonian $H(t_f)$. Below, we introduce a lemma, which is a generalization of \Cref{lemma:purity_formula}. This lemma is instrumental in deriving explicit equations in \Cref{theorem:purity_and_renyi_coherent_info} and \Cref{theorem:purity_and_renyi_coherent_info_complement}.
\begin{lemma}\label{lemma:generalized_purity}
    Let $\ket{s}$, $\ket{s'}$, $\ket{s''}$ and $\ket{s'''}$ be eigenstates defined as in (\ref{eq:definition_of_ket_s}). Then the following holds:
    \begin{multline}\label{eq:generalized_purity}
        \Tr\Big[ \Tr_B[\ketbra{s}{s'}] \Tr_B[\ketbra{s''}{s'''}]\Big] =\\
        \gamma \gamma' \gamma'' \gamma''' \sum_{ijkp} \frac{v_{ij}}{(s-d_{ij})} \frac{v^*_{kj}}{(s'-d_{kj})} \frac{v_{kp}}{(s''-d_{kp})}\frac{v^*_{ip}}{(s'''-d_{ip})}
    \end{multline}
\end{lemma}
Therefore, whenever $s=s',s'',s'''$, the equation in (\ref{eq:generalized_purity}) becomes $P(\rho_A)$ in (\ref{eq:purity}). Now, using \Cref{lemma:purity_formula} or \Cref{lemma:generalized_purity} it is straightforward to derive explicit equations for computing the quantities $\Tr[\Tr_B[\ketbra{s_{i0}}{s_{j0}}]]$ and $\Tr[\Tr_B[\ketbra{s_{i0}}{s_{i0}}]]$ that appear in \Cref{theorem:purity_and_renyi_coherent_info} and \Cref{theorem:purity_and_renyi_coherent_info_complement}. This, in turn, yields explicit non-perturbative equations for purity-based coherent information $PI^d$ and $PI^c$.

\subsection{Maximum and minimum of coherent information}
In this section, we show that the construction of the edge case $\ket{v}$ in \Cref{theorem:maximizing_transfer_preservation} that accomplishes entanglement transfer or preservation 
is just a special case of \Cref{theorem:purity_and_renyi_coherent_info} and \Cref{theorem:purity_and_renyi_coherent_info_complement} that maximizes/minimizes $PI^d$ and $PI^c$.

We commence by examining the initial quantum state specified by equation (\ref{eq:max_ent_state_tildeAAB_for_purity}), with the coefficients $\alpha_{ii}$ and $\alpha_{jj}$ each set to $1/\sqrt{2}$. A straightforward analysis reveals that initially (at $\mu=0$), the coherent information of the direct quantum channel from $A$ to itself stands at the maximum $PI_{\text{init}}^d = 1/2$. Concurrently, the initial coherent information associated with the complementary channel from $A$ to $B$ is at its minimum $PI_{\text{init}}^c = -1/2$. Given that $PI_{\text{init}}^d$ is positive, it is inferred that the system $\tilde{A}A$ is entangled, a fact that is directly observable from the initial state configuration. The derivation that $PI_{\text{init}}^d = 1/2$ is achieved through the application of \Cref{theorem:purity_and_renyi_coherent_info}, under the assumption that the system has not undergone evolution, resulting in $\ket{s_{i0}} = \ket{a_i}\ket{b_0}$ and $\ket{s_{j0}} = \ket{a_j}\ket{b_0}$. Therefore, the calculation proceeds as follows:
\begin{align*} 
    PI_{\text{init}}^d &= \nu \Tr\Big[ \Tr_B[\ket{a_i}\ket{b_0}\bra{a_j}\bra{b_0}]  \Tr_B[\ket{a_j}\ket{b_0}\bra{a_i}\bra{b_0}] \Big]\\
    & - \nu \Tr \Big [\Tr_B[\ket{a_i}\ket{b_0}\bra{a_i}\bra{b_0}]  \Tr_B[\ket{a_j}\ket{b_0}\bra{a_j}\bra{b_0}] \Big] \\
    &= \nu \Tr[\ketbra{a_i}{a_j} \ketbra{a_j}{a_i}] - \nu \Tr[\ketbra{a_i}\ketbra{a_j}]\\
    &= \nu \Tr[\ketbra{a_i}]\\
    &= 2 |\alpha_{ii}|^2|\alpha_{jj}|^2\\
    &= \frac{2}{4} = \frac{1}{2}
\end{align*}
The subsequent application of \Cref{theorem:purity_and_renyi_coherent_info_complement} (where we trace over $A$) yields $PI_{\text{init}}^c = -1/2$.

We now introduce an interaction term $\mu(t)\ketbra{v}$ to adiabatically evolve the initial state, where $\ket{v}$ is constructed as specified in \Cref{theorem:maximizing_transfer_preservation}, and $\mu(t_f)$ is selected based on the interval defined in equation (\ref{eq:mu_open_interval}). Specifically, the dominant component of $\ket{v}$, $v_{i0}$, is set to $1 - (nm-1)\epsilon^2$, and $\mu(t_f)$ is chosen from the interval $((d_{j0} - d_{i0})/|v_{i0}|^2, (d_{j1} - d_{i0})/|v_{i0}|^2)$.

By employing \Cref{theorem:eigenstates_dynamics} and \Cref{theorem:purity_and_renyi_coherent_info_complement}, we demonstrate that the coherent information of the complementary channel from $A$ to $B'$, denoted as $PI_{\text{final}}^c$, tends to its maximum value $1/2$ as $\epsilon$ tends to zero. This signifies that the entanglement initially present between $A$ and $\tilde{A}$ has been almost completely transferred to $B$ over the course of adiabatic evolution. The calculation for the coherent information of the complementary channel from $A$ to $B'$ is as follows:
\begin{align}
    PI_{\text{final}}^c &= \nu \Tr\Big[ \Tr_A[\ketbra{s_{i0}}{s_{j0}}]  \Tr_A[\ketbra{s_{j0}}{s_{i0}}] \Big] \nonumber \\ 
    &- \nu \Tr\Big[ \Tr_A[\ketbra{s_{i0}}{s_{i0}}]  \Tr_A[\ketbra{s_{j0}}{s_{j0}}] \Big]
\end{align}
Due to the construction of $\ket{v}$ and the choice of $\mu(t_f)$, by \Cref{theorem:eigenstates_dynamics} we know that as $\epsilon \rightarrow 0$, we have:
\begin{align*}
    \ket{s_{i0}} &= U(t_f)\ket{a_i}\ket{b_0} \rightarrow \ket{a_i}\ket{b_1},\\
    \ket{s_{j0}} &= U(t_f)\ket{a_j}\ket{b_0} \rightarrow \ket{a_i}\ket{b_0}
\end{align*}
Therefore, in the limit of $\epsilon \rightarrow 0$, we get:
\begin{align*} 
    PI_{\text{final}}^c &= \nu \Tr\Big[ \Tr_A[\ket{a_i}\ket{b_1}\bra{a_i}\bra{b_0}]  \Tr_A[\ket{a_i}\ket{b_0}\bra{a_i}\bra{b_1}] \\
    &- \nu \Tr \Big [\Tr_A[\ket{a_i}\ket{b_1}\bra{a_i}\bra{b_1}]  \Tr_A[\ket{a_i}\ket{b_0}\bra{a_i}\bra{b_0}] \Big]\\
    &= \nu \Tr[\ketbra{b_1}{b_0} \ketbra{b_0}{b_1}] -\nu \Tr[\ketbra{b_1} \ketbra{b_0}] \\
    &= \nu \Tr[\ketbra{b_1}]\\
    &= 2 |\alpha_{ii}|^2|\alpha_{jj}|^2\\
    &= \frac{2}{4} = \frac{1}{2}
\end{align*}
Correspondingly, it can be shown that in the limit 
$\epsilon \rightarrow 0$, we get $PI_{\text{final}}^d=-1/2$.

Hence, we establish that the value of coherent information of the complementary channel $PI_{\text{final}}^c$ at the end of the adiabatic evolution matches the initial maximum value of coherent information of the direct channel $PI_{\text{init}}^d$ of the system prior to evolution, confirming the nearly complete transfer of entanglement from $A$ to $B$ during the process.

% In other words, let $PI^c_{\text{final}}$ be a function of an arbitrary interaction term $\mu(t_f)\ketbra{v}$. Then we have
% \begin{equation}
%      \sup PI^c_{\text{final}}(\mu(t_f)\ketbra{v}) = PI^c_{\text{final}}(\mu^*(t_f)\ketbra{v^*}),
% \end{equation}
% where $\ket{v^*}$ has a dominant component $v_{i0} = 1 - (nm-1)\epsilon^2$ with $\epsilon \rightarrow 0$, and $\mu^*(t_f) \in ((d_{j0} - d_{i0})/|v_{i0}|^2, (d_{j1} - d_{i0})/|v_{i0}|^2)$. A similar bound can be attained for $PI^d_{\text{final}}$.

\section{\label{sec:outlook} Conclusions and Outlook}
%Entanglement, as a phenomenon without classical analog, is not only of foundational interest but also of practical importance as a key resource in quantum technologies. We developed here new nonperturbative methods to analyze the creation and transmission of entanglement which possess potential for wide applicability. Below, we outline a few directions for further research and applications.

\textbf{Adiabatic dynamics of entanglement in large $n$ multi-partite systems.} We decomposed adiabatic evolutions into a sequence of projector activations, and for each of these activations, we traced the changes of entanglement to eigenstate swaps that occur at avoided level crossings. 
We found that the speed and efficiency of the underlying swaps of eigenstates depend on how narrowly the level crossings are avoided, and this can be traced back to how closely the projector $\vert v\rangle\langle v \vert$ that is currently being activated is aligned to any prior eigenvector, $\vert d_k\rangle$.

It is worth noting that all of our main theorems and results are independent of the specific choice of adiabatic schedule and remain valid for general adiabatic computation or its discretized digital versions; i.e., gate-based quantum circuits. The sequential introduction of rank-one projectors serves not as a computational method but as an analytically tractable tool that isolates the fundamental mechanisms between entanglement behaviour and the spectral structure of a Hamiltonian.

While we showed this here explicitly for bi-partite and tri-partite systems, it is straightforward to extend this analysis to the dynamics of the more complex entanglement that can arise in $n$-partite systems for $n>3$, for example, in $n$-qubit adiabatic quantum computation.

\textbf{Entanglement Dynamics, Gaps, and Quantum Advantage}. 
%We applied our approach to adiabatic quantum computing, where we considered  the example of combinatorial optimization. There, we found a mechanism that translates the ruggedness of a cost function landscape, as characterized by the magnitude of the Hamming distances among low energy states such as local minima, into a correspondingly large peak of entanglement during the corresponding AQC. 

An interesting direction for future investigation concerns the relationship between the energy landscape structure — specifically the presence of Hamming-distant local minima — and the entanglement properties of the instantaneous ground state during adiabatic evolution. As we discussed above, the existence of such distant minima implies that, at some intermediate stage, after the high cost computational basis states have been eliminated from the ground state, the ground state $\ket{0(t)} $ must be a coherent superposition of regions of low-cost configurations that differ on a large number of qubits.  Such a cat-like state is fragile, see \cite{kendall2020dynamics} and, crucially, it cannot be factorized across any bipartition that splits the differing bits, implying the presence of large amounts of genuine multipartite entanglement. 

It will be very interesting to explore this connection further, particularly in relation to the nature of a potential quantum phase transition \cite{Farhi2000,Roland2002,Amin2009,Young2010,albash2018adiabatic} that the system may undergo during the adiabatic evolution, in the thermodynamic limit, as the regions of energy below a threshold that exist around local minima may link up or separate, analogously to Anderson localization. On Anderson localization, see, e.g., \cite{evers2008anderson}. In particular, the entanglement structure necessitated by the cost landscape might serve as a diagnostic for distinguishing first-order transitions that would come with exponentially small gaps and abrupt changes in ground state structure from second-order transitions, where the gap could close polynomially, potentially making the adiabatic quantum computation more efficient. A more detailed understanding of how the cost function landscape and corresponding entanglement build-up correlates with gap behavior and transition order could offer valuable insights into the design of more efficient adiabatic algorithms. 

%Let us now consider the set of those computational problems that are hard in the sense that the ruggedness of their cost function landscape leads to an entanglement peak that grows too fast to be classically simulatable. Within this set, of particular interest is the subset of computational problems that possess a quantum advantage, i.e., whose entanglement peak can nevertheless be efficiently (i.e., without undue slowdown) generated through adiabatic quantum dynamics. 

\bf New tools. \rm In the context of this set of questions, the tools and insights that we developed in this paper can be useful since they describe the elementary mechanism behind the adiabatic dynamics of entanglement: the effectively successive activation
of the projectors that make up the problem Hamiltonian, $H_p$, leading to 
the swaps of eigenvectors at the avoided level
crossings that change the entanglement structure.

In particular, we found in \Cref{sec:eigengap}, that the efficiency of these entanglement-changing processes is intrinsically linked to the size of their corresponding energy gaps.  
Specifically, edge cases, where the activated projector $\ket{v_k}$ is nearly aligned with a prior eigenstate, lead to the most efficient entanglement changes. But edge cases imply narrowly avoided crossings (\Cref{theorem:min_gap_theorem}) and narrow gaps $g_{min}$ necessitate a longer evolution time $T \propto 1/g_{min}^2$ (or potentially $1/g_{min}^3$, see \cite{amin2009consistency, albash2018adiabatic}) to maintain adiabaticity. Conversely, less narrow gaps can be traversed faster, but they generate less entanglement dynamics.

Therefore, at every avoided level crossing of the ground state, the link between the efficiency of entanglement redistribution and the size of the gap involves a nontrivial tradeoff: narrower gaps enable more effective redistribution of entanglement but necessitate slower evolution to preserve adiabaticity.

This trade-off needs to be understood to establish criteria for a quantum advantage. To this end, it will be useful 
to investigate further the non-edge cases. For this purpose, it will be very interesting to focus on the Cauchy interlaced values and their states. This is because, as discussed above, we can already infer from \cite{vsoda2022newton} that for all projector activations, edge case or not, the energy eigenvalues converge towards the interlaced eigenvalues of the Cauchy interlacing theorem. This means that for all projector activations, the instantaneous ground state $\vert E_1(\mu)\rangle$ converges toward the state $\vert E_1^*\rangle$. The edge cases are the special cases where $\vert E_2(0)\rangle \approx \vert E_1^*\rangle$.  

Beyond the analysis of eigenvector swaps (\Cref{sec:main_theorems,sec:permutations}), and the relationship between entanglement transfer and gaps (\Cref{sec:eigengap}), also the combinatorial observations and the quantitative measures of coherent information (\Cref{sec:entanglement_measures}) that we analyzed here, help provide a non-perturbative framework to investigate this intricate relationship between classical problem structure, quantum resource utilization (entanglement), and the potential for quantum advantage in AQC.

Further, also in quantum annealing, which has already found industrial applications \cite{inoue2021traffic, gabbassov2022transit, kitai2020designing, phillipson2024quantum, li2024optimisation}, it should be very interesting to use the present results to similarly explore the relationship between computational complexity, in the sense of the maximum speed of the computation, and the usage of the resource of entanglement. 

\textbf{Non-adiabatic quantum computation and Landau-Zener.}
More generally, the role of entanglement in adiabatic and adiabatic-inspired algorithms is an active area of research \cite{farhi2014quantum, zhou2020quantum, gabbassov2025lagrangian, nakhl2024calibrating, dupont2022entanglement, diez2021quantum, wiersema2020exploring, lanting2014entanglement} for which our results provide new tools. For example, the group theoretic approach in \Cref{sec:permutations} could enable qualitative tracking of entanglement, also possibly using graphical strategies, \cite{wen2023quantivine,bley2024visualizing},
without the need for full numerical simulations. 

In particular, it will be interesting to explore how the relationship between computational speed and the usage of the resource of entanglement that we found here generalizes to adiabatic-inspired and 
diabatic quantum computing algorithms such as QAOA \cite{farhi2014quantum} and adiabatic/diabatic quantum algorithms, see, e.g., \cite{crosson2021prospects, gabbassov2025lagrangian}, as well as to counter-diabatic algorithms, see, e.g., \cite{chandarana2022digitized, xu2024benchmarking}. To this end, it will be straightforward, for example, to generalize our present results by strategically permitting diabatic state transitions, namely by using the Landau-Zener transition probabilities to allow controlled non-adiabaticity at chosen avoided level crossings.

\textbf{Quantitative tracking of entanglement with coherent information.} Independently of AQC, in order to quantitatively track the creation and transfer of entanglement during any adiabatic interactions, we calculated coherent informations based on the 2-Renyi entropy. This allowed us to use our nonpertubative approach to track the dynamics of entanglement during the full duration of even strong interactions. For prior results on the perturbative tracking of the dynamics of entropies and entanglement during interactions, see,\cite{wen2022transfer,kendall2022transmission,kendall2020dynamics}. It will be interesting to apply our nonperturbative methods to the $1$-Renyi entropy-based coherent information since this will provide lower bounds for the quantum channel capacities of the direct and complementary channels.

\textbf{Quantum Error Correction.} 
In quantum processors, an important source of quantum errors is the inadvertent creation or transfer of entanglement due to unintended interactions among qubits or among qubits and their environment. Quantum error correction aims to protect quantum information through encoding and decoding techniques. Numerous studies, including those in adiabatic quantum computation, have focused on this goal \cite{mohseni2021error, amin2023quantum, young2013error, shingu2024quantum, bunyk2021error}. 

In this context, the theory presented in this work demonstrates, for example, that if a qubit $A$  is entangled with an ancilla $\tilde{A}$, it is possible to surgically disentangle $A$ from $\tilde{A}$ by having a qubit $B$ interact with $A$ using for example a tailored edge-case-like Hamiltonian. Equally, it is possible to choose the interaction between $A$ and $B$ such that the entanglement with $\tilde{A}$ becomes distributed between $A$ and $B$, see e.g., the tri-partite entangled state (\ref{eq:entanglement_distribution}). This ability to surgically entangle, disentangle and transfer entanglement, Hilbert space dimension by Hilbert space dimension, i.e., on the most elementary level, 
will be useful for quantum error prevention and correction in adiabatic, adiabatic-inspired and algorithmic quantum computation, see, e.g.,  \cite{mohseni2021error, amin2023quantum, young2013error, shingu2024quantum, bunyk2021error}. To this end, it should be very interesting to explore training machine learning models to use these surgical tools for quantum error prevention and correction adapted to the hardware at hand. For examples of machine learning applied to quantum computation (including adiabatic computation) and error correction, see e.g., \cite{zeng2020quantum, ding2021breaking, lin2020quantum, convy2022machine, kim2020quantum, henson2018approaching, khalid2023sample, ma2023adiabatic, mohseni2023deep}. 

\textbf{New entanglement measures.} Based on our finding that the dynamics of entanglement can be traced to eigenstate swaps at avoided level crossings, the development of new measures of multi-partite entanglement may be possible. For example, it will be interesting to explore measures based on the minimum number of projector activations, avoided level crossings and associated state swaps that are needed to create a given entangled state from an unentangled state. Such measures are nontrivial since, for example, for a projector activation to arrive at a desired critical $\mu$ value, it may first have to pass through other critical $\mu$ values that could impart collateral damage to the effort of generating the entangled target state. Similar to Rubik’s cube, a group theoretic analysis could be developed here, along with mnemonic devices. Inspired by the categorical approach to knot invariants 
\cite{gruenberg2006cohomological,kauffman2001knots,majid2000foundations}, it may also be possible to use category-theoretic methods to develop entanglement measures or entanglement invariants (i.e., functors) based on our findings. To this end, the over and under crossings in knots would be replaced by the different but reminiscent structure of eigenvector swaps at avoided level crossings. While the Reidemeister moves of knots lead to the quantum Yang Baxter equation whose solutions can be used to generate knot invariant polynomials, as is well known, see, e.g., \cite{majid2000foundations}, here it will be very interesting to pursue an analog approach by exploring possible analogs of Reidemeister moves for avoided level crossings.    

\bf Acknowledgements \rm
\newline AK and EG acknowledge 
support through a grant from the National Research Council of Canada (NRC). AK also acknowledges support through a Discovery Grant from the National Science and Engineering Council of Canada (NSERC) and a Discovery Project grant of the Australian Research Council (ARC). EG acknowledges partial support through a Mitacs Accelerate Fellowship. EG also acknowledges the support provided by NSERC through a Canada Graduate Scholarship – Doctoral Program (CGS-D).
$$$$

\bibliography{refs}% Produces the bibliography via BibTeX.

\newpage
\appendix

\section{\label{sec:permutations}Group theoretic view of
the special construction of $\ket{v}$}
In the previous section, we showed that the evolution with the edge case $\ket{v}$ constructed as in \Cref{theorem:eigenstate_swap} results in two eigenvalues swapping their eigenvectors. It is possible to represent this swapping as a transposition of two elements in a set comprised of eigenvectors of $H(t)$. A transposition in a set of elements is a permutation that exchanges two elements and keeps all the rest unchanged.

It is well known that the permutation group $S_N$ on $N$ elements can be generated by composing transpositions. This means that using multiple interaction terms, which are constructed according to \Cref{theorem:eigenstate_swap}, we can obtain any permutation of eigenvectors of the Hamiltonian $H(t)$. By this, we mean that upon addition of the specially constructed interaction terms $\mu^{(j)}(t)\vert v^{(j)}\rangle \langle v^{(j)} \vert$ to the initial Hamiltonian $D$, the eigenvalues of $H(t) = D + \sum_j \mu^{(j)}(t)\vert v^{(j)}\rangle \langle v^{(j)} \vert$ will have the eigenvectors of the initial Hamiltonian $D$ but permuted. To see this remarkable relation, consider the Hamiltonian
\begin{equation}\label{eq:first_iteration_hamiltonian}
    H(t) = D + \mu(t)\ketbra{v},
\end{equation}
where $\mu(0) = 0$. Let 
\begin{equation}
    M_0 := \{\ket{d_0}, \ket{d_1}, \ldots, \ket{d_k}, \ldots, \ket{d_{N-1}}\}
\end{equation}
denote an ordered set of eigenvectors of $H(0) = D$ with the corresponding eigenvalues ordered as
\begin{equation}
    d_0 < d_1 < \cdots  < d_k < \cdots < d_{N-1}.
\end{equation}
We now construct $\ket{v}$ in accordance with \Cref{theorem:eigenstate_swap}. Specifically, we set the $k$th component of $\ket{v}$ to be dominant, $v_k = 1 - (N-1)\epsilon^2$ where $\epsilon \ll 1$. The rest of the components are set to $\epsilon$. Let $t_f$ denote the final time. Set $\mu(t_f) = \mu_{k+1} + \Delta \mu$ where $\mu_{k+1} $ is the critical $\mu$-value defined as
\begin{equation}
    \mu_{k+1} = \frac{d_{k+1} - d_k}{|v_k|^2},
\end{equation}
and $0 < \Delta \mu < \mu_{k+2}$.

By \Cref{theorem:eigenstate_swap}, in the limit $\epsilon \rightarrow 0$, the final Hamiltonian $H(t_f)$ will have the following ordered set of eigenvectors
\begin{equation}
    M_{t_f} = \{\ket{d_0}, \ket{d_1}, \ldots, \ket{d_{k+1}}, \ket{d_k}, \ldots , \ket{d_{N-1}}\},
\end{equation}
with the corresponding eigenvalues
\begin{equation}
    s_0 < s_1 < \cdots < s_k < s_{k+1} < \cdots < s_{N-1}.
\end{equation}
We note that the final Hamiltonian $H(t_f)$ has the same eigenvectors as the initial Hamiltonian $H(0) = D$, but the $k$th and $k+1$ eigenvalues have traded eigenvectors, i.e., the eigenvalues $s_k < s_{k+1}$ have the associated eigenvectors $\ket{d_{k+1}}$ and $\ket{d_k}$, respectively. We have illustrated this behaviour in \Cref{two-vecs}.

% We remark that in order to perform such a clean eigenstate swap using adiabatic evolution, we need to surpass the critical point $\mu_{k+1}$. At this critical point, the gap between eigenvalues $s_{k}$ and $s_{k+1}$ is proportional to $\epsilon^2$ (see \Cref{theorem:min_gap_theorem}).

In the construction above, we have shown that choosing the $k$th component of $\ket{v}$ to be dominant and choosing the critical point $\mu_{k+1}$ accomplishes the transposition of eigenvectors $\ket{d_k}$ and $\ket{d_{k+1}}$. Using the group theoretic cycle notation, we let $(k, k+1)$ denote a transposition of elements $k$ and $k+1$, and $*$ denote the action of transposition on a set of elements. For example, $(3,4) * \{1,2,3,4\}= \{1,2,4,3\}$ signifies that $3$ is mapped to $4$, and $4$ is mapped to $3$. Therefore, the choice of the dominant component $v_k$ and the critical point $\mu_{k+1}$ can be succinctly represented as a transposition $(k,k+1)$ which acts on a set of eigenvectors $M_0$. Hence, we can write
\begin{equation}
    (k,k+1)*M_0 = M_{t_f}.
\end{equation}

It is now trivial to see how adding multiple rank-one projectors affects the eigenstates of $H(t)$. Let $D^{(1)}:= H(t_f)$ and add another interaction term $\mu^{(2)}(t) \vert v^{(2)}\rangle \langle v^{(2)} \vert$ such that $\mu^{(2)}(t) = 0$ for $t \leq t_f$. Then, the updated Hamiltonian is
\begin{align}
    H(t) &= D^{(1)} + \mu^{(2)}(t) \vert v^{(2)}\rangle \langle v^{(2)} \vert,
\end{align}
such that
\begin{align}
    H(t) &= D^{(1)}, \text{ for } t=t_f,\\
    H(t) &= D^{(1)} + \mu^{(2)}(t) \vert v^{(2)}\rangle \langle v^{(2)} \vert , \text{ for } t_f < t \leq t_f^{(2)}.
\end{align}
We, construct $\vert v^{(2)} \rangle$ with a dominant component $v^{(2)}_j$, choose the critical point $\mu_{j+1}^{(2)}$ and let $\mu(t_f^{(2)}) = \mu_{j+1}^{(2)} + \Delta \mu^{(2)}$. Using the cyclic notation for transposition, we can easily obtain the set $M_{t_f^{(2)}}$ of eigenvectors of $H(t_f^{(2)})$. To this end, perform the following computation:
\begin{equation}
    (j,j+1)(k,k+1)*M_0 = (j,j+1)*M_{t_f} = M_{t_f^{(2)}}
\end{equation}
That is, we first swap elements $k$ and $k+1$, and then we swap elements ${j,j+1}$.

This abstraction allows us to deduce several interesting properties about the addition of specially constructed rank-one projectors. For example, if $\mu^{(2)}(t) \vert v^{(2)}\rangle \langle v^{(2)} \vert$ is introduced before $\mu(t)\ketbra{v}$, how does the ordering of eigenstates change? We can easily answer this question by recalling that the transpositions $(j,j+1)$ and $(k,k+1)$ commute if $j \neq k$. Therefore, the ordering of eigenvectors of the final Hamiltonian $H(t_f^{(2)})$ is the same regardless of which interaction term is introduced first as long as $j \neq k$.

Recalling that transpositions generate the permutation group $S_N$, we can deduce that adding multiple specially constructed rank-one projectors is equivalent to composing transpositions, and by composing transpositions, we can achieve any permutation. Therefore, it is possible to induce any ordering of eigenstates of the final Hamiltonian.

We would like to point out that transpositions are not the only types of permutation that can be achieved. It is also possible to choose the critical point $\mu_k$ such that the associated permutation is a shift. To see this, choose the dominant component to be $v_k$ for $k < N-1$. Then choose the critical point $\mu_{N-1}$ and set $\mu(t_f) = \mu_{N-1} + \Delta \mu$. Then, the associated permutation is a shift to the right of all elements that are greater or equal to $k$. This shift in cycle notation is $(k, k+1, k+2,\ldots, N-1)$. The rest of the elements remain unchanged.

Since the set of permutations $S_N$ forms a group, it contains a neutral element, and each element in $S_N$ has an inverse. We can identify the neutral element of $S_N$ with the interaction term $\mu(t)\ketbra{v}$ where $\ket{v}$ is simply an eigenvector of $D$. Adding such an interaction term will keep the order of eigenvectors unchanged. To find the inverse of a permutation, it is sufficient to identify inverses of transpositions $(k,k+1)$. In other words, we want to find $\mu^{(2)}(t)\vert v^{(2)}\rangle \langle v^{(2)} \vert$, which reverses the action of $\mu(t)\ketbra{v}$ associated with the transposition $(k,k+1)$. To achieve this, we work in the eigenbasis of $D^{(1)} = D + \mu(t_f)\ketbra{v}$. Next, we choose a critical point $\mu^{(2)}_{k+1}$, set $\mu(t_f^{(2)}) = \mu^{(2)}_{k+1} + \Delta \mu^{(2)}$, and choose the dominant component $\ket{v}$ to be $v_k$. This can be again written as $(k,k+1)^{(2)}$, where the superscript indicates that transposition acts on eigenvectors of $D^{(1)} = D + \mu(t_f)\ketbra{v}$.

Having generalized the behaviour of eigenstates of the Hamiltonian under the addition of specially constructed rank-one projects, we can revisit \Cref{theorem:maximizing_transfer_preservation} and view it from the perspective of permutations. We again consider the tripartite system $\tilde{A}AB$, where $\tilde{A}A$ are entangled, and $B$ is in a pure, unentangled state.  For some fixed indices $i < j$, suppose eigenvalues of $D$ are ordered as
    \begin{equation}
        d_{i0} < d_{i1} < \cdots < d_{i\ell} < \cdots < d_{j0} < d_{j1},
    \end{equation}
and suppose that the combined system is initialized in the following state:
\begin{equation}\label{eq:init_max_ent_state_tildeAAB_permut}
    \ket{\psi_{\tilde{A}AB}} = \alpha_{ii}\ket{\tilde{a}_i}\ket{a_i}\ket{b_0} + \alpha_{jj}\ket{\tilde{a}_j}\ket{a_j}\ket{b_0}
\end{equation}
We construct $\ket{v}$ according to \Cref{theorem:maximizing_transfer_preservation}, namely, we let $v_{i\ell}$ to be the dominant component, and we choose the critical point $\mu_{j0} = (d_{j0}-d_{i\ell})/|v_{i\ell}|^2$ and set $\mu(t_f) = \mu_{j0} + \Delta \mu$. Then the initial state in (\ref{eq:init_max_ent_state_tildeAAB_permut}) will adiabatically evolve into
\begin{align}\label{eq:final_max_ent_state_permut}
    \ket{\psi_{\tilde{A}'A'B'}} &= \eta_{ii}\alpha_{ii}\ket{\tilde{a}_i}\ket{a_i}\ket{b_0} + \eta_{jj}\alpha_{jj}\ket{\tilde{a}_j}\ket{a_i}\ket{b_\ell} \nonumber \\
    &= \ket{a_i} \otimes (\eta_{ii}\alpha_{ii}\ket{\tilde{a}_i}\ket{b_0} + \eta_{jj}\alpha_{jj}\ket{\tilde{a}_j}\ket{b_\ell}).
\end{align}
In the above $\eta_{ii}, \eta_{jj}$ are dynamical phases. Importantly, $\ket{a_i}$ can be factored out. Therefore, $A$ is now unentangled and pure, whereas $\tilde{A}$ and $B$ are entangled. Essentially, we performed a shift and a swap in the eigenvectors of $H(t_f)$. Specifically, the eigenvalue $s_{j0}$ of $H(t_f)$ now has eigenvector $\ket{a_i}\ket{b_\ell}$, whereas the eigenvalue $s_{i\ell}$ has a acquired the eigenvector of the neighbouring larger eigenvalue  $d_{i,\ell+1}$. Having assigned (permuted) the eigenvectors of $H(t_f)$ to different eigenvalues, we made sure that the initial state in (\ref{eq:init_max_ent_state_tildeAAB_permut}) adiabatically evolves into (\ref{eq:final_max_ent_state_permut}).

\section{Proof of \Cref{lemma:eigenstate_s}}
Each activation of a projector is described by a time-dependent Hamiltonian of the form:
\begin{equation}\label{ap_eq:general_hamiltonian_rank1_proj}
    H(t) := D + \mu(t)\ketbra{v}{v}
\end{equation}

\setcounter{theorem}{0}
\begin{lemma}
   Let $H(t)$ in (\ref{ap_eq:general_hamiltonian_rank1_proj}) be the Hamiltonian of a quantum system with a non-degenerate spectrum. Let $\ket{v}$ be an arbitrary generic vector and $d_k$ be the eigenvalues of $D$. For any eigenvalue $s := s(\mu(t)) \neq d_k$ of $H(t)$, the corresponding eigenstate $\ket{s}$ is given by
    \begin{align}
        \ket{s} &= \gamma \big( sI - D \big)^{-1}\ket{v},
    \end{align}
   where $\gamma$ is a normalization constant defined as
   \begin{align}
       \gamma = \frac{1}{\| (sI - D)^{-1}\ket{v} \|}.
   \end{align}
\end{lemma}
Our goal is to derive an explicit expression for the eigenstate $\ket{s}$ of $H(t)$. Let $s := s(\mu(t))$ and $\mu := \mu(t)$. We want to consider the following eigenvector equation:
\begin{align}
    H(t)\ket{s} &= s\ket{s}\\
    \left(D + \mu \ketbra{v}{v} \right)\ket{s} &= s\ket{s}
\end{align}
Rearranging terms, we get:
\begin{align}
    (sI - D)\ket{s} &= \mu\ket{v}\braket{v}{s}\\
    \ket{s} &= \mu\braket{v}{s} (sI - D)^{-1}\ket{v}
\end{align}
Finally, we set $\gamma = 1/\|(sI - D)^{-1}\ket{v}\|$ and redefine $\ket{s}$ as
\begin{equation}
    \ket{s} := \gamma (sI -D)^{-1}\ket{v}.
\end{equation}

% \section{Proof of \Cref{lemma:eigenstate_s_adiabatic}}

\section{Proof of \Cref{lemma:explicit_eigenvalue_s}}
Explicit expression for $\mu$ as a function of the eigenvalue $s$ of $H(t)$ for all real $s$. Concretely, working in the eigenbasis $\{\ket{d_k}\}_{k=0}^{N-1}$ of the initial Hamiltonian $D$, we have:
\begin{equation}\label{ap_eq:function_mu}
    \mu(s) = \left(\sum_{k=0}^{N-1}\frac{|v_{k}|^2}{s - d_{k}} \right)^{-1}
\end{equation}

\setcounter{theorem}{2}
\begin{lemma}
    Let $s$ be an arbitrary real number. Consider the Hamiltonian defined by $H(s) = D + \mu(s)\ketbra{v}{v}$, where $\mu(s)$ is as in (\ref{ap_eq:function_mu}) and $\ket{v}$ is arbitrary. Then, $s$ is an eigenvalue of $H(s)$.
\end{lemma}
Consider an $N \times N$ Hermitian matrix $H(s) = D + \mu(s) \ketbra{v}$, where $\mu(s)$ is a scalar function to be determined. Let $\{d_k\}$ and $\{\ket{d_k} \}$ be the eigenvalues and eigenvectors of $D$. Define $v_k = \braket{d_k}{v}$. For a real number $s$, the characteristic polynomial $p(s)$ of $H(s)$ can be expressed as:
\begin{align}
    p(s) &= \det\left(H(s) - sI\right) \nonumber \\
    &= \det\left(D + \mu(s)\ketbra{v} - sI\right)
\end{align}
Using Cauchy’s formula for the determinant of a rank-one perturbation \cite[Section 0.8]{horn2012matrix}, we have:
\begin{equation}\label{eq:determinant_expression}
    p(s) = \det(D - sI)\left(1 - \mu(s) \sum_{k=0}^{N-1} \frac{|v_k|^2}{s - d_k}\right)
\end{equation}
Define $\mu(s)$ as:
\begin{equation}
    \mu(s) = \left(\sum_{k=0}^{N-1}\frac{|v_{k}|^2}{s-d_{k}}\right)^{-1}
\end{equation}
This choice of $\mu(s)$ ensures that the factor within the parentheses in the determinant expression [\ref{eq:determinant_expression}] becomes zero, leading to $p(s) = 0$. Therefore, $s$ is an eigenvalue of $H(s)$.

% Extending this result to a time-dependent Hamiltonian $H(t) = D + \mu(t) \ketbra{v}{v}$, consider $t_f$ to be the total time of an adiabatic evolution. Then, final Hamiltonian $H(t_f)$ is given by $D + \mu(t_f) \ketbra{v}$. Let $s^*$ be a fixed real number. Then by setting $\mu(t_f)$ to the value $\mu(s^*)$, we ensure that $s^*$ is an eigenvalue of the final Hamiltonian $H(t_f)$.

\section{Proof of \Cref{theorem:maximizing_transfer_preservation}}
\begin{theorem}
    Let $n = \dim A$ and $m = \dim B$. For some fixed indices $i < j$, suppose the eigenvalues of $D$ are ordered as
    \begin{equation}
        d_{i0} < d_{i1} < \cdots < d_{j0} < d_{j1}.
    \end{equation}
    Let system $\tilde{A}A$ be entangled and $B$ be in a pure, unentangled state:
    \begin{equation}\label{ap_eq:init_max_ent_state_tildeAAB}
        \ket{\psi_{\tilde{A}AB}} = \alpha_{ii}\ket{\tilde{a}_i}\ket{a_i}\ket{b_0} + \alpha_{jj}\ket{\tilde{a}_j}\ket{a_j}\ket{b_0}
    \end{equation}
    Fix an index $\ell \geq 0$ such that $d_{i0} \leq d_{i\ell} < d_{j0}$. For $\epsilon > 0$, if  $\ket{v}$ is such that 
    \begin{align}\label{ap_eq:choice_of_v}
        v_{kp} = 
        \begin{cases}
            1 - (nm - 1) \epsilon^2 & \text{for } k = i, p = \ell, \\
            \epsilon & \text{otherwise},
        \end{cases}
    \end{align}
   and
   \begin{equation}\label{ap_eq:mu_open_interval}
       \mu_{t_f} \in \left(\frac{d_{j0}-d_{i\ell}}{|v_{i\ell}|^2}, \frac{d_{j1}-d_{i\ell}}{|v_{i\ell}|^2}\right).
   \end{equation}
  Then, as $\epsilon \to 0$, the entanglement of the evolved system is such that $\tilde{A}$ becomes entangled with $B$, and $A$ becomes pure and unentangled.
    Furthermore, if 
    \begin{equation}\label{ap_eq:values_of_mu_larger}
        \mu_{t_f} > \frac{d_{j1}-d_{i\ell}}{|v_{i\ell}|^2},
    \end{equation}
    then as $\epsilon \to 0$, the entanglement of the evolved system is such that $\tilde{A}$ becomes again entangled with $A$, and $B$ becomes again pure and unentangled.
\end{theorem}
The proof is the direct application of \Cref{theorem:eigenstates_dynamics} or \Cref{theorem:eigenstates_dynamics_alter}. Let the system $\tilde{A}A$ be entangled and $B$ be in a pure and unentangled state:
\begin{equation}\label{eq:proof_init_max_ent_state_tildeAAB}
    \ket{\psi_{\tilde{A}AB}} = \alpha_{ii}\ket{\tilde{a}_i}\ket{a_i}\ket{b_0} + \alpha_{jj}\ket{\tilde{a}_j}\ket{a_j}\ket{b_0}.
\end{equation}
For $\epsilon > 0$ and fixed index $i$, define $\ket{v}$ in the eigenbasis $\{\ket{a_k}\ket{b_p}\}_{kp}$ of $H_A$ and $H_B$ such that
\begin{align}\label{eq:proof_choice_of_v}
        v_{kp} = 
        \begin{cases}
            1 - (nm - 1) \epsilon^2 & \text{for } k = i, p = 0, \\
            \epsilon & \text{otherwise}.
        \end{cases}
\end{align}

Suppose that $\mu_{t_f}$ is in the interval $\left(\frac{d_{j0}-d_{i0}}{|v_{i0}|^2}, \frac{d_{j1}-d_{i0}}{|v_{i0}|^2}\right)$. Then, by \Cref{theorem:eigenstates_dynamics}, at the limit $\epsilon \rightarrow 0$, the state
$\ket{a_i}\ket{b_0}$ evolves into the next higher-energy state $\ket{a_i}\ket{b_1}$, while the state $\ket{a_j}\ket{b_0}$ evolves into $\ket{a_i}\ket{b_0}$. Hence, the final state is:
\begin{align}\label{eq:proof_final_max_ent_state_tildeAAB}
    \ket{\psi_{\tilde{A}'A'B'}} &= \eta_{ii}\alpha_{ii}\ket{\tilde{a}_i}\ket{a_i}\ket{b_1} + \eta_{jj}\alpha_{jj}\ket{\tilde{a}_j}\ket{a_i}\ket{b_0} \nonumber \\
    &= \ket{a_i} \otimes \left(\eta_{ii}\alpha_{ii}\ket{\tilde{a}_i}\ket{b_1} + \eta_{jj}\alpha_{jj}\ket{\tilde{a}_j}\ket{b_0}\right)
\end{align}
In the above, $\eta_{ii}$ and $\eta_{jj}$ are dynamical phases that were acquired during the evolution. Importantly, we immediately see that $\ket{a_i}$ can be factored out, which implies that system $A$ is now untangled and pure, whereas $\tilde{A}$ and $B$ are entangled.

Suppose that $\mu_{t_f} > \frac{d_{j1}-d_{i0}}{|v_{i0}|^2}$. Then, by \Cref{theorem:eigenstates_dynamics}, at $\epsilon \rightarrow 0$, the states $\ket{a_i}\ket{b_0}$ and $\ket{a_j}\ket{b_0}$ evolve to their next respective higher-energy states: $\ket{a_i}\ket{b_1}$ and $\ket{a_j}\ket{b_1}$. Hence, the final state of the total system is:
\begin{equation}
    \ket{\psi_{\tilde{A}AB}} = \eta'_{ii}\alpha_{ii}\ket{\tilde{a}_i}\ket{a_i}\ket{b_1} + \eta'_{jj}\alpha_{jj}\ket{\tilde{a}_j}\ket{a_j}\ket{b_1}
\end{equation}
In the above, $\eta'_{ii},\eta'_{jj}$ are dynamical phases acquired during the evolution. Furthermore, the state $\ket{b_1}$ can be factored out. We deduce that the system $B$ is now pure and unentangled, whereas $\tilde{A}$ and $A$ are entangled as in the initial state. As such, the entanglement structure is preserved.

\section{Proof of \Cref{corollary:epsilon_square_convergence}}
\setcounter{theorem}{8}
\begin{lemma}
    Let $\ket{v}$ be constructed as in (\ref{ap_eq:choice_of_v}) and suppose that for $k \geq i,$ we have
    \begin{equation}
        \mu(t) > \frac{d_{k+1} - d_{i}}{|v_{i}|^2}.
    \end{equation}
    Then, there exists $C_1>0$ such that
    \begin{equation}
        d_{k+1} - s_{k}(\mu(t)) < C_1 \epsilon^2.
    \end{equation}
    Consequently,
    \begin{equation}
        \frac{\epsilon}{d_{k+1} - s_{k}(\mu(t))} > \frac{1}{C_1\epsilon}.
    \end{equation}
    Furthermore, if
    \begin{equation}
        0 \leq \mu(t) < \frac{d_{k}-d_{i}}{|v_{i}|^2},
    \end{equation}
    then there exists $C_2 \geq 0$ such that
    \begin{equation}
        s_{k} - d_{k} \leq C_2 \epsilon^2.
    \end{equation}
\end{lemma}
We first prove the \Cref{corollary:epsilon_square_convergence} and use it to prove \Cref{theorem:eigenstates_dynamics}.
Let $N$ be the dimension of the Hilbert space $H(t)$ acts on. Let $\{\ket{d_j}\}_{j=0}^{N-1}$ be the eigenbasis of $D$ with eigenvalues $d_0 < \cdots  < d_{N-1}$. Consider the characteristic polynomial of $H(t)$ in the eigenbasis $D$ and $\mu \geq 0$:
\begin{align}
    p(s) &= \det(D + \mu \ketbra{v} - sI)\\
    &= \det(D - sI) \left (1 + \mu \sum_{j=0}^{N-1} \frac{|v_j|^2}{d_j - s} \right )\\
    &= \prod_{j=0}^{N-1} (d_j - s) \left (1 + \mu \sum_{j=0}^{N-1} \frac{|v_j|^2}{d_j - s} \right ).
\end{align}
Fix index $i$ and construct $\ket{v}$ according to (\ref{ap_eq:choice_of_v_dynamics}) in \Cref{theorem:eigenstates_dynamics}. Then,
\begin{equation}
    v_j = \begin{cases}
        1 - (N-1)\epsilon^2 &\text{ for } j = i,\\
        \epsilon &\text{ otherwise.}
    \end{cases}
\end{equation}
Without loss of generality, assume that $i=0$, then $v_0 = 1 - (N-1)\epsilon^2$ and $v_j = \epsilon$ for $j > 0$. The characteristic polynomial becomes:
\begin{equation}
    p(s) = \prod_{j=0}^{N-1} (d_j - s) \left (1 + \mu\frac{|v_0|^2}{d_0 -s} + \mu \epsilon^2 \sum_{j=1}^{N-1} \frac{1}{d_j - s} \right )
\end{equation}
For any index $p$, due to the interlacing inequality, there exists eigenvalue $s$ of $H(t)$ such that $d_p \leq s < d_{p+1}$. Without loss of generality and for simplicity, we consider $p=0$. Then, $d_0 \leq s < d_{1}$ for $\mu \geq 0$. Hence, we have:
\begin{equation}
    p(s) = \prod_{j=0}^{N-1} (d_j - s) \left (1 + \mu\frac{|v_0|^2}{d_0 -s} + \mu \epsilon^2 \sum_{j=1}^{N-1} \frac{1}{d_j - s} \right ) = 0
\end{equation}
Assuming $\mu > 0$, we get $d_0 < s < d_{1}$. Therefore, we must have
\begin{equation}
    1 + \mu\frac{|v_0|^2}{d_0 -s} + \mu \epsilon^2 \sum_{j=1}^{N-1} \frac{1}{d_j - s}  = 0.
\end{equation}
Rewrite the above as
\begin{equation}
    1 + \mu\frac{|v_0|^2}{d_0 -s} + \mu \epsilon^2 \frac{1}{d_1 - s} + \mu \epsilon^2 \sum_{j=2}^{N-1} \frac{1}{d_j - s}  = 0.
\end{equation}
Since $d_0 < s < d_1 < \cdots < d_{N-1}$, we have $d_j - s > d_j - d_1$ for $j \geq 2$. It follows that
\begin{equation}
    \sum_{j=2}^{N-1} \frac{1}{d_j - s} < \sum_{j=2}^{N-1} \frac{1}{d_j - d_1}.
\end{equation}
Define $Z := \sum_{j=2}^{N-1} \frac{1}{d_j - d_1}$ and note that it is independent of $s$ and hence independent of $\mu$. Thus,
\begin{equation}
    1 + \mu\frac{|v_0|^2}{d_0 -s} + \mu \epsilon^2 \frac{1}{d_1 - s} + \mu \epsilon^2 Z > 0.
\end{equation}
Rearranging yields:
\begin{equation}
    \epsilon^2 \frac{1}{d_1 - s} > -\frac{1}{\mu} + \frac{|v_0|^2}{s - d_0} - \epsilon^2 Z.
\end{equation}
Since $d_0 < s < d_1$, we have $s - d_0 < d_1 - d_0$. Therefore, $|v_0|^2/(s - d_0) > |v_0|^2/(d_1 - d_0)$. Using this, we get
\begin{equation}
    \epsilon^2 \frac{1}{d_1 - s} > -\frac{1}{\mu} + \frac{|v_0|^2}{d_1 - d_0} - \epsilon^2 Z.
\end{equation}
Rearranging terms yields:
\begin{equation}
    \epsilon^2 + \epsilon^2 (d_1 - s) Z > (d_1 - s) \left (-\frac{1}{\mu} + \frac{|v_0|^2}{d_1 - d_0} \right)
\end{equation}
Since $d_1 - s < d_1 - d_0$, we can write:
\begin{equation}
    \epsilon^2 + \epsilon^2 (d_1 - d_0) Z > (d_1 - s) \left (-\frac{1}{\mu} + \frac{|v_0|^2}{d_1 - d_0} \right)
\end{equation}
The above can be rewritten as:
\begin{equation}\label{eq:proof_ineq_eps_sqr}
    \frac{\epsilon^2 (1 + (d_1 - d_0)Z)}{-\frac{1}{\mu} + \frac{|v_0|^2}{d_1 - d_0}} > d_1 - s
\end{equation}
Note that the denominator can be re-expressed as:
\begin{equation}
    -\frac{1}{\mu} + \frac{|v_0|^2}{d_1 - d_0} = \frac{|v_0|^2 \mu + d_0 - d_1}{(d_1 - d_0)\mu}
\end{equation}
Hence, (\ref{eq:proof_ineq_eps_sqr}) can be written as:
\begin{equation}
    \frac{\epsilon^2 (1 + (d_1 - d_0)Z)(d_1 - d_0)\mu}{|v_0|^2\mu + d_0 - d_1} > d_1 - s
\end{equation}
Since $\mu > (d_1 - d_0)/|v_0|^2$, we have $|v_0|^2\mu + d_0 - d_1 > 0$. Finally, we note that $|v_0|^2 < 1$, therefore $|v_0|^2 \mu < \mu < 2\mu$. Thus, we have:
\begin{equation}
    \frac{\epsilon^2 (1 + (d_1 - d_0)Z)(d_1 - d_0) 2 \mu}{\mu + d_0 - d_1} > d_1 - s
\end{equation}
We can eliminate the dependence on $\mu$ by letting $\mu \rightarrow +\infty$. In the limit $\mu \rightarrow +\infty$, we get a tighter bound:
\begin{equation}
    \epsilon^2 \left (1 + (d_1 - d_0)Z \right)(d_1 - d_0) 2 > d_1 - s
\end{equation}
Using similar arguments as above, it is straightforward to show that for $0 < \mu < d_1 - d_0 < (d_1 - d_0)/|v_0|^2$ and $d_1 < s < d_2$, we have
\begin{equation}
    \frac{\epsilon^2 \mu (d_1 - d_0)}{d_1 - \mu - d_0} \geq s - d_1.
\end{equation}
We note that the bound can be saturated for $\mu = 0$. In this case, we have $H(t) = D$, which means $s = d_1$.
This proves \Cref{corollary:epsilon_square_convergence}.

\section{Proof of \Cref{theorem:eigenstates_dynamics}}
\setcounter{theorem}{4}
\begin{theorem}\label{ap_theorem:evolution}
    Let $N$ be the dimension of the Hilbert space that $H(t)$ acts on and $\{d_j\}_{j=0}^{N-1}$ be the increasing sequence of the eigenvalues of $D$.
    
    Fix index $i$ and let $\epsilon > 0$. Suppose $\ket{v}$ is a vector characterized by its components $v_{j} = \braket{v}{d_{j}}$ as follows:
    \begin{equation}\label{ap_eq:choice_of_v_dynamics}
        v_{j} = 
        \begin{cases}
            1 - (N - 1) \epsilon^2 & \text{if } j = i\\
            \epsilon & \text{otherwise}
        \end{cases}
    \end{equation}
    For $k> i$, if $\mu_{t_f}$ lies within the interval
    \begin{equation}
        \left( -\infty, \frac{d_{k}-d_{i}}{|v_{i}|^2} \right),
    \end{equation}
    then the eigenstate $\ket{d_k}$ of $D$ adiabatically evolves as
    \begin{equation}\label{ap_eq:conversion_dk_to_dk}
        U_{t_f}\ket{d_k} \rightarrow \ket{d_k} \text{ as } \epsilon \rightarrow 0.
    \end{equation}
    If $\mu_{t_f}$ lies within the interval
    \begin{equation}\label{ap_eq:open_interval_mu}
       \left( \frac{d_{k}-d_{i}}{|v_{i}|^2},\frac{d_{k+1} - d_{i}}{|v_{i}|^2} \right),
    \end{equation}
    then
    \begin{equation}\label{ap_eq:conversion_dk_to_di}
        U_{t_f}\ket{d_k} \rightarrow \ket{d_i} \text{ as } \epsilon \rightarrow 0.
    \end{equation}
    If $\mu_{t_f}$ lies within the interval
    \begin{equation}\label{ap_eq:third_interval}
       \left( \frac{d_{k+1} - d_{i}}{|v_{i}|^2}, +\infty \right),
    \end{equation}
    then
    \begin{equation}\label{ap_eq:conversion_dk_to_dk+1}
        U_{t_f}\ket{d_k} \rightarrow \ket{d_{k+1}} \text{ as } \epsilon \rightarrow 0.
    \end{equation}
\end{theorem}
We now prove \Cref{theorem:eigenstates_dynamics}. By \Cref{lemma:eigenstate_s} and \Cref{lemma:eigenstate_s_adiabatic}, we have
\begin{equation}
    U(t_f)\ket{d_k} = \gamma (sI - D)^{-1}\ket{v},
\end{equation}
where eigenvalue $d_k < s < d_{k+1}$. For $0 < \mu < (d_k - d_i)/|v_i|^2$, by \Cref{corollary:epsilon_square_convergence}, we get that there exists $C_2 \geq 0$ such that
\begin{equation}
    s - d_k \leq C_2 \epsilon^2.
\end{equation}
Furthermore, we have:
\begin{equation}
    \frac{\epsilon}{s - d_k} \geq \frac{\epsilon}{C_2 \epsilon^2} = \frac{1}{C_2 \epsilon}
\end{equation}
It follows that the component $\braket{s}{d_k}$ of $\ket{s}$ tends to dominate all other components:
\begin{equation}
    \braket{s}{d_k} = \gamma \frac{v_k}{s - d_k} = \gamma \frac{\epsilon}{s - d_k} \rightarrow 1 \text{ as } \epsilon \rightarrow 0
\end{equation}
We remark that the normalization constant $\gamma$ prevents the equation above blow up by trending toward zero together with $\epsilon$. Therefore, $\ket{s} \rightarrow \ket{d_k}$ as $\epsilon \rightarrow 0$. This proves (\ref{ap_eq:conversion_dk_to_dk}).

Now, suppose that $(d_{k+1} - d_i)/|v_i|^2 < \mu$ and as before $d_k < s < d_{k+1}$. Then by \Cref{corollary:epsilon_square_convergence}, there exists $C_1$ such that:
\begin{equation}
    d_{k+1} - s < C_1 \epsilon^2
\end{equation}
It follows that
\begin{equation}
    \braket{s}{d_{k+1}} = \gamma \frac{v_{k+1}}{s - d_{k+1}} = \gamma\frac{\epsilon}{s - d_{k+1}} < - \gamma \frac{1}{C_1 \epsilon}.
\end{equation}
Hence, the component $\braket{s}{d_{k+1}}$ of $\ket{s}$ approaches $-1$ as $\epsilon$ tends to zero. Also, as $\epsilon$ tends to zero, so does $\gamma$. This proves (\ref{ap_eq:conversion_dk_to_dk+1}).

For a fixed $\mu \in \left(\frac{d_k - d_i}{|v_i|^2}, \frac{d_{k+1}-d_i}{|v_i|^2} \right)$, we have $d_k < s < d_{k+1}$. Therefore, the component 
\begin{equation}
\braket{s}{d_i} = \gamma \frac{v_i}{s - d_i} = \gamma \frac{1 - (N-1)\epsilon^2}{s - d_i}
\end{equation}
must converge to $1$ as $\epsilon$ tends to zero and $\gamma$ tends to $s - d_i$. This proves (\ref{ap_eq:conversion_dk_to_di}).

\section{Proof of \Cref{theorem:eigenstates_dynamics_alter}}
\begin{theorem}\label{ap_theorem:alternative}[Alternative]
    Let $\{d_j\}_{j=0}^{N-1}$ be an increasing sequence of eigenvalues of $D$.
     
    Fix index $i$ and let $\epsilon \in (0,1)$. Suppose $\ket{v}$ is a vector characterized by its components $v_{j}= \braket{v}{d_j}$ as follows:
    \begin{equation}\label{ap_eq:choice_of_v_alternative}
        v_{j} = 
        \begin{cases}
            1 - (N - 1) \epsilon^2 & \text{if } j = i \\
            \epsilon & \text{otherwise}
        \end{cases}
    \end{equation}
    For each $k \geq i$, define $\mu_{k+1}$ as:
    \begin{align}\label{ap_eq:critical_mu_k+1}
        \mu_{k+1} := \frac{d_{k+1}-d_{i}}{|v_{i}|^2}
    \end{align}
    Let $U(t)$ be the adiabatic evolution operator of $H(t)$. Then, in the limit as $\epsilon \to 0$, the following transitions occur:
    \begin{align}
        &U\left( t(\mu_{k+1} - \Delta \mu) \right)\ket{d_k} \rightarrow \ket{d_i}\\
        &U\left( t(\mu_{k+1} + \Delta \mu) \right)\ket{d_k} \rightarrow \ket{d_{k+1}} \\
        &U\left( t(\mu_{k+1} - \Delta \mu) \right)\ket{d_{k+1}} \rightarrow \ket{d_{k+1}} \\
        &U\left( t(\mu_{k+1} + \Delta \mu) \right)\ket{d_{k+1}} \rightarrow \ket{d_{i}}
    \end{align}
    In the above, $\Delta \mu$ is chosen such that
    \begin{equation}\label{ap_eq:delta_mu}
        0 < \Delta \mu < \min\{\mu_{k+2} - \mu_{k+1}, \mu_{k+1}-\mu_{k}\},
    \end{equation}
    where $\mu_{k} < \mu_{k+1} < \mu_{k+2}$.
\end{theorem}
\Cref{theorem:eigenstates_dynamics_alter} is a restatement of \Cref{theorem:eigenstates_dynamics} with a focus on the critical points $\mu_{k+1}:= (d_{k+1}-d_k)/|v_i|^2$ instead of the intervals given in \Cref{theorem:eigenstates_dynamics}. We only note that $\mu(t)$ is a monotonous function of time and hence is invertible. Therefore, for $\mu(t) = \mu_{k+1} \pm \Delta \mu$, we can solve for $t$ to get the critical time point $t(\mu_{k+1} \pm \Delta \mu)$ around which the eigenstate transitions are happening.

\section{Proof of \Cref{theorem:eigenstate_swap}}
\begin{theorem}[Eigenstates Swap]
    Let $d_{k+1}$ be an eigenvalue of $D$. Suppose that $\ket{v}$ is constructed as in (\ref{ap_eq:choice_of_v_alternative}).
    Let $\mu_{k+1}$ and $\Delta \mu >0$ be the critical point and increment of $\mu_{k+1}$ as described in (\ref{ap_eq:critical_mu_k+1}) and (\ref{ap_eq:delta_mu}), respectively. Then, in the limit $\epsilon \rightarrow 0$, we have the following relation:
    \begin{align*}
        \ket{s_{k}(\mu_{k+1} - \Delta \mu)} = \ket{v} \ \text{ and } \ \ket{s_{k+1}(\mu_{k+1} - \Delta \mu)} = \ket{d_{k+1}} \\
        \ket{s_{k}(\mu_{k+1} + \Delta \mu)} = \ket{d_{k+1}} \ \text{ and } \ \ket{s_{k+1}(\mu_{k+1} + \Delta \mu)} = \ket{v}
    \end{align*}
    Here, for the adiabatic evolution operator $U(t(\mu))$ of $H(t)$, we have:
    \begin{equation}
        \ket{s_k(\mu)} := U(t(\mu))\ket{d_k}
    \end{equation} 
\end{theorem}
This theorem is a succinct restatement of \Cref{ap_theorem:evolution,ap_theorem:alternative}.

\section{Proof of \Cref{theorem:min_gap_theorem}}
\begin{theorem}
    Let $\ket{v}$ be constructed as in (\ref{ap_eq:choice_of_v}).
    For index $k$, let $s_k(\mu(t)) < s_{k+1}(\mu(t))$ be any two consecutive eigenvalues of $H(t)$.
    Then, there exists $C_0>0$ such that to second order in $\epsilon$, we have: 
    \begin{equation}
        s_{k+1}(\mu(t)) - s_k(\mu(t)) \geq C_0 \epsilon^2  \text{ for all } t>0
    \end{equation}
\end{theorem}
For an index $k$, define $s_k:= s_k(\mu)$. Recall that for $\mu \geq 0$, $s_k \leq d_{k+1} \leq s_{k+1}$. We will show that for the construction of $\ket{v}$ as in \Cref{ap_theorem:evolution} there exists $C_0 > 0$ such that
\begin{align}
    d_{k+1} - s_k &> C_0 \epsilon^2, \label{eq:bound_1}\\
    s_{k+1} - d_{k+1} &\geq 0.\label{eq:bound_2}
\end{align}
Therefore, $d_{k+1} - s_k + s_{k+1} - d_{k+1} = s_{k+1} - s_{k} > C_0 \epsilon^2$.

For the case $\mu < 0$, we will simply have a vertically mirrored behaviour. That is
\begin{equation}
    s_{k-1}  \leq d_{k-1} \leq s_k \leq d_k,
\end{equation}
and it will follow that
\begin{align}
    s_k - d_{k-1} &> C_0 \epsilon^2,\\
    d_{k-1} - s_{k-1} &> 0.
\end{align}
Hence, $s_k - s_{k-1} > C_0 \epsilon^2$.

Since the case $\mu < 0$ is the mirrored version of the case $\mu \geq 0$, we prove the case for the positive $\mu$ for which $s_k \leq d_{k+1} \leq s_{k+1}$.

Without loss of generality, fix $i = 0$, then $v_0 = 1 - (N-1)\epsilon^2$ and $v_j = \epsilon$ for $j > 0$. Also, for notation clarity, fix $k = 0$. Then we show $s_1 - s_0 > C_0 \epsilon^2$.

Consider a characteristic polynomial $H(t) = D + \mu \ketbra{v}$ given by:
\begin{equation}
    p(s) = \prod_{j=0}^{N-1} (d_j - s) \left (1 + \mu\frac{|v_0|^2}{d_0 -s} + \mu \epsilon^2 \sum_{j=1}^{N-1} \frac{1}{d_j - s} \right )
\end{equation}
Due to the construction of $\ket{v}$ and $\mu \geq 0$, the eigenvalue $s_0$ of $H(t)$ is such that $d_0 \leq s_0 < d_{1}$. Hence, $p(s_0) = 0$. Since, the case $\mu=0$ is trivial, we consider $\mu > 0$. Then it follows that:
\begin{equation}\label{eq:char_poly_s0}
    1 + \mu\frac{|v_0|^2}{d_0 -s_0} + \mu \epsilon^2 \frac{1}{d_1 - s_0} + \mu \epsilon^2 \sum_{j=2}^{N-1} \frac{1}{d_j - s_0}  = 0
\end{equation}
We want to show that there exists $C_0 > 0$, such that $d_1 - s_0 > C_0 \epsilon^2$. To this end, rearrange (\ref{eq:char_poly_s0}) as follows:
\begin{equation}
     \frac{\epsilon^2}{d_1 - s_0} = - \frac{1}{\mu} - \frac{|v_0|^2}{d_0 - s_0} - \epsilon^2 \sum_{j=2}^{N-1}\frac{1}{d_j - s_0}
\end{equation}
Then,
\begin{equation}
    \frac{\epsilon^2}{d_1 - s_0} < -\frac{1}{\mu} -\frac{|v_0|^2}{d_0 - s_0} \nonumber = -\frac{1}{\mu} +\frac{|v_0|^2}{s_0 - d_0}.
\end{equation}
Suppose $s_0$ is closer to $d_1$, i.e., $s_0 > \frac{d_1 + d_0}{2}$, then
\begin{align}
    \frac{|v_0|^2}{s_0 - d_0} < \frac{|v_0|^2}{(d_1+d_0)/2 - d_0} = \frac{2|v_0|^2}{d_1 - d_0}.
\end{align}
It follows that
\begin{equation}
     \frac{\epsilon^2}{d_1 - s_0} < - \frac{1}{\mu} + \frac{2|v_0|^2}{d_1 - d_0} = \frac{2|v_0|^2 \mu - d_1 + d_0}{\mu (d_1 - d_0)}.
\end{equation}
Rearranging yields:
\begin{equation}
    \frac{\epsilon^2 \mu (d_1 - d_0)}{2 |v_0|^2 \mu + d_0 - d_1} < d_1 - s_0
\end{equation}
Note that for $\mu > (d_1 - d_0)/2|v_0|^2$, the denominator is finite and positive. Also, since $|v_0|^2 \mu < \mu$, we can write:
\begin{equation}
    \frac{\epsilon^2 \mu (d_1 - d_0)}{2 \mu + d_0 - d_1} < d_1 - s_0
\end{equation}
This shows that we can choose $C_0$ to be a function of $\mu > (d_1 - d_0)/2|v_0|^2$. However, we can also find a simpler but less tight bound which is independent of $\mu$. Note that $s_0$ is at its closest to $d_1$ for $\mu \rightarrow +\infty$. Therefore, we have
\begin{equation}
    \lim_{\mu \rightarrow +\infty} \frac{\epsilon^2 \mu (d_1 - d_0)}{2 \mu + d_0 - d_1} = \frac{\epsilon^2 (d_1 - d_0)}{2},
\end{equation}
and it follows that for $\mu >0$,
\begin{equation}
    \frac{\epsilon^2 (d_1 - d_0)}{2} < d_1 - s_0(\mu).
\end{equation}
Therefore, we can choose $C_0 = (d_1 - d_0)/2$.

We now show $s_1 - d_1 \geq 0$. Note that for $\mu = 0$, we get $s_1 = d_1$, and for $\mu > 0$, $s_1 - d_1 > 0$ because $v_j >0$ for all $j$. Hence, we have
\begin{align}
    s_1 - d_1 &\geq 0, \\
    d_1 - s_0 &> C_0 \epsilon^2.
\end{align}
Adding two inequalities together gives
\begin{equation}
    s_1 - s_0 > C_0 \epsilon^2.
\end{equation}

\section{Proof of Theorem~\ref{theorem:purity_and_renyi_coherent_info}}\label{proof:direct_channel}
We assume that the following Hamiltonian interaction governs the adiabatic evolution:
\begin{equation}\label{ap_eq:tripartite_hamiltonian}
    H(t) = I \otimes H_A \otimes I + I \otimes I \otimes H_B + \mu(t) I \otimes \ketbra{v}{v}
\end{equation}
The joint initial state of $\tilde{A}AB$ is:
\begin{equation}\label{ap_eq:max_ent_state_tildeAAB_for_purity}
    \ket{\psi_{\tilde{A}AB}} = \alpha_{ii}\ket{\tilde{a}_i}\ket{a_i}\ket{b_0} + \alpha_{jj}\ket{\tilde{a}_j}\ket{a_j}\ket{b_0}
\end{equation}

\setcounter{theorem}{9}
\begin{theorem}\label{ap_theorem:purity_and_renyi_coherent_info}
    Consider a tripartite quantum system $\tilde{A}AB$ initially prepared in a state described by (\ref{ap_eq:max_ent_state_tildeAAB_for_purity}). Assume that the system undergoes adiabatic evolution according to the Hamiltonian $H(t)$, as defined by (\ref{ap_eq:tripartite_hamiltonian}) with $\ketbra{v}$ being arbitrary. Then the purity-based coherent information for the direct channel $A \rightarrow A'$ is given by
    \begin{align}
        PI^d &= \nu \Tr\Big[ \Tr_B[\ketbra{s_{i0}}{s_{j0}}]  \Tr_B[\ketbra{s_{j0}}{s_{i0}}] \Big] \\ \nonumber
        &- \nu \Tr\Big[ \Tr_B[\ketbra{s_{i0}}{s_{i0}}]  \Tr_B[\ketbra{s_{j0}}{s_{j0}}] \Big],
    \end{align}
    where $\nu = 2 |\alpha_{ii}|^2 |\alpha_{jj}|^2$.
\end{theorem}
We commence with an initial state $\ket{\psi_{\tilde{A}AB}} \in \mathcal{H}_{\tilde{A}} \otimes \mathcal{H}_A \otimes \mathcal{H}_B$. Let $\{\tilde{a}_k\}$ be an arbitrary basis of $\mathcal{H}_{\tilde{A}}$. Let $\{\ket{a_m}\ket{b_n}\}$ be an eigenbasis given by Hamiltonians $H_A$ and $H_B$, which act on the Hilbert spaces $\mathcal{H}_A$ and $\mathcal{H}_B$, respectively. Then, for some fixed indices $i$ and $j$, the initial state is defined as
\begin{equation}
    \ket{\psi_{\tilde{A}AB}} = \left(\alpha_{ii}\ket{\tilde{a}_i}\ket{a_i} + \alpha_{jj}\ket{\tilde{a}_j}\ket{a_j}\right)\ket{b_0}.
\end{equation}
Here $\alpha_{ii}$ and $\alpha_{jj}$ are complex coefficients such that $|\alpha_{ii}|^2 + |\alpha_{jj}|^2 = 1$. We express the initial state in matrix density formalism, i.e., $\rho_{\tilde{A}AB} = \ketbra{\psi_{\tilde{A}AB}}{\psi_{\tilde{A}AB}}$. The goal is to compute the purity of the final state $\rho_{\tilde{A}'A'}$ of the evolved subsystem $\tilde{A}'A'$. 
Recall that given an initial eigenstate $\ket{a_k}\ket{b_0}$, the final eigenstate $\ket{s_{k0}}$ is given by Lemma~\ref{lemma:eigenstate_s}. That is, we have:
\begin{equation}
    \ket{s_{k0}} = \gamma (s_{k0}I - D)^{-1}\ket{v}
\end{equation}
Therefore, by Lemma~\ref{lemma:eigenstate_s}, the final state $\rho_{\tilde{A}'A'}$ is given by:
\begin{align}\label{eq:density_matrix_tildeAA}
    \rho_{\tilde{A}'A'} &= |\alpha_{ii}|^2 \ketbra{\tilde{a}_i}{\tilde{a}_i} \otimes \Tr_B[\ketbra{s_{i0}}{s_{i0}}] \\ \nonumber
    &+\alpha_{ii}\alpha_{jj}^* \ketbra{\tilde{a}_i}{\tilde{a}_j}\otimes \Tr_B[\ketbra{s_{i0}}{s_{j0}}]\\ \nonumber
    &+\alpha_{jj}\alpha_{ii}^* \ketbra{\tilde{a}_j}{\tilde{a}_i}\otimes \Tr_B[\ketbra{s_{j0}}{s_{i0}}]\\ \nonumber
    &+|\alpha_{jj}|^2 \ketbra{\tilde{a}_j}{\tilde{a}_j} \otimes \Tr_B[\ketbra{s_{j0}}{s_{j0}}]
\end{align}
In the above, we absorbed dynamical phases that arise during the adiabatic evolution into the definition of eigenbasis vectors $\ket{a_k}$. Alternatively, we could carry the dynamical phases throughout the analysis to find that they eventually simplify unity. Therefore, for ease of computation and clarity, we go with the former approach.
The square of the final state is:
\begin{align}
    \rho_{\tilde{A}'A'}^2 &= |\alpha_{ii}|^4 \ketbra{\tilde{a}_i}{\tilde{a}_i} \otimes \Tr_B[\ketbra{s_{i0}}{s_{i0}}]^2 \\ \nonumber
    &+|\alpha_{ii}|^2 \alpha_{ii}\alpha_{jj}^* \ketbra{\tilde{a}_i}{\tilde{a}_j} \otimes \Tr_B[\ketbra{s_{i0}}{s_{i0}}]\Tr_B[\ketbra{s_{i0}}{s_{j0}}] \\ \nonumber
    &+ |\alpha_{ii}|^2|\alpha_{jj}|^2\ketbra{\tilde{a}_i}{\tilde{a}_i} \otimes \Tr_B[\ketbra{s_{i0}}{s_{j0}}]\Tr_B[\ketbra{s_{j0}}{s_{i0}}] \\ \nonumber
    &+|\alpha_{jj}|^2 \alpha_{ii}\alpha_{jj}^* \ketbra{\tilde{a}_i}{\tilde{a}_j} \otimes \Tr_B[\ketbra{s_{i0}}{s_{j0}}]\Tr_B[\ketbra{s_{j0}}{s_{j0}}] \\ \nonumber
    &+|\alpha_{ii}|^2 \alpha_{jj}\alpha_{ii}^* \ketbra{\tilde{a}_j}{\tilde{a}_i} \otimes \Tr_B[\ketbra{s_{j0}}{s_{i0}}]\Tr_B[\ketbra{s_{i0}}{s_{i0}}] \\ \nonumber
    &+ |\alpha_{ii}|^2|\alpha_{jj}|^2\ketbra{\tilde{a}_j}{\tilde{a}_j} \otimes \Tr_B[\ketbra{s_{j0}}{s_{i0}}]\Tr_B[\ketbra{s_{i0}}{s_{j0}}] \\ \nonumber
    &+|\alpha_{jj}|^2 \alpha_{jj} \alpha_{ii}^* \ketbra{\tilde{a}_j}{\tilde{a}_i} \otimes \Tr_B[\ketbra{s_{j0}}{s_{j0}}] \Tr_B[\ketbra{s_{j0}}{s_{i0}}] \\ \nonumber
    &+|\alpha_{jj}|^4 \ketbra{\tilde{a}_j}{\tilde{a}_j} \otimes \Tr_B[\ketbra{s_{j0}}{s_{j0}}]^2 
\end{align}
Upon computing the trace of $\rho_{\tilde{A}'A'}^2$, and recognizing that $\Tr[\ketbra{\tilde{a}_m}{\tilde{a}_n}]$ yields the Kronecker delta $\delta_{mn}$, we can derive the expression for $P((\tilde{A}A)')$:
\begin{align}\label{eq:purity_evolved_joint_sys_proof}
    P\left((\tilde{A}A)'\right) &= |\alpha_{ii}|^4 \Tr\Big[ \big(\Tr_B[\ketbra{s_{i0}}{s_{i0}}]\big)^2\Big] \nonumber \\
    & + |\alpha_{jj}|^4 \Tr\Big[\big(\Tr_B[\ketbra{s_{j0}}{s_{j0}}]\big)^2\Big] \nonumber \\
    & + 2 |\alpha_{ii}|^2 |\alpha_{jj}|^2 \Tr\Big[ \Tr_B[\ketbra{s_{i0}}{s_{j0}}] \Tr_B[\ketbra{s_{j0}}{s_{i0}}]\Big]
\end{align}
Given (\ref{eq:density_matrix_tildeAA}) it is straightforward to compute $P(A')$. We first trace out $\tilde{A}$ and obtain:
\begin{multline}    \label{eq:density_matrix_A}
    \rho_{A'} = |\alpha_{ii}|^2 \Tr_B[\ketbra{s_{i0}}{s_{i0}}]
    +|\alpha_{jj}|^2 \Tr_B[\ketbra{s_{j0}}{s_{j0}}]
\end{multline}
Squaring $\rho_{A'}$ and taking its trace yields:
\begin{align}
    P(A') &= |\alpha_{ii}|^4 \Tr\Big[\big(\Tr_B[\ketbra{s_{i0}}{s_{i0}}]\big)^2\Big] \nonumber \\
    & + |\alpha_{jj}|^4 \Tr\Big[\big(\Tr_B[\ketbra{s_{j0}}{s_{j0}}]\big)^2\Big] \nonumber \\
    & + 2 |\alpha_{ii}|^2 |\alpha_{jj}|^2 \Tr\Big[\Tr_B[\ketbra{s_{i0}}{s_{i0}}] \Tr_B[\ketbra{s_{j0}}{s_{j0}}]\Big]
\end{align}
Subtracting $P(A')$ from $P((\tilde{A}A)')$ yields the result.

\section{Proof of \Cref{theorem:purity_and_renyi_coherent_info_complement}}
\begin{theorem}\label{ap_theorem:purity_and_renyi_coherent_info_complement}
    Under the assumptions stated in \Cref{theorem:purity_and_renyi_coherent_info}, the purity-based coherent information for the complementary channel $A \rightarrow B'$ is given by
    \begin{align}
        PI^c &= \nu \Tr\Big[ \Tr_A[\ketbra{s_{i0}}{s_{j0}}]  \Tr_A[\ketbra{s_{j0}}{s_{i0}}] \Big] \nonumber \\ 
        &- \nu \Tr\Big[ \Tr_A[\ketbra{s_{i0}}{s_{i0}}]  \Tr_A[\ketbra{s_{j0}}{s_{j0}}] \Big],
    \end{align}
    where $\nu = 2 |\alpha_{ii}|^2 |\alpha_{jj}|^2$.
\end{theorem}
The proof of \Cref{theorem:purity_and_renyi_coherent_info_complement} follows the exact steps of the proof in \Cref{proof:direct_channel} except we trace over $A$ instead of $B$. Therefore, the purity of $P((\tilde{A}B)')$ reads as:
    \begin{align}\label{eq:purity_evolved_joint_sys_AB}
        P\left((\tilde{A}B)'\right) &= |\alpha_{ii}|^4 \Tr\Big[ \big(\Tr_A[\ketbra{s_{i0}}{s_{i0}}]\big)^2\Big] \nonumber \\
        & + |\alpha_{jj}|^4 \Tr\Big[\big(\Tr_A[\ketbra{s_{j0}}{s_{j0}}]\big)^2\Big] \nonumber \\
        & + 2 |\alpha_{ii}|^2 |\alpha_{jj}|^2 \Tr\Big[ \Tr_A[\ketbra{s_{i0}}{s_{j0}}] \Tr_A[\ketbra{s_{j0}}{s_{i0}}]\Big]
    \end{align}
    Additionally, the marginal purity of \(B'\) is given by:
    \begin{align}
        P(B') &= |\alpha_{ii}|^4 \Tr\Big[\big(\Tr_A[\ketbra{s_{i0}}{s_{i0}}]\big)^2\Big] \nonumber \\
        & + |\alpha_{jj}|^4 \Tr\Big[\big(\Tr_A[\ketbra{s_{j0}}{s_{j0}}]\big)^2\Big] \nonumber \\
        & + 2 |\alpha_{ii}|^2 |\alpha_{jj}|^2 \Tr\Big[\Tr_A[\ketbra{s_{i0}}{s_{i0}}] \Tr_A[\ketbra{s_{j0}}{s_{j0}}]\Big]
    \end{align}
Subtracting $P(B')$ from $P((\tilde{A}B)')$ yields the final result.

\section{Proof of \Cref{lemma:purity_formula}}\label{proof:purity_formula}
\begin{lemma}\label{ap_lemma:purity_formula}
    Let $\rho_{AB} := \ketbra{s}{s}$ and $\rho_A := \Tr_B \left[\rho_{AB}\right]$, then the purity of $\rho_A$ is given by
    \begin{align}\label{ap_eq:purity}
        P(\rho_A) = \gamma^4 \sum_{ijkp} \frac{v_{ij}}{(s-d_{ij})} \frac{v^*_{kj}}{(s-d_{kj})} \frac{v_{kp}}{(s-d_{kp})}\frac{v^*_{ip}}{(s-d_{ip})},
    \end{align}
    where $v_{ij} = \bra{v}\ket{a_i, b_j}$ is a coefficient matrix of $\ket{v}$ in the eigenbasis $\{\ket{a_i} \ket{b_j}\}_{ij}$ of $D$ and
    \begin{equation}\label{ap_eq:gamma_normalization}
        \gamma = \left(\sum_{ij} \left|\frac{v_{ij}}{s - d_{ij}}\right|^2\right)^{-1/2}.
    \end{equation}
\end{lemma}
By \Cref{lemma:eigenstate_s} we have
\begin{equation}
    \ket{s} = \gamma \sum_{ij}\frac{v_{ij}}{s - d_{ij}} \ket{a_i}\ket{b_j},
\end{equation}
where $\{\ket{a_i}\ket{b_j}\}_{ij}$ is the eigenbasis of $D = H_A \otimes I + I \otimes H_B$.
It follows that:
\begin{equation}
    \rho_{AB} = \ketbra{s}{s} = \gamma^2 \sum_{ijkp}\frac{v_{ij}}{s-d_{ij}}\frac{v_{kp}^*}{s-d_{kp}}\ket{a_i}\ket{b_j}\bra{a_k}\bra{b_p}
\end{equation}
Tracing over $B$ yields:
\begin{equation}
    \rho_A := \Tr_B(\ketbra{s}) = \gamma^2 \sum_j \sum_{ik}\frac{v_{ij}}{s-d_{ij}}\frac{v_{kj}^*}{s-d_{kj}}\ketbra{a_i}{a_k}
\end{equation}
Squaring $\rho_A$ and tracing yields the result.

\section{Proof of \Cref{lemma:generalized_purity}}
The eigenvector of $H(t)$ is given by:
\begin{equation}\label{ap_eq:definition_of_ket_s}
    \ket{s} = \gamma (sI -D)^{-1}\ket{v}
\end{equation}

\begin{lemma}\label{ap_lemma:generalized_purity}
    Let $\ket{s}$, $\ket{s'}$, $\ket{s''}$ and $\ket{s'''}$ be eigenstates defined as in (\ref{ap_eq:definition_of_ket_s}). Then the following holds:
    \begin{multline}\label{ap_eq:generalized_purity}
        \Tr\Big[ \Tr_B[\ketbra{s}{s'}] \Tr_B[\ketbra{s''}{s'''}]\Big] =\\
        \gamma \gamma' \gamma'' \gamma''' \sum_{ijkp} \frac{v_{ij}}{(s-d_{ij})} \frac{v^*_{kj}}{(s'-d_{kj})} \frac{v_{kp}}{(s''-d_{kp})}\frac{v^*_{ip}}{(s'''-d_{ip})}
    \end{multline}
\end{lemma}
The proof follows the same steps as the proof in \Cref{proof:purity_formula} except we use \Cref{lemma:eigenstate_s} to compute $\ketbra{s}{s'}$ and $\ketbra{s''}{s'''}$  where $s, s', s'', s'''$ are eigenvalues of final Hamiltonian:
\begin{align*}
    \ketbra{s}{s'} &= \gamma\gamma' \sum_{ijkp}\frac{v_{ij}}{s-d_{ij}}\frac{v_{kp}^*}{s'-d_{kp}}\ket{a_i}\ket{b_j}\bra{a_k}\bra{b_p},\\
    \ketbra{s''}{s'''} &= \gamma''\gamma''' \sum_{ijkp}\frac{v_{ij}}{s''-d_{ij}}\frac{v_{kp}^*}{s'''-d_{kp}}\ket{a_i}\ket{b_j}\bra{a_k}\bra{b_p}
\end{align*}
Tracing over $B$ yields:
\begin{align*}
    \Tr_B(\ketbra{s}{s'}) &= \gamma \gamma' \sum_j \sum_{ik}\frac{v_{ij}}{s-d_{ij}}\frac{v_{kj}^*}{s'-d_{kj}}\ketbra{a_i}{a_k},\\
    \Tr_B(\ketbra{s''}{s'''}) &= \gamma'' \gamma''' \sum_j \sum_{ik}\frac{v_{ij}}{s''-d_{ij}}\frac{v_{kj}^*}{s'''-d_{kj}}\ketbra{a_i}{a_k}
\end{align*}
Computing the product $\Tr_B(\ketbra{s}{s'})\Tr_B(\ketbra{s''}{s'''})$ and tracing yields the result.

\end{document}